\documentclass[12pt]{article}
\usepackage[a4paper,margin=1in]{geometry}

\usepackage{verbatim, color, amssymb, amsmath}					
\usepackage{amsthm}		
\usepackage{bm}
\usepackage{dsfont}			
\usepackage{multirow}
\usepackage[mathscr]{euscript}
\usepackage{fancyhdr}
\usepackage{enumitem}
\usepackage{graphicx}
\usepackage{xcolor}
\usepackage{hyperref}
\usepackage{bbm}
\usepackage{mathtools}
\mathtoolsset{showonlyrefs=true}
\usepackage[section]{placeins}
\usepackage{tikz} 
\usetikzlibrary{bayesnet} 
\usetikzlibrary{arrows}

\usepackage{algorithm}
\usepackage{algpseudocode}

\usepackage{tabularx, booktabs}
\newcolumntype{Y}{>{\centering\arraybackslash}X}

\usepackage{subfig}
\usepackage{float}
\usepackage{lineno}

\usepackage[T1]{fontenc}
\usepackage{fix-cm}
\usepackage{setspace}
\usepackage{wrapfig}

\newif\ifblind 
\blindfalse

\usepackage{csquotes}
\usepackage[style=ext-authoryear, maxcitenames=1, giveninits=true, maxbibnames=99, sortcites=false, natbib, uniquename=init, uniquelist=false]{biblatex} 
\DeclareNameAlias{author}{family-given}
\DeclareNameAlias{editor}{family-given}

\renewbibmacro{in:}{%
	\ifentrytype{article}
	{}
	{\bibstring{in}%
		\printunit{\intitlepunct}}}
\DeclareFieldFormat[article]{pages}{#1}

\usepackage{lscape}

\newtheorem{Thm}{\underline{\bf Theorem}}

\newtheorem{Prop}{\underline{\bf Proposition}}
\newtheorem{Lem}{\underline{\bf Lemma}}

\def\bbR{\mathbb{R}}
\def\bbE{\mathbb{E}}

\def\var{\mathbb{V}\hbox{ar}}
\def\cov{\mathbb{C}\hbox{ov}}

\def\diag{\hbox{diag}}

\def\bone{{\mathbf 1}}

\def\ep{{\textsc{ep}}}
\def\adf{{\textsc{adf}}}
\def\new{{\textsc{new}}}

\DeclareRobustCommand{\sun}{%
  \ifmmode
    \text{\normalfont\textsc{sun}}%
  \else
    SUN%
  \fi
}

\def\Cat{\hbox{Cat}}

\def\Normal{\textsc{n}}

\def\Poi{\hbox{Poisson}}

\def\Unif{\hbox{Unif}}

\def\ba{{\mathbf a}}
\def\bA{{\mathbf A}}
\def\bb{{\mathbf b}}

\def\bc{{\mathbf c}}

\def\bD{{\mathbf D}}
\def\be{{\mathbf e}}
\def\b1f{{\mathbf f}}

\def\bG{{\mathbf G}}

\def\bI{{\mathbf I}}

\def\bJ{{\mathbf J}}

\def\bM{{\mathbf M}}

\def\bP{{\mathbf P}}
\def\bQ{{\mathbf Q}}
\def\br{{\mathbf r}}

\def\bs{{\mathbf s}}
\def\bS{{\mathbf S}}
\def\bt{{\mathbf t}}
\def\bu{{\mathbf u}}
\def\bU{{\mathbf U}}

\def\bV{{\mathbf V}}
\def\bw{{\mathbf w}}
\def\bW{{\mathbf W}}
\def\bx{{\mathbf x}}
\def\bX{{\mathbf X}}
\def\by{{\mathbf y}}
\def\bY{{\mathbf Y}}
\def\bz{{\mathbf z}}
\def\bZ{{\mathbf Z}}

\def\bzero{{\mathbf 0}}

\newcommand{\boldeta}{\mbox{\boldmath $\eta$}}
\newcommand{\bmu}{\mbox{\boldmath $\mu$}}
\newcommand{\bDelta}{\mbox{\boldmath $\Delta$}}

\newcommand{\bxi}{\mbox{\boldmath $\xi$}}
\newcommand{\bepsilon}{\mbox{\boldmath $\epsilon$}}

\newcommand{\bbeta}{\mbox{\boldmath $\beta$}}
\newcommand{\bgamma}{\mbox{\boldmath $\gamma$}}

\newcommand{\bSigma}{ \mathbf{\Sigma} }

\newcommand{\bomega}{\mbox{\boldmath $\omega$}}
\newcommand{\blambda}{\mbox{\boldmath $\lambda$}}
\newcommand{\bLambda}{\mbox{\boldmath $\Lambda$}}
\newcommand{\bOmega}{\mbox{\boldmath $\Omega$}}

\newcommand{\bGamma}{\mbox{\boldmath $\Gamma$}}

\def\simind{\stackrel{\mbox{\scriptsize{ind}}}{\sim}}
\def\simiid{\stackrel{\mbox{\scriptsize{iid}}}{\sim}}

\renewcommand\footnoterule{\kern-3pt \hrule \textwidth 2in \kern 2.6pt}

\def\boxit#1{\vbox{\hrule\hbox{\vrule\kern6pt \vbox{\kern6pt \textcolor{blue}{#1}\kern6pt}\kern6pt\vrule}\hrule}}

\def\authorfootnote#1{{\let\thefootnote\relax\footnotetext{#1}}}

\allowdisplaybreaks

\title{Online Inference for Dynamic Categorical Regression}
\date{}
\ifblind
\author{}
\else 
\author{Lorenzo Rimella \\
        \small{Economics Dept., University of Bergamo, Bergamo, IT}\\
        Augusto Fasano and Giovanni Rebaudo\\
        \small{ESOMAS Dept., University of Torino and Collegio Carlo Alberto, Torino, IT}\\
        }
\fi
        
\begin{document}

\def\spacingset#1{\renewcommand{\baselinestretch}%
{#1}\small\normalsize} \spacingset{1}

 \maketitle
\begin{refsection}

\begin{abstract}
\baselineskip=12pt
    The multinomial probit is a widely used model for categorical data in many applications, especially in econometrics and the discrete-choice literature.
    The computational challenges encountered when fitting this model still motivate ongoing research from the frequentist and Bayesian viewpoints.
    We consider a dynamic formulation based on a state-space model, where at each time one observes a sample from a multinomial probit whose parameter evolves according to Markovian dynamics, thereby inducing dependence across time.
    We show that the filtering and predictive distributions of the latent parameter belong to the unified skew-normal family, developing an associated i.i.d.\ sampler to approximate quantities of interest via Monte Carlo.
    Motivated by the computational bottlenecks of the sampler encountered for larger sample sizes, we also develop approximate methods for online inference based on assumed density filtering and expectation propagation.
    This gives more scalable, yet accurate, algorithms for online inference about the latent state and prediction of future observations.
    Finally, we derive a novel expectation maximization procedure to learn from the data the covariance matrix underlying the latent-utility representation of the multinomial probit.
    We investigate the performance of the proposed methods on simulated data and on a real dataset of reservations from an online booking platform.
\end{abstract}
\vskip 8mm


\spacingset{1.8}

\section{Introduction} 
\label{sec:1}
Discrete-choice models represent a popular and widely used tool to study preferences of agents.
Among them, the multinomial probit (MNP) model \citep{hausman1978conditional, daganzo1979multinomial} has gained widespread popularity across many fields, extending the original range of the social sciences and economics applications also to machine learning and related fields \citep{girolami2006variational, riihimaki2013nested, kindo2016multinomial, fasano2022class, loaiza2023fast}.
One of the most appreciated benefits of the MNP is its enhanced interpretability, arising from the latent covariate-dependent Gaussian utilities employed in the construction of the model.
This comes together with great flexibility, since the aforementioned latent Gaussian utilities can be correlated, avoiding the restrictive assumption of \textit{independence of irrelevant alternatives} (IIA) \citep{hausman1978conditional}.
See also \citet{greene2017econometric} and \citet{agresti2012categorical} for more details on the MNP and discrete-choice models in general.
When preferences of individuals are observed at different time instants, classical MNP constructions would fail to capture the time-dependence of the samples.
Using such a construction in this case would translate into the strong assumption that the effect of the predictors remains constant in time, which may be too restrictive.
We thus focus on dynamic formulations of the MNP, leveraging state-space models, where dependence across time is driven by some latent parameter evolving according to Markovian dynamics.
In the considered construction, at each time one observes a sample from an MNP model with time-specific parameter $\bbeta_{t}$, which is allowed to vary across time to consider possible variations of the effect of predictors on the class-probabilities for the output.
A key task is the ability to perform efficiently online inference, namely filtering and prediction, which boils down to the study of the conditional distributions $p(\bbeta_{t}\mid \by_{1:t})$ and $p(\bbeta_{t+1}\mid \by_{1:t})$, where $\by_{1:t}$ represents the set of all observed samples $\by_{1},\ldots,\by_{t}$ from time $1$ to $t$ and $\by_{s}=(y_{s,1},\ldots,y_{s,m_{s}})$ represents the sample of observations for time $s=1,\ldots,t$.
We then first show that the aforementioned distributions belong to the unified skew-normal (SUN) family \citep{arellano2006unification,azzalini2014skew}, developing an associated i.i.d.\ Monte Carlo sampler.
However, due to the form of the SUN distribution characterizing the filtering and predictive densities, such an approach becomes computationally unfeasible when the number of observations and/or possible choices, or classes, increases.
The form of the likelihood poses severe computational challenges also to classical particle filter approaches, as they would require the computation of a multivariate Gaussian cumulative distribution function (cdf) of size equal to the number of alternatives reduced by one, for each particle.
Moreover, as the dimension of the parameter $\bbeta_{t}$ increases, the curse of dimensionality arises \citep{rebeschini2015can}.

To overcome the above-mentioned problems encountered in sampling approaches, we then turn to scalable approximate methods whose key idea is to preserve as much as possible of the exact analytical structure delivered by the SUN filter, while replacing the increasingly costly exact recursions with tractable simpler SUN or Gaussian approximations.
This leads to two online procedures for dynamic MNP models. First, we propose an assumed density filter (ADF) \citep{sorenson1968non, minka2001family}, which projects the current filtering distribution onto a Gaussian approximation before moving to the next time step, resulting in a simpler SUN approximation for the subsequent filtering distribution.
To further improve computational scalability when the number of observations at each time point is large, we then develop a sequential expectation propagation (EP) scheme \citep{minka2001family, minka2001expectation}, which further approximates the likelihood contributions to obtain a fully Gaussian recursive procedure.
Taken together, these methods provide a toolkit ranging from exact Monte Carlo inference to approximations whose per-time cost grows only linearly in the number of current observations, making online inference feasible in regimes where the exact filter is not.

The above online filters are developed with a fixed covariance matrix of the latent utilities.
Since this quantity plays a crucial role in MNP models, we also derive a novel expectation maximization (EM) routine that estimates such a parameter by maximizing the marginal likelihood under the trace identifiability constraint defined in \citet{burgette2012trace}.
In the experiments, the estimation is done offline on a training horizon: the proposed EM procedure is shown to effectively retrieve the structure of the covariance matrix in simulation studies and is then used in the real-world application.

The paper is organized as follows.
Section~\ref{sec:2} provides notation and background on SUN distributions and the MNP model.
Section~\ref{sec:3} introduces the dynamic MNP model and derives the exact filtering, predictive, and marginal likelihood recursions.
Section~\ref{sec:4} introduces the Monte Carlo sampler for the exact filtering algorithm and the two novel algorithms along with their derivations.
Parameter inference via EM is then discussed in Section~\ref{sec:5}.
Synthetic data experiments and the real-world experiment on Expedia data are available in Section~\ref{sec:6}, while Section~\ref{sec:7} concludes with final comments.
Substantive additional details, including proofs of all theoretical results, are provided in the Supplementary Materials.

\section{Background}
\label{sec:2}

In this section, after introducing the notation, we briefly review two topics playing a central role in the manuscript, namely the SUN distribution and the MNP model.

\subsection{Notation}
\label{subsec:2.1}
    We denote by $\bI_{k}$ the $k\times k$ identity matrix.
    We use $\bzero_{p}$ for the $p$-dimensional column vector of all zeros, $\bone_{p}$ for the $p$-dimensional vector of all ones, and $\be_{l}^{L}=(e_{l,1}^{L},\ldots,e_{l,L}^{L})^\top$ for the $L$-dimensional column vector with components $e_{l,i}^{L}=0$ if $i\neq l$ and $e_{l,l}^{L}=1$.
    We use $\phi_{p}(\bbeta; \bSigma)$ for the probability density function (pdf) of a $p$-multivariate Gaussian with mean $\bzero_{p}$ and variance-covariance matrix $\bSigma$ evaluated at $\bbeta$, while we use $\Phi_{p}(\bbeta; \bSigma)$ for the cdf of a $p$-multivariate Gaussian with mean $\bzero_{p}$ and variance-covariance matrix $\bSigma$ evaluated at $\bbeta$.
    We denote by $\textsc{n}_{p}(\bmu,\bSigma)$ the distribution of a $p$-dimensional Gaussian random variable with mean $\bmu$ and variance-covariance matrix $\bSigma$ and by $\textsc{tn}_{p}(\bmu,\bSigma;\bc)$ the distribution of a $p$-variate Gaussian random variable with mean $\bmu$ and variance-covariance matrix $\bSigma$, truncated below $\bc\in \bbR^{p}$
    componentwise, so that the components are constrained to lie above $\bc$.
    We use
    $$\begin{bmatrix}
        \bx_{1} & \bx_{2} 
    \end{bmatrix} \text{ and } 
    \begin{bmatrix}
        \bx_{1} \\
        \bx_{2} 
    \end{bmatrix}
    $$
    for the operations of stacking two (column) vectors $\bx_{1}, \bx_{2}$ with the same dimension on the columns or rows of a matrix.
    The same notation is used when stacking together matrices, and the two notations can be combined to build block matrices.
    Finally, we use $\odot$ to denote the element-wise Hadamard product between vectors or matrices of conformable dimensions.

\subsection{The Multinomial Probit Model}
\label{subsec:2.2}

We start focusing on the static MNP model where the parameter $\bbeta$ is not indexed by time and we have a sample of observations $y_{i}\in\{0,\ldots,L\}$, $i=1,\ldots,m$, each one with a corresponding matrix of covariates $\bX_{i} = \left[\bx_{i1},\ldots,\bx_{iL}\right]^{\intercal}\in \bbR^{L\times p}$.
Details about the MNP model and its interpretation in terms of underlying latent utilities can be found, for instance, in \citet{hausman1978conditional, daganzo1979multinomial, fasano2022class, loaiza2023fast}.
According to this interpretation, each subject $i=1,\ldots,m$ receives a utility $z_{i\ell}$ associated with alternative $\ell=0,\ldots,L$, with $z_{i0}=0$ for identifiability purposes. Thus, the default alternative has null utility, while, for $\ell=1,\ldots,L$, $z_{i\ell}$ represents the utility of alternative $\ell$ relative to the default alternative.
This utility can depend on both class- and individual-specific quantities, encoded in the vector $\bx_{i\ell}$.
Specifically, for $i=1,\ldots, m$ and $\ell=1,\ldots, L$, the utility $z_{i\ell}$ takes the form $z_{i \ell}=\bx_{i \ell}^{\intercal} \bbeta + \epsilon_{i\ell}$, with $\bepsilon_{i} = (\epsilon_{i 1}, \ldots, \epsilon_{i L})^{\intercal} \sim \textsc{n}_{L} \left(\bzero_{L}, \bSigma \right)$ and $\bSigma$ positive definite.
When joint inference on $\bbeta$ and $\bSigma$ is desired, an identifiability constraint should also be put on $\bSigma$: although multiple options are possible, we follow \citet{burgette2012trace} and use a trace constraint on $\bSigma$ (see Section~\ref{sec:5}).
Denoting $\bz_{i} = (z_{i 1}, \ldots, z_{i L})^{\intercal} \in \bbR^{L}$, we can write the latent utilities in matrix notation as $\bz_{i} = \bX_{i} \bbeta + \bepsilon_{i}$.

Subject $i$ then chooses the alternative with maximal utility, i.e., $ y_{i} = \text{argmax}_{\ell \in \{0, \ldots, L\}} z_{i \ell}$.
Thus, for alternatives $\ell =1,\dots, L$, this entails the following likelihood:
\begin{equation}
    \begin{split}
    	 &\Pr(y_{i} = \ell \mid \bbeta)=\Pr(\bx_{i \ell}^{\intercal} \bbeta+ \epsilon_{i \ell}>\bx_{i k}^{\intercal}\bbeta+ \epsilon_{i k} \,\ \forall k \in\{1,\ldots,L\}\setminus\{\ell\} \,\ \& \,\ \bx_{i \ell}^{\intercal} \bbeta+ \epsilon_{i \ell}>0)\\
    	&\quad = \Pr(-\bV_{[\ell]} \bepsilon_{i} \le \bV_{[\ell]} \bX_{i} \bbeta \mid \bbeta) =\Phi_{L}\left ( \bV_{[\ell]} \bX_{i} \bbeta; \bV_{[\ell]} \bSigma \bV_{[\ell]}^{\intercal} \right ),
    \label{eq:compact_lik}
    \end{split}
\end{equation}
where $\bV_{[\ell]} =\left[\be_{\ell}^{L} - \be_{1}^{L}, \ldots, \ \be_{\ell}^{L} - \be_{\ell-1}^{L}, \ \be_{\ell}^{L} - \be_{\ell+1}^{L}, \ldots, \ \be_{\ell}^{L} - \be_{L}^{L}, \ \be_{\ell}^{L} \right]^{\intercal} \in \{0,-1,1\}^{L \times L}$, while for alternative $\ell=0$ one has
\begin{equation}
	\begin{split}
		\Pr(y_{i} = 0 \mid \bbeta)=\Pr(\bx_{i k}^{\intercal}\bbeta+ \epsilon_{i k} < 0, \,\ \forall k \in\{1,\ldots,L\}) =\Phi_{L}\left ( - \bX_{i} \bbeta; \bSigma \right ).
		\label{eq:compact_lik 0}
	\end{split}
\end{equation}

In essence, we consider a categorical likelihood whose class probabilities are computed according to \eqref{eq:compact_lik} and \eqref{eq:compact_lik 0}, meaning that, calling $\bV_{[0]}=-\bI_{L}$, we can write compactly
\begin{equation}
\begin{split}
 y_{i}\mid \bbeta \simind \Cat \left( \pi_{i,0},\ldots,\pi_{i,L} \right ), \text{ with } \pi_{i,\ell} = \Phi_{L}(\bV_{[\ell]}\bX_{i}\bbeta;\bV_{[\ell]}\bSigma\bV_{[\ell]}^{\intercal}),\quad \ell=0,\ldots,L.
\end{split}
\label{eq:compact_lik_full}
\end{equation}
When a model has likelihood \eqref{eq:compact_lik_full}, we write, for short, $y_{i}\mid\bbeta \simind \text{MNP}(\bbeta; \bX_{i}, \bSigma)$, $i=1,\ldots,m$.
The modeling advantages of the MNP model, like the above-mentioned avoidance of IIA and the high interpretability, come, however, with some computational difficulties associated with parameter inference.
These difficulties mainly arise from the form of the likelihood \eqref{eq:compact_lik_full}, which is a multivariate Gaussian cdf: while multiple relevant contributions are available on the topic \citep{genz1992numerical, trinh2015bivariate, botev2017normal,genton2018hierarchical, cao2019hierarchical, cao2021exploiting, fasano2025multivariate, cao2026linear}, dealing with multivariate Gaussian integrals still poses computational challenges.
As a consequence, since its development, the MNP has motivated extensive research on associated computational methods, both from frequentist \citep{mcfadden1989method, stern1992method,borsch1993smooth, geweke1994alternative, natarajan2000monte, ding2024computationally} and Bayesian standpoints.
The latter are of particular interest for the present setting, where the predictive distribution $p(\bbeta_{t}\mid \by_{1:t-1})$ naturally plays the role of the prior, which is then updated after observing the sample $\by_{t}$ to obtain the filtering distribution $p(\bbeta_{t}\mid \by_{1:t})$, which plays the role of the posterior instead.
Among this vast literature, early contributions mainly exploited MCMC sampling \citep{albert1993bayesian, mcculloch1994exact, nobile1998hybrid, mcculloch2000bayesian, chen2002discrete, imai2005bayesian}, while subsequent research considered both deterministic approximations \citep{girolami2006variational, girolami2006data, knowles2011non, riihimaki2013nested, loaiza2023fast} and sampling approaches \citep{burgette2012trace, johndrow2019mcmc, burgette2021symmetric, loaiza2022scalable}.
Recently, \citet{fasano2022class} showed that Bayesian formulations with Gaussian or SUN priors lead to a posterior distribution belonging to the SUN family, developing both an i.i.d.\ sampler and variational approximations.
See also \citet{anceschi2023bayesian} for a review of Bayesian models, including the MNP, leading to a SUN posterior and associated computational methods.
Due to its centrality for the subsequent results, the SUN distribution and its main properties are briefly recalled in Section~\ref{subsec:2.3} below.
The MNP model is then generalized to dynamic settings in Section~\ref{sec:3}.

\subsection{The Unified Skew-normal Distribution}
\label{subsec:2.3}
Following \citet{arellano2006unification}, a random vector $\bbeta \in \bbR^{p}$ has a SUN distribution, $\bbeta \sim \sun_{p,q}(\bxi,\bOmega,\bDelta,\bgamma,\bGamma)$, if its density function $p(\bbeta)$ can be expressed as
\begin{equation}
\label{eq:dSUN}
    \mathrm{d} \sun_{p,q}(\bbeta;\bxi,\bOmega,\bDelta,\bgamma,\bGamma) = \phi_{p}(\bbeta -\bxi; \bOmega) \frac{\Phi_{q}\left(\bgamma+\bDelta^{\intercal} \bar{\bOmega}^{-1} \bomega^{-1}(\bbeta-\bxi){;} \bGamma{-}\bDelta^{\intercal}\bar{\bOmega}^{-1}\bDelta \right)}{\Phi_{q}(\bgamma;\bGamma)},
\end{equation}
where the covariance matrix $\bOmega$ can be decomposed as $\bOmega=\bomega \bar{\bOmega} \bomega$, obtained by re-scaling the $p \times p$ correlation matrix $\bar{\bOmega}$ via the positive diagonal scale matrix $ \bomega=(\bOmega \odot \bI_{p})^{1/2}$.
Moreover, both $\bGamma$ and the matrix $\bOmega^{\ast}$, defined by the blocks $\bOmega_{[11]}^{\ast}=\bar{\bOmega}$, $\bOmega_{[12]}^{\ast}=\bOmega_{[21]}^{*\intercal}=\bDelta$, $\bOmega_{[22]}^{\ast}=\bGamma$,
are correlation matrices, and $\bDelta\in\bbR^{p\times q}$, $\bgamma\in\bbR^{q}$, and $\bGamma\in\bbR^{q\times q}$.

The SUN distribution has some appealing properties, like the closure under linear transformations and the following additive representation, which allows one to probabilistically characterize a random variable distributed as a $\sun_{p,q}$ with a linear combination of a $p$-variate Gaussian and a $q$-variate truncated Gaussian \citep{azzalini2014skew}.
Indeed, if $\bU_{0}\sim \textsc{n}_{p}(\bzero_{p}, \bar{\bOmega} - \bDelta \bGamma^{-1} \bDelta^{\intercal})$ and $\bU_{1}\sim\textsc{tn}_{q}(\bzero_{q},\bGamma; -\bgamma)$, with $\bU_{0}$ independent of $\bU_{1}$, then the random variable defined as
\begin{equation} \label{eq:additiveRepresentationSUN}
    \bU = \bxi + \bomega(\bU_{0} + \bDelta\bGamma^{-1}\bU_{1})
\end{equation}
has density \eqref{eq:dSUN}.
As clarified in the following, such an additive representation is particularly useful to develop i.i.d.\ samplers.
    
\section{Dynamic Multinomial Probit Online Inference}
\label{sec:3}

\subsection{Model} \label{subsec:model}
\begin{figure}
    \centering
    \begin{tikzpicture}
    
    \begin{scope}[scale=1, transform shape]
    \node[latent, minimum size=1cm] (theta0) {$\bbeta_{0}$};
    \node[latent, right=of theta0, minimum size=1cm] (theta1) {$\bbeta_{1}$};
    \node[latent, right=of theta1, minimum size=1cm, draw=none, fill=none] (dots1) {$\dots$};
    \node[latent, right=of dots1, minimum size=1cm] (thetat) {$\bbeta_{t}$};
    \node[latent, right=of thetat, minimum size=1cm, draw=none, fill=none] (dots2) {$\dots$};
    \node[latent, right=of dots2, minimum size=1cm] (thetan1) {$\bbeta_{T-1}$};
    \node[latent, right=of thetan1, minimum size=1cm] (thetan) {$\bbeta_{T}$};

    \node[obs, below=of theta1, minimum size=1cm] (y1) {$\by_{1}$};
    \node[obs, below=of thetat, minimum size=1cm] (yt) {$\by_{t}$};
    \node[obs, below=of thetan1, minimum size=1cm] (yn1) {$\by_{T-1}$};
    \node[obs, below=of thetan, minimum size=1cm] (yn) {$\by_{T}$};
    
    \edge {theta0} {theta1};
    \edge[dashed] {theta1} {dots1};
    \edge[dashed] {dots1} {thetat};
    \edge[dashed] {thetat} {dots2};
    \edge[dashed] {dots2} {thetan1};
    \edge {thetan1} {thetan};

    \edge {theta1} {y1};
    \edge {thetat} {yt};
    \edge {thetan1} {yn1};
    \edge {thetan} {yn};
    
    \end{scope}
    \end{tikzpicture}
    \caption{Graphical representation of the dynamic MNP model in \eqref{eq:DMNP_lik}--\eqref{eq:init_dist_MP}.
    Shaded nodes represent observed quantities.}
    \label{fig:DAG}
\end{figure}
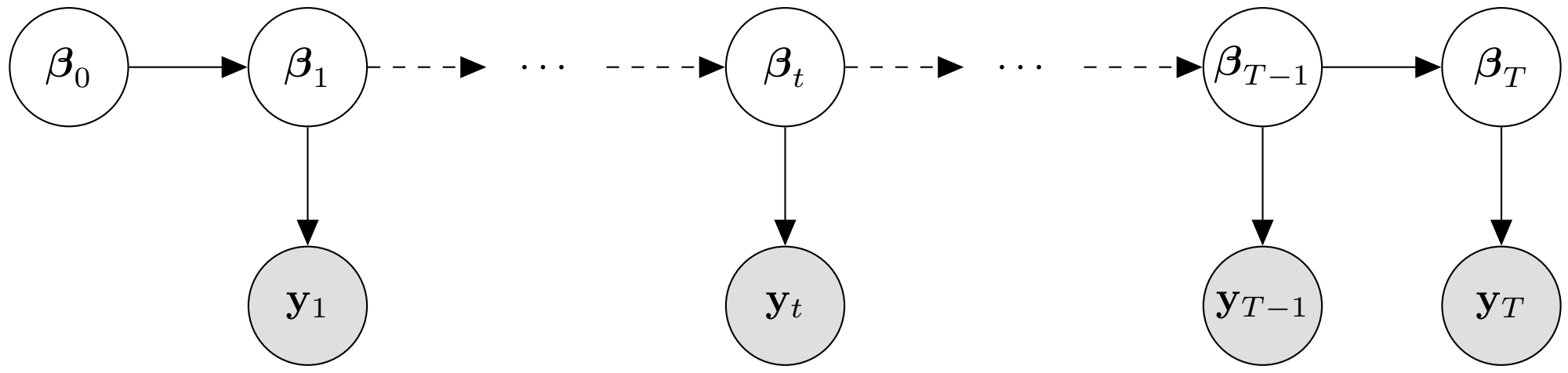

When observations are available for different time instants $t=1,\ldots,T$, a classical MNP model might not be able to capture the evolution of the effect of the covariates on the latent utilities, and hence on the output, across time.
Pooling all the observations together and fitting a single MNP model would not capture any dynamic behavior of the underlying parameters, while fitting separate MNP models at each time would not borrow any information from the observations across time.
This motivates the development of an extension of the static MNP model that is able to deal with dynamic coefficients, leveraging state-space models \citep{west2006bayesian, petris2009dynamic, durbin2012time, chopin2020introduction}, where the dependence of the observations across time is induced by a Markovian dynamic for the coefficients.
To this purpose, we consider the following dynamic formulation of the MNP, hereafter referred to as dynamic MNP:
\begin{align}
    &y_{t,i} \mid \bbeta_{t} \simind \text{MNP}(\bbeta_{t};\bX_{t,i},\bSigma), \quad i=1,\ldots,m_{t}, \label{eq:DMNP_lik}\\
    &\bbeta_{t}\mid \bbeta_{t-1} \sim \textsc{n}_{p}(\bG_{t}\bbeta_{t-1}, \bW_{t}), \quad t = 1, \ldots, T \label{eq:trans_kern_MP}\\
    &\bbeta_{0} \sim \textsc{n}_{p}\left ( \ba_{0}, \bP_{0}\right ), \label{eq:init_dist_MP}
\end{align}
where the dependence structure is defined by the directed acyclic graph in Figure~\ref{fig:DAG}.
In~\eqref{eq:DMNP_lik}, $\bX_{t,i}$ is the $L\times p$ covariate matrix associated to the $i$-th observation of time $t$, while $\ba_{0},\bP_{0},(\bG_{t})_{t= 1}^{T}$ and $(\bW_{t})_{t= 1}^{T}$ in \eqref{eq:trans_kern_MP}--\eqref{eq:init_dist_MP} are hyper-parameters of the model, with $\ba_{0} \in \bbR^{p}$, $\bG_{t} \in \bbR^{p\times p}$ non-singular, and $\bP_{0}$ and $\bW_{t}$ positive definite covariance matrices in $\bbR^{p\times p}$.
Notice that the sample size at time $t$ is $m_{t}$ and can vary across time.
Calling $\by_{t} = (y_{t,1},\ldots, y_{t,m_{t}})^{\intercal}$ and exploiting properties of products of multivariate Gaussian cdfs, the emission distribution \eqref{eq:DMNP_lik} equivalently translates to the following density:
\begin{align}
    &p\left ( \by_{t}\mid \bbeta_{t} \right ) = \prod_{i=1}^{m_{t}} \Phi_{L}\left ( \bV_{[y_{t,i}]}\bX_{t,i} \bbeta_{t}; \bV_{[y_{t,i}]} \bSigma \bV_{[y_{t,i}]}^{\intercal} \right )= \Phi_{m_{t} L}\left( \bar{\bX}_{t,\by_{t}} \bbeta_{t}; \bLambda_{t} \right), \label{eq:emission_MP}
\end{align}
where $\bar{\bX}_{t,\by_{t}} = \left[ \bX_{t,1}^{\intercal} \bV_{[y_{t,1}]}^{\intercal} \ \dots \ \bX_{t,m_{t}}^{\intercal} \bV_{[y_{t,m_{t}}]}^{\intercal}\right]^{\intercal}$ is an $(m_{t} L) \times p$ matrix obtained by stacking the $L\times p$ matrices $\bV_{[y_{t,1}]}\bX_{t,1},\ldots,\bV_{[y_{t,m_{t}}]}\bX_{t,m_{t}}$, and $\bLambda_{t}$ denotes an $( m_{t} L) \times (m_{t} L)$ block diagonal covariance matrix with $L \times L$ diagonal blocks given by $\bV_{[y_{t,i}]} \bSigma \bV_{[y_{t,i}]}^{\intercal}$, for every $i = 1, \ldots ,m_{t}$.
A particular role is played by $\bSigma$, the covariance matrix of the latent utility errors, for which, as anticipated, an EM algorithm is developed in Section~\ref{sec:5}.

As $(\bbeta_{t})_{t=0}^{T}$ arises from an unobserved Markov process and the $y_{t,i}$'s are conditionally independent of all other variables given $\bbeta_{t}$, we can conclude that $(\bbeta_{t})_{t\geq 0},(\by_{t})_{t\geq 1}$ is a state-space model, which motivates the development of tools for online inference.

\subsection{Exact Filtering and Marginal Likelihood}

Online inference in state-space models generally boils down to the sequential estimation of the filtering distribution $p(\bbeta_{t}\mid \by_{1:t})$ and related quantities, like, e.g., the one-step-ahead predictive distributions for the parameter $p(\bbeta_{t+1}\mid \by_{1:t})$ or the observations $p(\by_{t+1}\mid \by_{1:t})$ \citep{durbin2012time}.
One-step-ahead prediction is of key interest to devise sequential procedures, so we will mainly focus on this in the following, but the methodology naturally extends to the $k$-step-ahead prediction setting.
The filtering distribution is updated sequentially thanks to a recursive application of a \textit{prediction} and a \textit{correction} step when a new sample of observations $\by_{t}$ is observed.
In the prediction step the filtering distribution $p(\bbeta_{t-1}\mid\by_{1:t-1})$ is updated via the Markov transition kernel $p\left ( \bbeta_{t} \mid \bbeta_{t-1} \right )$ to the predictive distribution $p(\bbeta_{t}\mid \by_{1:t-1})$:
\begin{equation*}
    p(\bbeta_{t}\mid\by_{1:t-1}) = \int p\left ( \bbeta_{t} \mid \bbeta_{t-1} \right ) p(\bbeta_{t-1}\mid\by_{1:t-1}) d \bbeta_{t-1}.
\end{equation*}
The correction step consists of the computation of the marginal likelihood increment $p(\by_{t}\mid\by_{1:t-1})$ and an application of Bayes' theorem to update $p(\bbeta_{t}\mid\by_{1:t-1})$:
\begin{equation*}
    p(\by_{t}\mid\by_{1:t-1}) = \int p(\by_{t}\mid\bbeta_{t}) p(\bbeta_{t}\mid\by_{1:t-1}) d \bbeta_{t} \text{ and } p(\bbeta_{t}\mid\by_{1:t})= \frac{p(\by_{t}\mid\bbeta_{t}) p(\bbeta_{t}\mid\by_{1:t-1})}{p(\by_{t}\mid\by_{1:t-1})}.
\end{equation*}
A crucial point for such an online learning procedure to be feasible is that the relevant quantities of interest must be available in closed form and easily computable.
A key contribution of this section is to show that, for the dynamic MNP model in \eqref{eq:DMNP_lik}--\eqref{eq:init_dist_MP}, both the prediction and correction steps admit closed-form expressions by exploiting the properties of the SUN family, extending the theoretical results for dynamic binary probit models in \citet{fasano2021closed}.
To inductively show the prediction and correction forms for generic time $t$, let us start from $t=1$.
The first prediction follows standard properties of multivariate Gaussian distributions, resulting in 
\begin{equation}\label{eq:iid_filter_pred_0}
    p(\bbeta_{1}) = \phi_{p} \left ( \bbeta_{1} - \ba_{1\mid0}; \bP_{1\mid0} \right ),
\end{equation}
with $\ba_{1\mid0} = \bG_{1} \ba_{0}$ and $\bP_{1\mid0} = \bG_{1} \bP_{0} \bG_{1}^{\intercal} + \bW_{1}$.

The first correction instead leads to a SUN filtering distribution, as shown in the following proposition, providing an inductive basis.

\begin{Prop}\label{prop:iid_filter_pdt_0}
    Consider the emission distribution \eqref{eq:emission_MP} for $t=1$ and the Gaussian prediction $p(\bbeta_{1})$ as in \eqref{eq:iid_filter_pred_0}.
    Then:
    \begin{equation*}
    p(\by_{1}) = \Phi_{m_{1} L}(\bgamma_{1}; \bGamma_{1} ) \quad \text{and} \quad p(\bbeta_{1}\mid\by_{1}) = \mathrm{d} \sun_{p,m_{1} L}(\bbeta_{1}; \bxi_{1}, \bOmega_{1}, \bDelta_{1}, \bgamma_{1}, \bGamma_{1}),
    \end{equation*}
    where $\bxi_{1}=\ba_{1\mid0}$, $\bOmega_{1}=\bP_{1\mid0}$, $\bDelta_{1}=\bar{\bOmega}_{1}{\bomega}_{1} \bar{\bX}_{1,\by_{1}}^{\intercal} \bs_{1}^{-1}$, $\bgamma_{1} = \bs_{1}^{-1} \bar{\bX}_{1,\by_{1}} \bxi_{1}$, $\bGamma_{1} = \bs^{-1}_{1} (\bar{\bX}_{1,\by_{1}} \bOmega_{1} \bar{\bX}_{1,\by_{1}}^{\intercal}+\bLambda_{1})\bs^{-1}_{1}$,
    with $\bs_{1}= \left[ \left( \bar{\bX}_{1,\by_{1}} \bOmega_{1} \bar{\bX}_{1,\by_{1}}^{\intercal}+\bLambda_{1} \right) \odot \bI_{m_{1} L} \right]^{1/2}$, $\bOmega_{1} = \bomega_{1} \bar{\bOmega}_{1} \bomega_{1}$ and $\bomega_{1} =\left(\bOmega_{1} \odot \bI_{p}\right)^{1/2}$.
\end{Prop}

Since the first filtering distribution is SUN, if we can recursively handle the prediction and correction starting from a SUN filtering distribution, we have an exact filtering algorithm.
This is indeed possible and stated in the following theorem.

\begin{Thm}\label{thm:iid_filter}
    Consider the dynamic MNP model  \eqref{eq:DMNP_lik}--\eqref{eq:init_dist_MP} and let $M_{t} = \sum_{s=1}^{t} m_{s}$. If $ p(\bbeta_{t-1}\mid \by_{1:t-1}) = \mathrm{d} \sun_{p,M_{t-1} L}(\bbeta_{t-1}; \bxi_{t-1}, \bOmega_{t-1}, \bDelta_{t-1}, \bgamma_{t-1}, \bGamma_{t-1})$, then after the prediction step one gets:
    \begin{equation*}
    	p(\bbeta_{t} \mid \by_{1:t-1}) = \mathrm{d} \sun_{p,M_{t-1}L}(\bbeta_{t} ; \bxi_{t\mid t-1}, \bOmega_{t\mid t-1}, \bDelta_{t\mid t-1}, \bgamma_{t\mid t-1}, \bGamma_{t\mid t-1}),
    \end{equation*}
    with $\bxi_{t\mid t-1}=\bG_{t} \bxi_{t-1}$, $\bOmega_{t\mid t-1}=\bG_{t}\bOmega_{t-1}\bG_{t}^{\intercal}+\bW_{t}$, $\bDelta_{t\mid t-1} = \bomega_{t\mid t-1}^{-1} \bG_{t}\bomega_{t-1}\bDelta_{t-1}$, $\bgamma_{t\mid t-1} = \bgamma_{t-1}$, $\bGamma_{t\mid t-1} = \bGamma_{t-1}$, and $\bomega_{t\mid t-1}= (\bOmega_{t\mid t-1}\odot\bI_{p})^{1/2}$.
    After the correction step, one gets:
    \begin{equation*}
        p(\by_{t} \mid \by_{1:t-1}) = \frac{\Phi_{M_{t} L }(\bgamma_{t}; \bGamma_{t} )}{\Phi_{M_{t-1} L}(\bgamma_{t\mid t-1}; \bGamma_{t\mid t-1} )} \text{ and } p(\bbeta_{t} \mid \by_{1:t}) = \mathrm{d} \sun_{p,M_{t}L}(\bbeta_{t}; \bxi_{t}, \bOmega_{t}, \bDelta_{t}, \bgamma_{t}, \bGamma_{t}),
    \end{equation*}
    where $\bxi_{t} = \bxi_{t\mid t-1}$, $\bOmega_{t} = \bOmega_{t\mid t-1}$, $\bDelta_{t} = \begin{bmatrix} \bDelta_{t\mid t-1} & \bar{\bOmega}_{t}\bomega_{t}\bar{\bX}_{t,\by_{t}}^{\intercal} \bs_{t}^{-1} \end{bmatrix}$,
    \begin{align*}
    &\bgamma_{t} = \begin{bmatrix} \bgamma_{t \mid t-1} \\
        \bs_{t}^{-1}\bar{\bX}_{t,\by_{t}}\bxi_{t} \end{bmatrix}, \quad \bGamma_{t} = 
        \begin{bmatrix}
        \bGamma_{t\mid t-1} & \bDelta_{t\mid t-1}^{\intercal} \bomega_{t} \bar{\bX}_{t,\by_{t}}^{\intercal}\bs_{t}^{-1}\\
        \bs_{t}^{-1} \bar{\bX}_{t,\by_{t}} \bomega_{t} \bDelta_{t\mid t-1} & \bs_{t}^{-1}(\bar{\bX}_{t,\by_{t}}\bOmega_{t}\bar{\bX}_{t,\by_{t}}^{\intercal} + \bLambda_{t}) \bs_{t}^{-1}
        \end{bmatrix}
    \end{align*}
    with $\bs_{t}= \left[ \left( \bar{\bX}_{t,\by_{t}} \bOmega_{t} \bar{\bX}_{t,\by_{t}}^{\intercal}+\bLambda_{t} \right) \odot \bI_{m_{t} L} \right]^{1/2}$, $\bOmega_{t} = \bomega_{t} \bar{\bOmega}_{t} \bomega_{t}$, and $\bomega_{t} =\left(\bOmega_{t} \odot \bI_{p}\right)^{1/2}$.
\end{Thm}

For a given time horizon $T$, the marginal likelihood of the whole sample $p(\by_{1:T})$ can then be computed as $p(\by_{1:T})=\prod_{t=1}^{T} p(\by_{t}\mid\by_{1:t-1})$.
Consequently, exploiting the resulting telescoping cancellations, the marginal likelihood boils down to 
$p(\by_{1:T})= \Phi_{M_{T} L }(\bgamma_{T}; \bGamma_{T})$.
 
\section{Approximate Statistical Inference}\label{sec:4}

Performing the calculations in Proposition~\ref{prop:iid_filter_pdt_0} and then the ones in Theorem~\ref{thm:iid_filter} recursively over time allows for exact computation of the predictive distribution, the filtering distribution, and the marginal likelihood.
In particular, for every $t$, both the densities $p(\bbeta_{t} \mid \by_{1:t-1})$ and $p(\bbeta_{t} \mid \by_{1:t})$ are available in closed form through analytically updated SUN parameters.
Therefore, the dynamic MNP model admits an exact solution to online inference, extending to the present setting the same type of analytical characterization that is provided by the Kalman filter \citep{kalman1960approach} in Gaussian state-space models.

A further implication of Theorem~\ref{thm:iid_filter} is that the moments of the exact filtering distribution can also be expressed analytically as functions of the SUN parameters defining $(\bbeta_{t} \mid \by_{1:t}) \sim \sun_{p,M_{t} L}(\bxi_{t},\bOmega_{t},\bDelta_{t},\bgamma_{t},\bGamma_{t}),$ see, e.g., \citet{azzalini2010prospective, azzalini2014skew, arellano2022some}.
However, these expressions involve multivariate Gaussian cdfs and their derivatives, with dimension of the order $M_{t} L$.
As a consequence, although the recursive filter is exact, its direct implementation rapidly becomes computationally demanding and potentially numerically unstable as either the number of time points $T$ or the number of observations per time point $m_{t}$ increases.
See, e.g., \citet{genz1992numerical, genton2018hierarchical, cao2021exploiting, fasano2025multivariate} for efficient methods to evaluate multivariate Gaussian cdfs.

These considerations, together with the practical need to evaluate arbitrary functionals of interest of the filtering and predictive distributions of the state parameters, motivate the inferential strategies developed in the next sections.
For simplicity, we focus on algorithms that approximate the filtering distribution $p(\bbeta_{t}\mid\by_{1:t})$.
These can be modified to provide approximations of the predictive distribution $p(\bbeta_{t}\mid\by_{1:t-1})$ exploiting \eqref{eq:trans_kern_MP}.

\subsection{Sampling i.i.d.\ from the Filtering Distribution}

The exact SUN characterization derived in Proposition~\ref{prop:iid_filter_pdt_0} and Theorem~\ref{thm:iid_filter} naturally induces an exact i.i.d.\ Monte Carlo sampler for the predictive and filtering distributions.
Focusing on the filtering as anticipated, for each time $t$, the distribution $p(\bbeta_{t} \mid \by_{1:t})$ is available in closed form within the SUN family, with parameters updated analytically through the exact recursion.
This makes it possible to generate independent draws directly from the corresponding target law by exploiting the additive representation in \eqref{eq:additiveRepresentationSUN}.

The resulting sampler is summarized in Algorithm~\ref{alg:IID}.
In practice, the computational bottleneck lies in sampling from the multivariate truncated Gaussian component appearing in the additive representation, for which we rely on the accept-reject sampler of \citet{botev2017normal}, as implemented in the R package \texttt{TruncatedNormal}.
This choice allows direct simulation from multivariate truncated Gaussian distributions, preserving the i.i.d.\ nature of the Monte Carlo output.
In our applications, this step remains practically feasible when the truncation dimension is of a few hundreds, but it can become unfeasible beyond that regime.

Once i.i.d.\ samples are available, arbitrary functionals of the filtering distribution can be approximated by standard Monte Carlo averages.
Specifically, for any integrable function $h(\cdot)$ and samples $\bbeta_{t}^{(1)}, \ldots, \bbeta_{t}^{(R)}$ drawn from $p(\bbeta_{t} \mid \by_{1:t})$, the corresponding expectation can be estimated as $\bbE\left[ h(\bbeta_{t}) \mid \by_{1:t}\right] \approx \tfrac{1}{R}\sum_{r=1}^{R} h \left(\bbeta_{t}^{(r)}\right)$, without introducing any approximation beyond the asymptotically (in $R$) vanishing Monte Carlo error.

\begin{algorithm}[httb!]
\caption{i.i.d.\ filter} \label{alg:IID}
    \begin{algorithmic}
    \Require $\ba_{0},\bP_{0},(\bG_{t})_{t=1}^{T},(\bW_{t})_{t=1}^{T},\bSigma$, and $R$ (number of Monte Carlo samples)\vspace{0.1cm}
    \For{$t \in 1,\dots,T$}\vspace{0.1cm}
            \State Compute $\bxi_{t}, \bOmega_{t}, \bDelta_{t}, \bgamma_{t}, \bGamma_{t}$ from Proposition~\ref{prop:iid_filter_pdt_0} if $t=1$ or Theorem~\ref{thm:iid_filter} otherwise \vspace{0.1cm}
            \State Sample i.i.d.\ from the filtering distribution via the representation in \eqref{eq:additiveRepresentationSUN}
        $$\bbeta_{t}^{(r)} \simiid \sun_{p,M_{t}L}(\bxi_{t},\bOmega_{t},\bDelta_{t},\bgamma_{t},\bGamma_{t}), \text{ for } r=1, \ldots, R$$
    \EndFor
    \State Compute the marginal likelihood $p(\by_{1:T})= \Phi_{M_{T} L }(\bgamma_{T}; \bGamma_{T})$
    \end{algorithmic}
\end{algorithm}
Algorithm~\ref{alg:IID} should therefore be regarded as an i.i.d.\ sampling procedure built on top of the analytical SUN filter.
Its limitation is computational.
As $t$ increases, the truncation dimension in the SUN additive representation grows as $M_{t}L$, making exact sampling increasingly demanding.
Computing the exact marginal likelihood additionally requires evaluating an $M_{t}L$-dimensional Gaussian cdf.
For this reason, the exact i.i.d.\ sampler is mainly useful in small-to-moderate scale settings, where it provides the benchmark for assessing the approximate methods developed below.

Finally, note that a natural alternative in state-space models is sequential Monte Carlo, and in particular the bootstrap particle filter \citep{gordon1993novel, chopin2020introduction}, which propagates particles through the transition kernel \eqref{eq:trans_kern_MP} and reweights them according to the emission density \eqref{eq:emission_MP}.
In the present setting, however, this still requires, for each particle $r$ and each time $t$, the evaluation of a likelihood contribution given by a product of $m_{t}$ Gaussian cdfs of dimension $L$.
Hence, particle filtering also inherits a substantial computational burden from the MNP observation equation while having a lower effective sample size compared to i.i.d.\ sampling from the exact filtering distribution, making it comparatively less appealing in our setting.
These computational issues would also appear if extensions of particle filters tailored for dynamic binary probit models are considered \citep{andrieu2002particle,fasano2021closed}.
See Section~\ref{sec:sis_supp} of the Supplementary Materials for implementation details and Section~\ref{sec:exp_filtering} for empirical comparisons.

\subsection{Assumed Density Filtering}
\label{subsec:ADF}

As noted by \citet{minka2001expectation}, assumed density filtering (ADF) was developed independently in multiple literatures, ranging from statistics \citep{lauritzen1992propagation}, to artificial intelligence \citep{boyen1998tractable, opper1999bayesian} and control \citep{maybeck1982stochastic}, possibly under different names.
Due to its sequential nature, it finds its best fit in dynamic models, although it can also be employed in static settings for approximate Bayesian posterior inference, where it can be seen as a single-pass expectation propagation: see \citet{minka2001expectation} for details.
Considering the filtering problem, the principle of ADF is to perform approximate filtering by processing the data sequentially in time, doing a Bayes update, and then projecting the resulting filtering back into a manageable form (which in our case will be a Gaussian density) before moving to the next time instant.
The projection is performed by minimizing a Kullback-Leibler divergence (KL) criterion, which reduces to moment-matching when the Gaussian class is considered as in the present case.

More specifically, at each time $t$ an approximate filtering distribution $h_{t}^{\adf}(\bbeta_{t})$ belonging to the $\sun_{p,m_{t}L}$ family is sequentially obtained.
This leads to a SUN-approximated filtering distribution which can account for skewness, but is much more tractable than the true filtering.
Considering the sequential prediction and correction recursion, Proposition~\ref{prop:iid_filter_pdt_0} guarantees that the filtering distribution at time $t=1$ is already of the required form, hence $h_{1}^{\adf}(\bbeta_{1})=p(\bbeta_{1}\mid\by_{1})$.
The ADF procedure is then applied sequentially from step $t=2$ onwards, as follows.

For $t\ge 2$, the ADF recursion first projects the previous approximate filtering distribution
$h_{t-1}^{\adf}(\bbeta_{t-1})$ onto the Gaussian family by matching its first two moments.
This gives $q_{t-1}^{\adf}(\bbeta_{t-1}) =\phi_p(\bbeta_{t-1}-\ba_{t-1}^{\adf};\bP_{t-1}^{\adf})$,
where $\ba_{t-1}^{\adf}$ and $\bP_{t-1}^{\adf}$ equal the mean and covariance matrix of
$h_{t-1}^{\adf}(\bbeta_{t-1})$.
Combining this Gaussian approximation with the transition kernel in \eqref{eq:trans_kern_MP} yields the approximate predictive distribution
\begin{equation}\label{eq:predADF}
    q_{t\mid t-1}^{\adf}(\bbeta_{t}) = \phi_{p}\left(\bbeta_{t} -\ba_{t\mid t-1}^{\adf}; \bP_{t\mid t-1}^{\adf} \right), \quad \ba_{t\mid t-1}^{\adf} = \bG_{t} \ba_{t-1}^{\adf}, \quad \bP_{t\mid t-1}^{\adf}= \bG_{t} \bP_{t-1}^{\adf} \bG_{t}^{\intercal} + \bW_{t}.
\end{equation}
The correction step then combines $q_{t\mid t-1}^{\adf}(\bbeta_{t})$ with the likelihood contribution
$p(\by_{t}\mid \bbeta_{t})$.
The following proposition shows that this update remains in the SUN family and provides the corresponding approximation to the marginal likelihood increment $p(\by_t\mid \by_{1:t-1})$.

\begin{Prop}\label{prop:ADFcorrection}
For any $t\ge 2$, let $h_{t-1}^{\adf}(\bbeta_{t-1})$ be the ADF approximation of the filtering distribution $p(\bbeta_{t-1}\mid\by_{1:t-1})$ obtained at time $t-1$, with $h_{1}^{\adf}(\bbeta_{1})=p(\bbeta_{1}\mid \by_{1})$.
Call its Gaussian approximation $q_{t-1}^{\adf}(\bbeta_{t-1}) = \phi_{p}\left (\bbeta_{t-1} -\ba_{t-1}^{\adf}; \bP_{t-1}^{\adf} \right )$, with $\ba_{t-1}^{\adf}=\mathbb{E}_{h_{t-1}^{\adf}(\bbeta_{t-1})}[\bbeta_{t-1}]$ and $\bP_{t-1}^{\adf}=\var_{h_{t-1}^{\adf}(\bbeta_{t-1})}[\bbeta_{t-1}]$, as above, and $q_{t\mid t-1}^{\adf}(\bbeta_{t})=\phi_{p}\left (\bbeta_{t} -\ba_{t\mid t-1}^{\adf}; \bP_{t\mid t-1}^{\adf} \right )$ the resulting approximate Gaussian predictive distribution for $\bbeta_{t}$, where $\ba_{t\mid t-1}^{\adf} = \bG_{t} \ba_{t-1}^{\adf}$ and $\bP_{t\mid t-1}^{\adf}= \bG_{t} \bP_{t-1}^{\adf} \bG_{t}^{\intercal} + \bW_{t}$.
Then, the resulting updated ADF approximate filtering  following from the correction step is
\begin{equation*}
    h_{t}^{\adf}(\bbeta_{t})=\mathrm{d} \sun_{p,m_{t} L}(\bbeta_{t}; \bxi_{t}^{\adf}, \bOmega_{t}^{\adf}, \bDelta_{t}^{\adf}, \bgamma_{t}^{\adf}, \bGamma_{t}^{\adf}),
\end{equation*}
where $\bxi_{t}^{\adf}=\ba_{t\mid t-1}^{\adf},\ \bOmega_{t}^{\adf}=\bP_{t\mid t-1}^{\adf},\ \bDelta_{t}^{\adf}=\bar{\bOmega}_{t}^{\adf}{\bomega}_{t}^{\adf} \bar{\bX}_{t,\by_{t}} ^{\intercal} (\bs_{t}^{\adf})^{-1},\ \bgamma_{t}^{\adf} = (\bs_{t}^{\adf})^{-1} \bar{\bX}_{t,\by_{t}} \bxi_{t}^{\adf}$, and $\bGamma_{t}^{\adf} = (\bs_{t}^{\adf})^{-1} (\bar{\bX}_{t,\by_{t}} \bOmega_{t}^{\adf}\bar{\bX}_{t,\by_{t}}^{\intercal}+\bLambda_{t})(\bs_{t}^{\adf})^{-1}$, with $\bs_{t}^{\adf}= \left[ \left( \bar{\bX}_{t,\by_{t}} \bOmega_{t}^{\adf} \bar{\bX}_{t,\by_{t}}^{\intercal}+\bLambda_{t} \right) \odot \bI_{m_{t} L} \right]^{1/2}$, $\bOmega_{t}^{\adf} = \bomega_{t}^{\adf} \bar{\bOmega}_{t}^{\adf} \bomega_{t}^{\adf}$ and $\bomega_{t}^{\adf} =\left(\bOmega_{t}^{\adf} \odot \bI_{p}\right)^{1/2}$.

Moreover, the ADF approximation of the marginal likelihood contribution $p(\by_{t}\mid \by_{1:t-1})$ equals $\widehat{p}_{\adf} (\by_{t}\mid \by_{1:t-1})=\Phi_{m_{t} L }(\bgamma_{t}^{\adf}; \bGamma_{t}^{\adf})$.
\end{Prop}

From the above, the crucial property of the SUN ADF filtering density $h_{t}^{\adf}(\bbeta_{t})$ is that its second dimension (the one associated with the multivariate truncated normal) involves only the number of observations \emph{at time} $t$, namely $m_{t}$, leading to significant computational advantages with respect to the exact filtering, which is a $\sun_{p,M_{t}L}$, where the second dimension depends on the number of observations \emph{up to time} $t$, $M_{t}$, instead.
The ADF thus avoids the increase in the dimension of the truncated normal component by projecting the approximate filtering of the previous step $h_{t-1}^{\adf}(\bbeta_{t-1})$ onto the Gaussian class of distributions, which in turn induces a Gaussian predictive distribution for $\bbeta_{t}$.
This projection is done via moment matching, where the first two moments of $h_{t-1}^{\adf}(\bbeta_{t-1})$ are estimated with Monte Carlo via sampling from a $p$-variate Gaussian and an $m_{t-1}L$ truncated normal.
The analytical computation of the moments could also be considered, but this would require multiple computations of Gaussian cdfs of size $m_{t-1}L - 1$ and $m_{t-1}L - 2$ \citep{arellano2022some}, making the Monte Carlo computation of the moments usually preferable.
The ADF filter is summarized in Algorithm~\ref{alg:ADF}.

\begin{algorithm}[httb!]
\caption{ADF filter }\label{alg:ADF}
    \begin{algorithmic}
    \Require $\ba_{0},\bP_{0},(\bG_{t})_{t=1}^{T},(\bW_{t})_{t=1}^{T},\bSigma$ and $R$ (number of Monte Carlo samples)\vspace{0.1cm}
    \State Initialize $h_{1}^{\adf}(\bbeta_{1})=p(\bbeta_{1}\mid \by_{1})$ from Proposition~\ref{prop:iid_filter_pdt_0};
    \ set $\widehat{p}_{\adf}(\by_{1})=\Phi_{m_{1} L}(\bgamma_{1}; \bGamma_{1} )$ \vspace{0.1cm}
    \State \mbox{Sample i.i.d.\ $\bbeta^{(r)}_{1} \simiid h_{1}^{\adf}(\bbeta_{1})$, $r=1,\dots,R$, using the additive representation \eqref{eq:additiveRepresentationSUN}}\vspace{0.1cm}
    \For{$t \in 2,\dots,T$}\vspace{0.1cm}
        \State Set $\ba_{t-1}^{\adf} = \tfrac{1}{R} \sum_{r=1}^{R} \bbeta_{t-1}^{(r)}$ and $\bP_{t-1}^{\adf} = \tfrac{1}{R} \sum_{r=1}^{R} \bbeta_{t-1}^{(r)}{\left(\bbeta_{t-1}^{(r)}\right)}^{\intercal} - \ba_{t-1}^{\adf} {\left(\ba_{t-1}^{\adf}\right)}^{\intercal}$\vspace{0.1cm}
        \State Compute $\ba_{t\mid t-1}^{\adf},\bP_{t\mid t-1}^{\adf}$ as in \eqref{eq:predADF}\vspace{0.1cm}
        \State Approximate $p(\bbeta_{t}\mid \by_{1:t})$ with $h_{t}^{\adf}(\bbeta_{t})=\mathrm{d}\sun_{p,m_{t} L}$ as in Proposition~\ref{prop:ADFcorrection}\vspace{0.1cm}
        \State \mbox{Sample i.i.d.\ $\bbeta^{(r)}_{t} \simiid h_{t}^{\adf}(\bbeta_{t})$, $r=1,\dots,R$, using the additive representation \eqref{eq:additiveRepresentationSUN}}\vspace{0.1cm}
        \State Set $\widehat{p}_{\adf}(\by_{t}\mid \by_{1:t-1})=\Phi_{m_{t} L }(\bgamma_{t}^{\adf}; \bGamma_{t}^{\adf})$
        \vspace{0.1cm}
    \EndFor\vspace{0.1cm}
    \State \mbox{Compute the marginal likelihood approximation $
    \widehat{p}_{\adf}(\by_{1:T}) =\widehat{p}_{\adf}(\by_{1}) \prod_{t=2}^{T} \widehat{p}_{\adf}(\by_{t}\mid \by_{1:t-1})$}
    \end{algorithmic}
\end{algorithm}

Notice that Algorithm~\ref{alg:ADF} requires linear algebra operations on vectors and matrices with dimensions that are proportional to $m_{t}$, and it also relies on the sampling of $m_{t}L$-dimensional truncated normals.
This leads to massive computational savings when compared to sampling from the exact filtering as presented in Algorithm~\ref{alg:IID}.
However, this computation can be inefficient when many observations are available at each time step, making ADF appealing only for moderate $m_{t}$.
While the main computational advantage comes from the Gaussian approximation of the filtering, the main computational inefficiency comes from the emission distribution \eqref{eq:emission_MP}, which results in a $\sun_{p,m_{t}L}$ approximate filtering distribution.
Since sampling from $m_{t}L$-variate truncated normals does not scale linearly in $m_{t}$, we thus introduce a more scalable approach, where the factors in \eqref{eq:emission_MP} are approximated with Gaussian kernels, in an EP fashion.
Consequently, instead of computing the moments of the $\sun_{p,m_{t}L}$ appearing in ADF, we need to evaluate the moments of $m_{t}$ so-called hybrid densities belonging to the $\sun_{p,L}$ family.
This results in a Gaussian filtering approximation, obtained with a computational cost that scales linearly in $m_{t}$, making it preferable when many observations are available at each time.

\subsection{Sequential Expectation Propagation} \label{sec:EP}
To obtain a Gaussian filtering approximation with a computational cost scaling linearly both in $m_{t}$ and $t$, we introduce an EP procedure, which is repeated at each time $t$.
The EP approximation leverages the fact that the filtering distribution can be written as
\begin{equation*}
    p(\bbeta_{t}\mid \by_{1:t}) \propto p(\bbeta_{t}\mid \by_{1:t-1}) \prod_{i=1}^{m_{t}} p(y_{t,i}\mid \bbeta_{t}),
\end{equation*}
where $p(y_{t,i}\mid \bbeta_{t})=\Phi_{L}\left ( \bV_{[y_{t,i}]}\bX_{t,i} \bbeta_{t}; \bV_{[y_{t,i}]} \bSigma \bV_{[y_{t,i}]}^{\intercal} \right )$.
The key principle of EP is then to approximate the filtering with a density admitting the same factorization form, namely
\begin{equation}\label{eq:EPapproximation}
    q_{t}^{\ep}(\bbeta_{t}) \propto q_{t,0}^{\ep}(\bbeta_{t}) \prod_{i=1}^{m_{t}} q_{t,i}^{\ep}(\bbeta_{t}),
\end{equation}
where $q_{t,0}^{\ep}(\bbeta_{t})$ is the Gaussian predictive approximation induced by the previous filtering step, while $q_{t,i}^{\ep}(\bbeta_{t})$ is an approximation of the likelihood contribution of the $i$-th observation of time $t$.
In order to obtain a tractable form for $q_{t}^{\ep}(\bbeta_{t})$, Gaussian kernels are used for the factors, also called \textit{sites}, in \eqref{eq:EPapproximation}, that is
\begin{equation*}
    q_{t,i}^{\ep}(\bbeta_{t})=\frac{1}{Z_{t,i}}\exp \{-\tfrac{1}{2} \bbeta_{t}^{\intercal} \bQ_{t,i} \bbeta_{t} + \bbeta_{t}^{\intercal} \br_{t,i}\},\quad i=0,1,\ldots,m_{t}.
\end{equation*}
See, e.g., \citet{vehtari2020expectation,zhou2026bayesian} for more details on EP.
Thus, $q_{t}^{\ep}(\bbeta_{t}) \propto \exp\left\{ -\tfrac{1}{2} \bbeta_{t}^{\intercal} \bQ_{t}^{\ep} \bbeta_{t} + \bbeta_{t}^{\intercal} \br_{t}^{\ep} \right\}$, with $\br_{t}^{\ep} = \sum_{i=0}^{m_{t}} \br_{t,i}$, $\bQ_{t}^{\ep} = \sum_{i=0}^{m_{t}} \bQ_{t,i}$, meaning
\begin{equation*}
    q_{t}^{\ep}(\bbeta_{t}) = \phi_p(\bbeta_{t}-\ba_{t}^{\ep};\bP_{t}^{\ep}), \quad \bP_{t}^{\ep}=(\bQ_{t}^{\ep})^{-1}, \quad \ba_{t}^{\ep}=(\bQ_{t}^{\ep})^{-1}\br_{t}^{\ep}.
\end{equation*}
The predictive site $q_{t,0}^{\ep}(\bbeta_{t})$ is obtained from the Gaussian filtering approximation at time $t-1$ and the transition kernel \eqref{eq:trans_kern_MP}.
Specifically, $q_{t,0}^{\ep}(\bbeta_{t}) = \phi_p(\bbeta_{t}-\ba_{t\mid t-1}^{\ep};\bP_{t\mid t-1}^{\ep})$, with $\ba_{t\mid t-1}^{\ep} = \bG_{t}\ba_{t-1}^{\ep}$, $\bP_{t\mid t-1}^{\ep} = \bG_{t}\bP_{t-1}^{\ep}\bG_{t}^{\intercal} + \bW_{t}$.
Equivalently, in natural parametrization, $\bQ_{t,0} = (\bP_{t\mid t-1}^{\ep})^{-1}$, $\br_{t,0} = \bQ_{t,0}\ba_{t\mid t-1}^{\ep}$, $Z_{t,0}=\Psi(\br_{t,0},\bQ_{t,0})$, where
\begin{equation}\label{eq:Psi_EP}
    \Psi(\br,\bQ) = \int \exp\left\{ -\frac{1}{2}\bbeta^{\intercal}\bQ\bbeta + \bbeta^{\intercal}\br \right\}d\bbeta = \sqrt{\frac{(2\pi)^p}{\det(\bQ)}} \exp\left\{ \frac{1}{2}\br^{\intercal}\bQ^{-1}\br \right\}.
\end{equation}

The likelihood site parameters $\bQ_{t,i},\ \br_{t,i}$, and $Z_{t,i}$, $i=1,\ldots,m_{t}$, are instead updated iteratively, enforcing the EP moment-matching conditions explained below.
For site $i$, define the cavity distribution $q_{t,-i}^{\ep}(\bbeta_{t}) \propto q_{t}^{\ep}(\bbeta_{t})\,/\,q_{t,i}^{\ep}(\bbeta_{t}) \propto \exp\left\{ -\tfrac{1}{2}\bbeta_{t}^{\intercal}\bQ_{t,-i}\bbeta_{t} + \bbeta_{t}^{\intercal}\br_{t,-i} \right\}$,
where $\bQ_{t,-i}=\sum_{j\neq i}\bQ_{t,j}$, $\br_{t,-i}=\sum_{j\neq i}\br_{t,j}$.
When $\bQ_{t,-i}$ is positive definite, the cavity density is $q_{t,-i}^{\ep}(\bbeta_{t}) = \phi_p(\bbeta_{t}-\bQ_{t,-i}^{-1}\br_{t,-i};\bQ_{t,-i}^{-1})$ and the updated site parameters are obtained by matching the normalizing constant, mean, and covariance matrix of the updated global approximation to those of the hybrid density
\[
    h_{t,i}^{\ep}(\bbeta_{t}) \propto p(y_{t,i}\mid\bbeta_{t})q_{t,-i}^{\ep}(\bbeta_{t}).
\]
Thus, the tractability of $h_{t,i}^{\ep}(\bbeta_{t})$ is crucial for the implementation of the EP updates.
The following proposition entails that all necessary quantities can be computed efficiently.
\begin{Prop}\label{prop:EPcorrection}
    For any $t\ge 1$ and $i=1,\ldots,m_{t}$, assume that $\bQ_{t,-i}$ is positive definite.
    Then $h_{t,i}^{\ep}(\bbeta_{t}) = \mathrm{d}\sun_{p,L}(\bbeta_{t};\bxi_{t,i}^{\ep},\bOmega_{t,i}^{\ep},\bDelta_{t,i}^{\ep},\bgamma_{t,i}^{\ep},\bGamma_{t,i}^{\ep})$, where
    \begin{equation}\label{eq:EP_site_i_SUN}
    \begin{split}
        &\bxi_{t,i}^{\ep} = \bQ_{t,-i}^{-1}\br_{t,-i}, \quad \bOmega_{t,i}^{\ep} = \bQ_{t,-i}^{-1}, \quad \bDelta_{t,i}^{\ep} = \bar{\bOmega}_{t,i}^{\ep}{\bomega}_{t,i}^{\ep} \bX_{t,i}^{\intercal}\bV_{[y_{t,i}]}^{\intercal} (\bs_{t,i}^{\ep})^{-1},\\
        &\bgamma_{t,i}^{\ep} = (\bs_{t,i}^{\ep})^{-1} \bV_{[y_{t,i}]}\bX_{t,i}\bxi_{t,i}^{\ep}, \quad \bGamma_{t,i}^{\ep} = (\bs_{t,i}^{\ep})^{-1} \bV_{[y_{t,i}]} \left( \bX_{t,i}\bOmega_{t,i}^{\ep}\bX_{t,i}^{\intercal} + \bSigma \right) \bV_{[y_{t,i}]}^{\intercal} (\bs_{t,i}^{\ep})^{-1},
    \end{split}
    \end{equation}
    with $\bs_{t,i}^{\ep} = [ \bV_{[y_{t,i}]} \left( \bX_{t,i}\bOmega_{t,i}^{\ep}\bX_{t,i}^{\intercal} + \bSigma \right) \bV_{[y_{t,i}]}^{\intercal} \odot \bI_L ]^{1/2}$, $\bOmega_{t,i}^{\ep} = \bomega_{t,i}^{\ep} \bar{\bOmega}_{t,i}^{\ep} \bomega_{t,i}^{\ep}$, $\bomega_{t,i}^{\ep} = (\bOmega_{t,i}^{\ep}\odot\bI_p)^{1/2}$.
\end{Prop}

Proposition~\ref{prop:EPcorrection} is the key computational ingredient of the sequential EP filter, showing that each EP update only requires the normalizing constant and moments of a $\sun_{p,L}$ distribution, independently of $m_{t}$.
Thus, increasing the number of observations at time $t$ increases the number of site updates, but not the dimension of the truncated normal component, equal to $L$, involved in each update.

Let $Z_{h_{t,i}^{\ep}} = \int p(y_{t,i}\mid\bbeta_{t})q_{t,-i}^{\ep}(\bbeta_{t})d\bbeta_{t}$, $\bmu_{h_{t,i}^{\ep}} = \mathbb{E}_{h_{t,i}^{\ep}}(\bbeta_{t})$, $\bSigma_{h_{t,i}^{\ep}} =\var_{h_{t,i}^{\ep}}(\bbeta_{t})$.
By Proposition~\ref{prop:EPcorrection}, $Z_{h_{t,i}^{\ep}}=\Phi_L(\bgamma_{t,i}^{\ep};\bGamma_{t,i}^{\ep})$, while the first two moments can be computed either using SUN moment formulae or by Monte Carlo sampling from the additive representation in \eqref{eq:additiveRepresentationSUN}.
In the numerical implementation below, we use Monte Carlo moment estimates.

The moment matching conditions of the EP updates thus give
\begin{equation}\label{eq:EPmomentsZ_update}
    \bQ_{t,i}^{\new} = \bSigma_{h_{t,i}^{\ep}}^{-1} - \bQ_{t,-i}, \ \br_{t,i}^{\new}= \left( \bQ_{t,-i} + \bQ_{t,i}^{\new} \right) \bmu_{h_{t,i}^{\ep}} -\br_{t,-i}, \ Z_{t,i}^{\new} = \frac{ \Psi(\br_{t}^{\ep,\new},\bQ_{t}^{\ep,\new}) }{ \Psi(\br_{t,-i},\bQ_{t,-i}) Z_{h_{t,i}^{\ep}} },
\end{equation}
where $\bQ_{t}^{\ep,\new}= \bQ_{t,-i} + \bQ_{t,i}^{\new}$, $\br_{t}^{\ep,\new} = \br_{t,-i} + \br_{t,i}^{\new}$.
After convergence of the site updates, the EP approximation of the marginal likelihood contribution is
\begin{equation}\label{eq:EPmarginalLikelihood}
    \widehat{p}_{\ep}(\by_{t}\mid\by_{1:t-1}) = \frac{ \Psi(\br_{t}^{\ep},\bQ_{t}^{\ep}) }{ \Psi(\br_{t,0},\bQ_{t,0}) \prod_{i=1}^{m_{t}} Z_{t,i} }.
\end{equation}
This expression follows by replacing the predictive and likelihood factors in the exact marginal likelihood identity with their EP approximations.
The derivations of \eqref{eq:EPmomentsZ_update} and \eqref{eq:EPmarginalLikelihood} are provided in the Supplementary Materials.

The resulting EP filter is designed for regimes complementary to those targeted by ADF.
The latter preserves a richer SUN approximation at each time point, but its cost increases with the number of observations at the current time through the dimension of the multivariate truncated normal in the SUN additive representation.
\begin{algorithm}[t]
\caption{EP filter} \label{alg:EP}
    \begin{algorithmic}
    \Require $\ba_{0},\bP_{0},(\bG_{t})_{t=1}^{T},(\bW_{t})_{t=1}^{T},\bSigma$, and $R$ (number of Monte Carlo samples) \vspace{0.1cm}
    \State Initialize $\ba_{0}^{\ep}= \ba_{0}$ and $\bP_{0}^{\ep}=\bP_{0}$\vspace{0.1cm}
    \For{$t \in 1,\dots,T$}\vspace{0.1cm}
        \State Set $\ba_{t\mid t-1}^{\ep}=\bG_{t}\ba_{t-1}^{\ep}$ and $\bP_{t\mid t-1}^{\ep}=\bG_{t}\bP_{t-1}^{\ep}\bG_{t}^{\intercal}+\bW_{t}$\vspace{0.1cm}
        \State Set $\bQ_{t,0}=(\bP_{t\mid t-1}^{\ep})^{-1}$, $\br_{t,0}=\bQ_{t,0}\ba_{t\mid t-1}^{\ep}$, and $Z_{t,0}=\Psi(\br_{t,0},\bQ_{t,0})$\vspace{0.1cm}
        \State Set $\bQ_{t,i}=\bzero_{p\times p}$, $\br_{t,i}=\bzero_p$, and $Z_{t,i}=1$, for $i=1,\ldots,m_{t}$\vspace{0.1cm}
        \While{convergence is not reached}\vspace{0.1cm}
        \For{$i \in 1,\dots,m_{t}$}\vspace{0.1cm}
            \State Compute $\bQ_{t,-i}$, $\br_{t,-i}$ and then $\bxi_{t,i}^{\ep}, \bOmega_{t,i}^{\ep}, \bDelta_{t,i}^{\ep}, \bgamma_{t,i}^{\ep}, \bGamma_{t,i}^{\ep}$ as in \eqref{eq:EP_site_i_SUN}\vspace{0.1cm}
            \State Sample i.i.d.\ $\bbeta_{t}^{(r)} \simiid \sun_{p,L}(\bxi_{t,i}^{\ep},\bOmega_{t,i}^{\ep},\bDelta_{t,i}^{\ep},\bgamma_{t,i}^{\ep},\bGamma_{t,i}^{\ep})$ for $r=1,\ldots,R$\vspace{0.1cm}
            \State Set $\bmu_{h_{t,i}^{\ep}} =R^{-1}\sum_{r=1}^{R}\bbeta_{t}^{(r)}$ and $\bSigma_{h_{t,i}^{\ep}} = R^{-1}\sum_{r=1}^{R}\bbeta_{t}^{(r)}(\bbeta_{t}^{(r)})^{\intercal} - \bmu_{h_{t,i}^{\ep}}\bmu_{h_{t,i}^{\ep}}^{\intercal}$\vspace{0.1cm}
            \State Set $Z_{h_{t,i}^{\ep}}=\Phi_L(\bgamma_{t,i}^{\ep};\bGamma_{t,i}^{\ep})$\vspace{0.1cm}
            \State Update $\bQ_{t,i}$, $\br_{t,i}$, and $Z_{t,i}$ as in \eqref{eq:EPmomentsZ_update}\vspace{0.1cm}
        \EndFor\vspace{0.1cm}
        \EndWhile\vspace{0.1cm}
        \State Set $\bQ_{t}^{\ep}=\sum_{i=0}^{m_{t}}\bQ_{t,i}$, $\br_{t}^{\ep}=\sum_{i=0}^{m_{t}}\br_{t,i}$, $\bP_{t}^{\ep}=(\bQ_{t}^{\ep})^{-1}$ and $\ba_{t}^{\ep}=\bP_{t}^{\ep}\br_{t}^{\ep}$\vspace{0.1cm}
        \State Approximate $p(\bbeta_{t}\mid \by_{1:t})$ with $\phi_p(\bbeta_{t} - \ba_{t}^{\ep};\, \bP_{t}^{\ep})$\vspace{0.1cm}
        \State Set $\widehat{p}_{\ep}(\by_{t}\mid\by_{1:t-1}) = \Psi(\br_{t}^{\ep},\bQ_{t}^{\ep}) / \left[ \Psi(\br_{t,0},\bQ_{t,0}) \prod_{i=1}^{m_{t}}Z_{t,i} \right]$\vspace{0.1cm}
    \EndFor\vspace{0.1cm}
    \State Compute the marginal likelihood approximation $\widehat{p}_{\ep}(\by_{1:T}) = \prod_{t=1}^{T} \widehat{p}_{\ep}(\by_{t}\mid\by_{1:t-1})$
    \end{algorithmic}
\end{algorithm}
Sequential EP instead returns a Gaussian filtering approximation, but it only requires more tractable SUN moment calculations at the site level.
This replaces one moment calculation for a $\sun_{p,m_{t}L}$ distribution with $m_{t}$ moment calculations for $\sun_{p,L}$ distributions, which is more scalable when $m_{t}$ is large.
The whole sequential EP procedure is summarized in Algorithm~\ref{alg:EP}.
As in standard EP algorithms, site updates may occasionally yield cavity or global precision matrices that are not positive definite.
In such cases, one can use damping or skip unstable updates \citep{vehtari2020expectation,zhou2026bayesian}.
We did not encounter this issue in the simulation studies and the application reported in Section~\ref{sec:6}.

\paragraph{Computational cost analysis.} The computational differences among the three filters are driven by the dimension of the truncated normal component in the SUN additive representation.
Motivated by the online inference setting, we focus on the dependence on the time index $t$ and on the number of observations $m_{t}$ available at each time point.
As derived in Section~\ref{sec:supp_computational_scaling} of the Supplementary Materials, the exact i.i.d.\ sampler has leading cost exponential in the cumulative sample size $M_{t}$.
ADF removes this accumulation over time, but its per time cost still grows exponentially with the current sample size $m_{t}$.
Sequential EP instead has constant cost per site update, with respect to both $t$ and $m_{t}$, and overall per-time cost linear in $m_{t}$.
Table~\ref{tab:main_computational_scaling} summarizes these scaling regimes.
\begin{table}[ht]
\centering
\caption{Computational scaling of the three filtering strategies.
Here $c$ is a positive constant not depending on $t$ and $m_{t}$.
Summing these per-update costs over $t=1,\ldots,T$ gives the total filtering cost up to time $T$.}
\label{tab:main_computational_scaling}
\begin{tabular}{ll}
\toprule
Method & Leading cost at time $t$ \\
\midrule
Exact i.i.d.\ sampler & $\mathcal{O}(\exp(cM_{t}))$ \\
ADF & $\mathcal{O}(\exp(cm_{t}))$ \\
Sequential EP & $\mathcal{O}(m_{t})$\\
\bottomrule
\end{tabular}
\end{table}

\section{EM Algorithm for Parameter Inference} \label{sec:5}

A crucial quantity of interest in MNP models is the covariance matrix $\bSigma$ of the error terms $\bepsilon_{t,i}\sim \Normal_L(\boldsymbol{0}_L,\bSigma)$ appearing in the latent utility vectors $\bz_{t,i}= \bX_{t,i}\bbeta_{t} + \bepsilon_{t,i}$ of the $i$-th individual at time $t$.
In this section, we derive a novel EM algorithm \citep{dempster1977maximum} to estimate such an important hyperparameter from the data via maximization of the marginal likelihood $p(\by_{1:T};\bSigma)$, where we write the dependence on $\bSigma$ explicitly.
This provides a principled two-step procedure to first estimate $\bSigma$ and then perform online inference on $\bbeta_{t}$ via the filters introduced above.
In the proposed EM, we consider the quantities $\bbeta_{0:T}=(\bbeta_0^{\intercal},\ldots,\bbeta_{T}^{\intercal})^{\intercal}$ and $\bepsilon_{1:T}=(\bepsilon_1^{\intercal},\ldots,\bepsilon_{T}^{\intercal})^{\intercal}$, with $\bepsilon_{t}=(\bepsilon_{t,1}^{\intercal},\ldots,\bepsilon_{t,m_{t}}^{\intercal})^{\intercal}$, $t=1,\ldots,T$, as latent variables.
The complete log likelihood can be written as
\begin{align*}
    &\log p(\bbeta_{0:T}, \bepsilon_{1:T}, \by_{1:T};\, \bSigma) = \log p(\bbeta_{0}) + \sum_{t=1}^{T} \log p(\bbeta_{t} \mid \bbeta_{t-1})\\
    & \quad + \sum_{t=1}^{T} \sum_{i=1}^{m_{t}} \left[ \log p(y_{t,i}\mid \bepsilon_{t,i}, \bbeta_{t}) + \log p(\bepsilon_{t,i}; \bSigma) \right]
    = \sum_{t=1}^{T} \sum_{i=1}^{m_{t}} \log p(\bepsilon_{t,i}; \bSigma) + const.
\end{align*}
Thus, calling ${\bSigma}^{\prime}$ the current estimate of $\bSigma$, in the E-step one computes
\begin{equation*}
    Q(\bSigma\mid \bSigma^{\prime})
    = \sum_{t=1}^{T} \sum_{i=1}^{m_{t}} \bbE_{p(\bepsilon_{t,i} \mid \by_{1:T};\bSigma^{\prime})} \left[ \log p(\bepsilon_{t,i}; \bSigma) \right],
\end{equation*}
which is then maximized in the M-step with respect to $\bSigma\in\bbR^{L\times L}$ symmetric and positive definite, subject to the identifiability trace constraint $\text{tr}(\bSigma)=L$ \citep{burgette2012trace}.
We now highlight the main advantages and key quantities needed for the proposed EM algorithm, deferring the details to Section~\ref{sec:app_param_inf} of the Supplementary Materials.

First, including the terms $\bepsilon_{t,i}$ among the latent variables is crucial for obtaining tractable first-order conditions.
This avoids the need to compute the derivatives of the likelihood terms $p(y_{t,i}\mid \bbeta_{t}; \bSigma)$ that would be needed if considering $\bbeta_{0:T}$ alone.
With the addition of $\bepsilon_{t,i}$, $y_{t,i}$ becomes a deterministic function of $\bbeta_{t}$ and $\bepsilon_{t,i}$, with the only terms depending on $\bSigma$ being the tractable multivariate Gaussian terms $p(\bepsilon_{t,i}; \bSigma)$.
Second, a key quantity for the EM implementation is $\bS = \sum_{t=1}^{T}\sum_{i=1}^{m_{t}} \bbE_{p(\bepsilon_{t,i}\mid\by_{1:T};\bSigma^{\prime})} \left[ \bepsilon_{t,i}\bepsilon_{t,i}^{\intercal} \right]$.
The required second moments are estimated by combining an approximate Gaussian smoothing distribution for $\bbeta_{t}$, obtained from either the ADF or EP filtering approximation, with the resulting $\sun_{L,L}$ approximation of $p(\bepsilon_{t,i}\mid\by_{1:T};\bSigma^{\prime})$.
The moments of the latter can be computed analytically or by Monte Carlo sampling, as done in the experiments.
See Section~\ref{sec:app_param_inf} of the Supplementary Materials for more details.
Finally, since the first-order conditions imply that the optimal $\bSigma$ and $\bS$ commute, they are simultaneously diagonalizable, so that the conditions translate into a set of $L+1$ scalar equations, making the system easily solvable numerically.
The finite-sample performance of the proposed EM procedure is evaluated in Section~\ref{sec:EM_constrained}, showing good recovery properties and supporting its use to get reliable estimates of $\bSigma$ before using the filters to perform online inference.

\section{Experiments}
\label{sec:6}

We evaluate the proposed methodology in two simulation studies and an application to Expedia booking data.
The first simulation compares the accuracy and running times of the i.i.d.\ sampler, ADF, EP, and the sequential Monte Carlo discussed in Section~\ref{sec:sis_supp}.
The second assesses recovery of the covariance matrix $\bSigma$ using the constrained EM algorithm proposed in Section~\ref{sec:5}.
\ifblind
\else
Code, data, and the workflow required to reproduce the numerical results are available in the \href{https://github.com/LorenzoRimella/dynamic_multinomial_probit}{project repository}.
\fi
The two simulation studies share the same state evolution.
Specifically, we set $L=3$, $p=12$, $\ba_{0}=\bzero_{p}$, $\bP_{0}=\bI_{p}$, $\bG=\alpha\bI_{p}$, and $\bW=(1-\alpha^{2})\bI_{p}$, with $\alpha=0.5$.
Consequently, $\Normal_{p}(\bzero_{p},\bI_{p})$ is the stationary distribution of the latent state.
In both studies, the same data-generating covariance matrix $\bSigma$, reported in Table~\ref{tab:recovery_sigma}, is used.
Details on the construction of the covariates and on the simulation settings are given in Section~\ref{sec:app_simulation_design} of the Supplementary Materials.

\subsection{Filtering Accuracy and Computational Performance}
\label{sec:exp_filtering}

We compare the exact i.i.d.\ filter, ADF, EP, and a bootstrap particle filter, denoted by SMC, in three regimes, $(T,\lambda_m)\in\{(10,10),(50,10),(50,50)\}$, where $\lambda_m$ is the Poisson intensity governing the number of observations per time point.
More precisely, $m_{t}$ is generated from a $\Poi(\lambda_m)$ censored to $[2,4\lambda_m]$, which avoids degenerate sample sizes of $0$ or $1$ and excludes large values with very low probability.

Within each regime, the covariate design is held fixed, and $48$ latent trajectories, together with their corresponding responses, are simulated independently (one per available core).
Each method is then applied to the same datasets, ensuring that all comparisons are paired.
We impose a time limit of 96 hours, which is the maximum permitted by the cluster.
The exact i.i.d.\ filter is run in the smallest regime, where its truncation dimension remains manageable, and provides the benchmark; in the two larger regimes it is computationally unfeasible and ADF replaces it as a common reference, so that the distances reported there measure agreement with ADF rather than error from the i.i.d.\ filter.
SMC is omitted for $(T,\lambda_m)=(50,50)$, where its cost becomes prohibitive.

Table~\ref{tab:rt_a} reports elapsed times together with two accuracy summaries: the paired absolute relative difference in approximate log marginal likelihood between each method and the reference on the same dataset, and the average, over time and state coordinates, of the univariate $2$-Wasserstein distance between the corresponding empirical filtering marginals.
Averaging marginal distances avoids estimating a $12$-dimensional transport distance while remaining sensitive to discrepancies in location, spread, and shape.

\begin{table}[t]
\caption{Filtering accuracy and computational cost across $48$ independently simulated datasets ($27$ when $(T,\lambda_m)=(10,10)$), reported as Monte Carlo means $\pm$ standard deviations.
The log marginal likelihood absolute relative difference and the Wasserstein distance are computed against the exact i.i.d.\ filter for $(T,\lambda_m)=(10,10)$ and against ADF in the two larger regimes.
A dash indicates that a method was not run.}
\label{tab:rt_a}
\centering
\small
\begin{tabular}{cclcccc}
\toprule
$T$ & $\lambda_m$ & Metric & ADF & EP & SMC & i.i.d.\ \\
\midrule
\multirow{3}{*}{10}
& \multirow{3}{*}{10}
& Time (sec.)
&  {7.11 $\pm$ 0.75}
&  {18.43 $\pm$ 2.07}
&  {2269.6 $\pm$ 21.5}
&  {198819 $\pm$ 72864} \\
&
& Log-lik.\ diff.
&  {0.007 $\pm$ 0.007}
&  {0.008 $\pm$ 0.007}
&  {0.008 $\pm$ 0.007}
& $0$ \\
&
& Distance
&  {0.021 $\pm$ 0.001}
&  {0.039 $\pm$ 0.002}
&  {0.100 $\pm$ 0.014}
& $0$ \\
\midrule
\multirow{3}{*}{50}
& \multirow{3}{*}{10}
& Time (sec.)
&  {33.58 $\pm$ 1.64}
&  {88.64 $\pm$ 4.49}
&  {12001 $\pm$ 92}
& \textemdash \\
&
& Log-lik.\ diff.
& $0$
&  {0.002 $\pm$ 0.000}
&  {0.001 $\pm$ 0.001}
& \textemdash \\
&
& Distance
& $0$
&  {0.038 $\pm$ 0.001}
&  {0.099 $\pm$ 0.008}
& \textemdash \\
\midrule
\multirow{3}{*}{50}
& \multirow{3}{*}{50}
& Time (sec.)
&  {6152.8 $\pm$ 1023.09}
&  {466.07 $\pm$ 34.72}
& \textemdash
& \textemdash \\
&
& Log-lik.\ diff.
& $0$
&  {0.001 $\pm$ 0.000}
& \textemdash
& \textemdash \\
&
& Distance
& $0$
&  {0.050 $\pm$ 0.001}
& \textemdash
& \textemdash \\
\bottomrule
\end{tabular}
\end{table}

In the exact-benchmark regime, only $27$ replicates were completed within the $96$-hour time limit.
In this regime, ADF and EP attain mean marginal 2-Wasserstein distances of $0.021 \pm 0.001$ and $0.039 \pm 0.002$, respectively, compared with $0.100 \pm 0.014$ for SMC.
Thus, the proposed approximate methods are not only substantially faster than SMC, but also more accurate in recovering the exact filtering distribution.
Their paired log marginal-likelihood absolute relative differences indicate that all three methods produce log marginal-likelihood estimates close to the corresponding reference values, but at significantly different running times.

When $T$ increases to $50$ with $\lambda_m=10$, EP and SMC remain in agreement with ADF, while the exact i.i.d.\ sampler is no longer computationally practical.
When $\lambda_m$ increases to $50$, EP is about $13$ times faster than ADF while retaining a mean marginal distance of $0.050 \pm 0.001$.
This reversal in their runtime ordering agrees with the scaling regimes in Table~\ref{tab:main_computational_scaling}: ADF requires simulation from an $m_{t}L$-dimensional truncated Gaussian distribution, whereas EP processes $m_{t}$ sites whose truncation dimension remains fixed at $L$.
Representative filtering trajectories are reported in Section~\ref{sec:app_filtering} of the Supplementary Materials.

\subsection{Recovery of \texorpdfstring{$\bSigma$}{Sigma} via Constrained Expectation Maximization}
\label{sec:EM_constrained}

We now assess recovery of the data-generating $\bSigma$ using the constrained EM algorithm of Section~\ref{sec:5}.
We set the Poisson intensity to $\lambda_m=15$ (and the usual censoring) and consider $T\in\{10,30,60\}$.
For each value of $T$, one covariate design is generated and held fixed, and $48$ latent state trajectories (one for each available core) and corresponding response datasets are independently simulated.
For every dataset, the EM algorithm is initialized at $\bSigma^{(0)}=\bI_{L}$ and the E-step is performed leveraging the sequential EP filter of Section~\ref{sec:EP} as detailed in the Supplementary Materials.
We run $200$ EM iterations and report the final estimates: as shown in Section~\ref{sec:app_em_convergence} of the Supplementary Materials, the approximate log marginal likelihood has stabilized by the end of the run.

Table~\ref{tab:recovery_sigma} reports the Monte Carlo means and standard deviations of the estimates, while Figure~\ref{fig:em_shrink} reports their central $95\%$ empirical intervals.
As $T$ increases, dispersion decreases for every entry and the estimates concentrate around the data-generating values, supporting the ability of the proposed EM procedure to recover both variances and cross-alternative dependence as the amount of information grows.
\begin{table}[t]
\centering

\begin{minipage}[t]{0.5\textwidth}
\vspace{0pt}
\caption{Recovery of $\bSigma$.
Entries are Monte Carlo means $\pm$ standard deviations across $48$ simulated datasets.}
\label{tab:recovery_sigma}
\centering
\scriptsize
\setlength{\tabcolsep}{2pt}
\begin{tabular}{ccccc}
\toprule
Par. & Truth & $T=10$ & $T=30$ & $T=60$ \\
\midrule
$\Sigma_{11}$ & $1.6$ & $1.44 \pm 0.46$ & $1.62 \pm 0.24$ & $1.55 \pm 0.23$ \\
$\Sigma_{12}$ & $-0.1$ & $-0.08 \pm 0.42$ & $-0.06 \pm 0.24$ & $-0.06 \pm 0.17$ \\
$\Sigma_{13}$ & $0$ & $-0.04 \pm 0.51$ & $0.04 \pm 0.35$ & $-0.1 \pm 0.3$ \\
$\Sigma_{22}$ & $0.2$ & $0.45 \pm 0.29$ & $0.32 \pm 0.15$ & $0.32 \pm 0.11$ \\
$\Sigma_{23}$ & $0.1$ & $0.11 \pm 0.31$ & $0.09 \pm 0.19$ & $0.11 \pm 0.16$ \\
$\Sigma_{33}$ & $1.2$ & $1.11 \pm 0.49$ & $1.06 \pm 0.24$ & $1.14 \pm 0.26$ \\
\bottomrule
\end{tabular}
\end{minipage}
\hfill
\begin{minipage}[t]{0.49\textwidth}
\vspace{0pt}
\centering
\includegraphics[width=\linewidth]{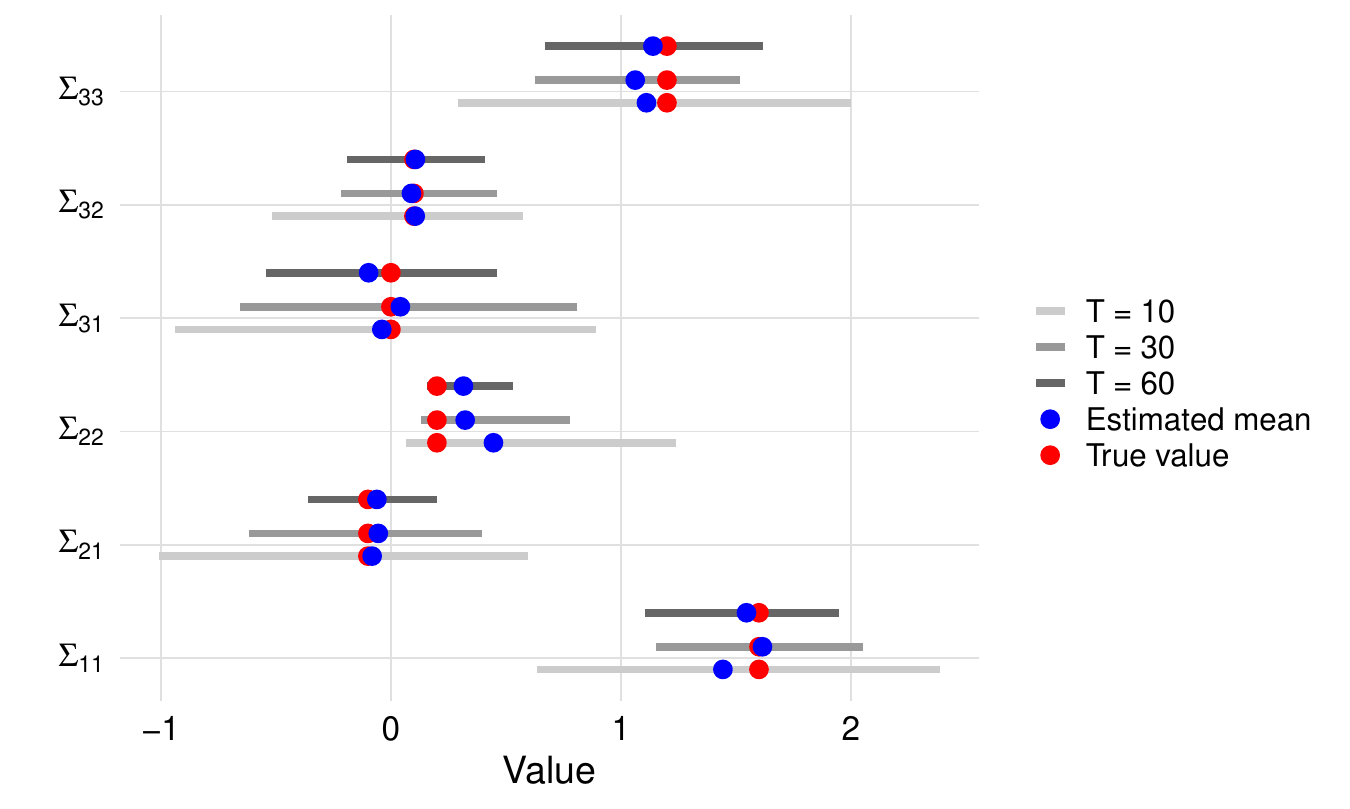}
\captionof{figure}{Sampling distributions of the EM estimates.
Points denote Monte Carlo means and true values; gray bands are central $95\%$ empirical intervals.}
\label{fig:em_shrink}
\end{minipage}
\end{table}

\subsection{Application to Expedia Data}
The publicly available Expedia data \citep{friedman2013personalized} combine customer and hotel characteristics with dated ``impressions'', namely ranked lists of hotels displayed in response to a search.
Such impressions can be sorted according to Expedia's ranking algorithm or randomly.
Since randomization removes the confounding induced by Expedia's ranking algorithm, we focus on this case to study how hotel attributes and display position are associated with booking choices \citep{ding2024computationally}.

More specifically, we retain impressions from March 1 to June 30, 2013, restricted to the most represented property country and customer country (USA, labeled 219), one-night stays, at most two guests, and searches resulting in a booking.
The resulting series contains $1,629$ impressions over $122$ days.
Each choice set contains the first $10$ displayed hotels; outcome $0$ denotes a booking at a hotel displayed below position $10$ and occurs in $32\%$ of the impressions.
Since we condition on bookings, all estimates are interpreted as associations conditional on booking.

Because the EM of Section~\ref{sec:5} requires a full filtering pass at every iteration, and the exact filter would require truncated normal simulation in dimension $M_{T}L=16,290$ by the end of the series, we use the proposed sequential EP, whose truncated normal simulations needed for hybrid moments calculation have fixed dimension $L=10$.

For each impression, $\bX_{t,i}$ is the $L\times p$ covariate matrix of Section~\ref{subsec:model}, with $L=10$ and $p=13$: its columns comprise three standardized hotel attributes, namely location score, log price and star rating, together with $10$ position-specific intercepts.
Following the specification adopted in the simulation study of Section~\ref{sec:EM_constrained}, we set $\ba_{0}=\bzero_{p}$, $\bP_{0}=\bI_{p}$, $\bG=\alpha\bP_{0}$ and $\bW=(1-\alpha^{2})\bP_{0}$, so that $\Normal_{p}(\bzero_{p},\bI_{p})$ is the stationary distribution of the latent coefficients and $\alpha\in(0,1)$ governs their persistence.
We fix $\alpha=0.99$, so that the effect of a state innovation decays as $0.99^{h}$ after $h$ days and is halved after approximately $69$ days, allowing the latent coefficients to evolve gradually over the four-month window.

\begin{figure}[ht!]
    \centering
    \includegraphics[width=0.9\linewidth]{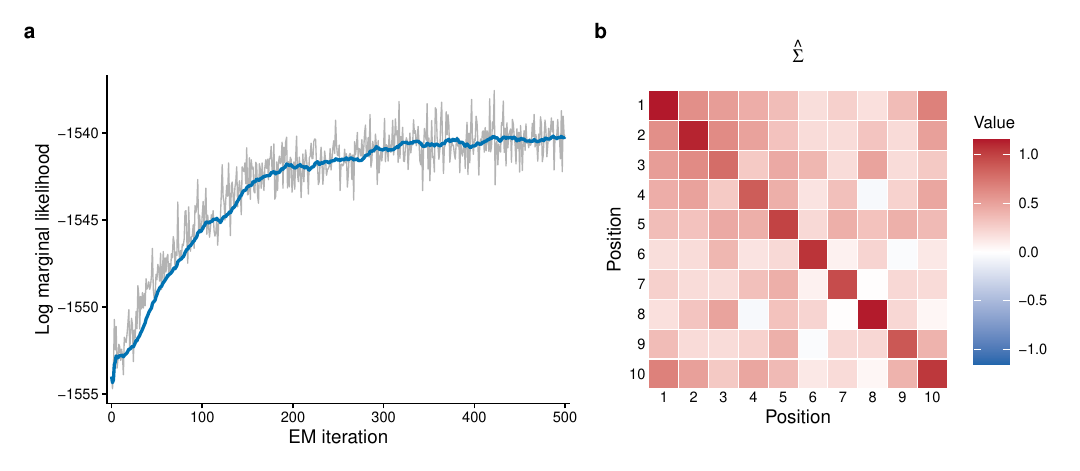}
    \caption{Estimation of $\bSigma$ from the first $50$ days.
    (a) EP approximations of the log marginal likelihood (gray) and their trailing 25-iteration average (blue) over 500 EM iterations.
    (b) Estimate of $\bSigma$.}
    \label{fig:expedia_training}
\end{figure}

We estimate the error covariance $\bSigma$ using the constrained EM procedure of Section~\ref{sec:5} on the first 50 days (732 impressions), initializing at $\bI_{10}$.
Figure~\ref{fig:expedia_training}(a) shows a clear upward trend in the approximate log marginal likelihood, which stabilizes over the last iterations; the residual variability is due to the Monte Carlo evaluation of the E-step, consistent with the approximate nature of the algorithm discussed in Section~\ref{sec:app_param_inf}.
The estimated covariance matrix in Figure~\ref{fig:expedia_training}(b) is markedly non-diagonal: the corresponding correlations among positions 1--3 are $0.53$, $0.55$ and $0.66$, while that between positions $1$ and $10$ is $0.62$.
These residual dependencies may reflect unobserved impression-level factors and illustrate the cross-alternative dependence accommodated by the MNP but excluded under independent errors.

\begin{figure}[ht!]
    \centering
    \includegraphics[width=0.8\linewidth]{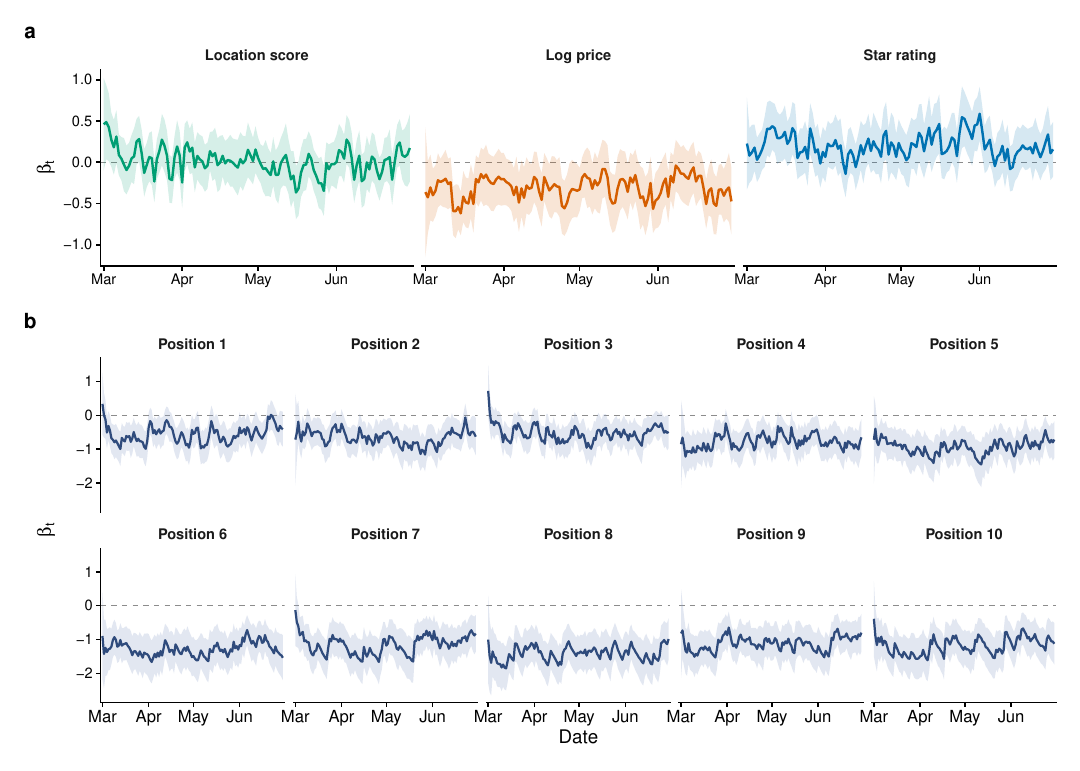}
    \caption{EP filtering means (solid) and 95\% credible intervals (shaded) for (a) standardized hotel attributes and (b) position-specific intercepts; dashed horizontal lines mark 0.}
    \label{fig:coef}
\end{figure}

We then fix $\bSigma$ at its final estimate and run the proposed sequential EP filter over all 122 days.
Figure~\ref{fig:coef} reports the filtering means and pointwise 95\% credible intervals.
The coefficient of log price is negative throughout the observation period, whereas star rating is positive over most of the period and location score is centred close to zero.
The signs are those expected of customers preferring cheaper and higher-quality hotels.
Since the attributes are standardized and the trace constraint fixes the average error variance to one, their coefficients are directly comparable: averaging the filtering means over the 122 days, a one-standard-deviation increase in log price changes the latent utility by $-0.30$, against $0.20$ for star rating and $0.02$ for location score.

The position intercepts are negative over most of the observation period, consistent with the reference alternative being the most frequent outcome, chosen in $32\%$ of the impressions.
On the same scale, the time-averaged intercepts of positions $1$--$4$ exceed those of positions $7$--$10$ by $0.58$; the position-1 versus position-10 gap is $0.62$ and remains positive on all $122$ days.
This suggests that, all else equal, hotels displayed near the top have a higher probability of being selected.
Finally, the filtered paths drift gradually over the four months, illustrating the smooth evolution captured by the dynamic formulation.

\section{Discussion}\label{sec:7}

In this contribution, we derived an exact analytical solution to online inference in dynamic multinomial probit models leveraging SUN distributions \citep{arellano2006unification}.
The characterization of the filtering and state predictive distributions as SUN yields an i.i.d.\ sampler and provides the analytical basis for the ADF- and EP-based proposed approximations so that online inference can be performed efficiently in different regimes.

Indeed, while the i.i.d.\ algorithm becomes rapidly more expensive as the cumulative sample size grows, the proposed ADF avoids this accumulation and has overall cost linear in $t$, but not in the time-specific sample size $m_{t}$.
Sequential EP overcomes this remaining issue by reducing each update to $m_{t}$ fixed-dimensional site calculations, resulting in a Gaussian approximation derived at a cost also linear in $m_{t}$.
The experiments show that ADF and EP closely reproduce the exact filtering distributions and log marginal likelihood when the exact benchmark is available.
In larger regimes, where the i.i.d.\ sampler becomes computationally unfeasible, the proposed EP remains close to ADF while providing substantial computational savings.

Several directions follow naturally.
The SUN conjugacy exploited here follows from combining Gaussian or SUN priors with Gaussian cumulative distribution function likelihoods, and is therefore not tied to Markovian dynamics: replacing the state equation by a Gaussian process prior over time, space, or both preserves the SUN form of the filtering and would extend the framework to spatio-temporal discrete choice models.
Therefore, the proposed algorithms can also be adapted to such a framework.
Learning the state-evolution parameters jointly with $\bSigma$ within the same EM scheme is a further natural future research direction.

\paragraph*{Disclosure of AI Use}
The authors used ChatGPT in order to proofread and polish the manuscript.
After using it, the authors reviewed and edited the content as needed and take full responsibility for the content of the article.

\paragraph*{Data Availability Statement}
The Expedia dataset is provided as Supplementary Material, and the code is available at
\url{https://github.com/LorenzoRimella/dynamic_multinomial_probit}.

\ifblind

\else

\section*{Acknowledgements}
The authors acknowledge the CINECA award under the ISCRA initiative for the availability of high-performance computing resources and support.
LR is supported by EPSRC Grant UKRI3641: ``Scalable statistical inference for disease contact networks''.
\fi

\spacingset{1.8}

\vspace*{-0.5cm}
\printbibliography

\end{refsection}

\newpage

\clearpage\pagebreak\newpage
\pagestyle{fancy}
\fancyhf{}
\rhead{\bfseries\thepage}
\lhead{\bfseries SUPPLEMENTARY MATERIALS}

\baselineskip=27pt
\begin{center}
{\LARGE{Supplementary Materials for\\} 
\bf Online inference for dynamic categorical regression
}
\end{center}

\baselineskip=12pt
\vskip 2mm

\ifblind
\else
\begin{center}
{\large
Lorenzo Rimella \\
Economics Dept., University of Bergamo, Bergamo, IT\\
Augusto Fasano and Giovanni Rebaudo\\
ESOMAS Dept., University of Torino and Collegio Carlo Alberto, Torino, IT
}
\end{center}
\fi

\setcounter{equation}{0}
\setcounter{page}{1}
\setcounter{table}{1}
\setcounter{figure}{0}
\setcounter{section}{0}
\numberwithin{table}{section}
\renewcommand{\theequation}{S.\arabic{equation}}
\renewcommand{\thesubsection}{S.\arabic{section}.\arabic{subsection}}
\renewcommand{\thesection}{S.\arabic{section}}

\renewcommand{\theHsection}{supp.\arabic{section}}
\renewcommand{\theHsubsection}{supp.\arabic{section}.\arabic{subsection}}

\renewcommand{\theThm}{S.\arabic{Thm}}
\renewcommand{\theCor}{S.\arabic{Cor}}
\renewcommand{\theProp}{S.\arabic{Prop}}
\renewcommand{\theLem}{S.\arabic{Lem}}
\renewcommand{\thepage}{S.\arabic{page}}
\renewcommand{\thetable}{S.\arabic{table}}
\renewcommand{\thefigure}{S.\arabic{figure}}

\renewcommand{\theHequation}{supp.eq.\arabic{equation}}
\renewcommand{\theHfigure}{supp.fig.\arabic{figure}}
\renewcommand{\theHtable}{supp.tab.\arabic{section}.\arabic{table}}

\baselineskip=14pt 
\vskip 10mm

\begin{refsection}
\spacingset{1.8}

\section{Proofs}
This section contains some useful lemmas on manipulations of SUN densities and the proofs of the theoretical results presented in the manuscript.

\subsection{Preliminary lemmas on SUN distributions}
In this section, we derive some results about the SUN distribution that we will exploit in the proofs of the main results of the manuscript.

\begin{Lem}\label{le: sun from gaussian x cdf}
Let 
\begin{equation*}
	p(\bbeta) \propto \phi_{p}(\bbeta -\bmu; \bLambda) \cdot \Phi_{n}(\bb + \bA \bbeta; \bSigma) 
\end{equation*}
with $\bLambda \succ 0 $ and $\bSigma \succ 0$ (full-rank covariance matrices).
Then,
\begin{equation*}
	p(\bbeta) = \mathrm{d} \sun_{p,n}(\bxi,\bOmega,\bDelta,\bgamma,\bGamma), \text{ with }
\end{equation*}
\begin{align*}
	&\bxi = \bmu, \quad \bOmega = \bLambda, \quad \bDelta = \bar\bLambda \blambda \bA^{\intercal} \bs^{-1},\\
	&\bgamma = \bs^{-1} \bb + \bs^{-1} \bA \bmu, \quad \bGamma = \bs^{-1} \left( \bA \bLambda \bA^{\intercal} + \bSigma \right) \bs^{-1},
\end{align*}
where $\bs = \left[ \left( \bA \bLambda \bA^{\intercal} + \bSigma \right) \odot\bI_{n} \right]^{1/2}$ and $\bLambda = \blambda \bar{\bLambda} \blambda$ with $\blambda = (\bLambda \odot \bI_{p})^{1/2}$, with $\bGamma$ and 
$\begin{bmatrix} 
    \bar{\bOmega}       & \bDelta\\
    \bDelta^{\intercal} & \bGamma
\end{bmatrix}$ 
full-rank correlation matrices (i.e., positive definite with unit diagonal), where $\bOmega = \bomega \bar{\bOmega} \bomega$ with $\bomega=(\bOmega\odot \bI_{p})^{1/2}$.
\end{Lem}

\begin{proof}
Write $\bOmega=\bLambda$ and decompose $\bOmega=\bomega\,\bar{\bOmega}\,\bomega$ with $\bomega=(\bOmega\odot\bI_{p})^{1/2}$ (diagonal scale matrix) and $\bar{\bOmega}=\bar{\bLambda}$ (correlation matrix).
Recall the SUN density \citep[Section~7.1]{azzalini2014skew} in the standard parametrization
\[
\mathrm{d}\sun_{p,n}(\bbeta;\bxi,\bOmega,\bDelta,\bgamma,\bGamma)
\ \propto\
\phi_{p}(\bbeta-\bxi; \bOmega) 
\Phi_{n} \Big(\bgamma+\bDelta^{\intercal} \bar{\bOmega}^{-1}\bomega^{-1}(\bbeta-\bxi);\ \bGamma-\bDelta^{\intercal}\bar{\bOmega}^{-1}\bDelta\Big),
\]
with normalizing constant $\Phi_{n}(\bgamma;\bGamma)$.

We now show that the product $\phi_{p}(\bbeta-\bmu;\bLambda)\, \Phi_{n}(\bb+\bA\bbeta;\bSigma)$ coincides with the kernel above for the stated choice of $(\bxi,\bOmega,\bDelta,\bgamma,\bGamma)$.
Set $\bxi=\bmu$ and $\bOmega=\bLambda$ so that the Gaussian pdf factor matches.
For the cdf argument, using $\bDelta=\bar{\bLambda}\,\blambda \bA^{\intercal}\bs^{-1}$ and $\bar{\bOmega}=\bar{\bLambda}$,
\begin{align*}
    &\bDelta^{\intercal}\bar{\bOmega}^{-1}\bomega^{-1}(\bbeta-\bxi) = \big(\bs^{-1}\bA\blambda\bar{\bLambda}\big)\,\bar{\bOmega}^{-1}\bomega^{-1}(\bbeta-\bmu) =\\
    &\bs^{-1}\bA\blambda\bar{\bLambda}\bar{\bLambda}^{-1}\bomega^{-1}(\bbeta-\bmu) = \bs^{-1}\bA\blambda\bomega^{-1}(\bbeta-\bmu) = \bs^{-1}\bA(\bbeta-\bmu),
\end{align*}
Hence
\[
    \bgamma+\bDelta^{\intercal}\bar{\bOmega}^{-1}\bomega^{-1}(\bbeta-\bxi) = \bs^{-1}(\bb+\bA\bmu)+\bs^{-1}\bA(\bbeta-\bmu) = \bs^{-1}(\bb+\bA\bbeta).
\]
For the cdf covariance, using $\bOmega=\bLambda$,
\[
    \bGamma-\bDelta^{\intercal}\bar{\bOmega}^{-1}\bDelta = \bs^{-1} \big(\bA\bLambda\bA^{\intercal}+ \bSigma\big)\bs^{-1} -\bs^{-1}\bA \bLambda \bA^{\intercal}\bs^{-1} = \bs^{-1}\bSigma \bs^{-1}.
\]
Therefore, the cdf factor in the SUN kernel becomes
\[
    \Phi_{n} \Big(\bs^{-1}(\bb+\bA\bbeta);\ \bs^{-1}\bSigma \bs^{-1}\Big).
\]
Since $\bs$ is diagonal with positive entries, the coordinatewise inequality $\bZ\le \bx$ is equivalent to $\bs^{-1}\bZ\le \bs^{-1}\bx$, and if $\bZ \sim \Normal_{n}(\bzero_{n},\bSigma)$ then $\bs^{-1}\bZ\sim\Normal_{n} \big(\bzero_{n}, \bs^{-1}\bSigma \bs^{-1}\big)$.
Hence
\[
    \Phi_{n}(\bb+\bA\bbeta;\bSigma) = \Phi_{n} \Big(\bs^{-1} (\bb+ \bA\bbeta); \bs^{-1}\bSigma \bs^{-1}\Big),
\]
so the cdf factor also matches.
Thus
\[
    \phi_{p}(\bbeta-\bmu;\bLambda) \Phi_{n}(\bb+\bA\bbeta;\bSigma) \propto \mathrm{d}\sun_{p,n}(\bbeta;\bmu,\bLambda,\bDelta,\bgamma,\bGamma),
\]
with proportionality constant equal to $\Phi_{n}(\bgamma;\bGamma)$, which is the SUN normalizing constant.
Indeed,
\[
    \int \phi_{p}(\bbeta-\bmu;\bLambda) \Phi_{n}(\bb+\bA\bbeta;\bSigma)\,d\bbeta =\ \Pr \big(\bZ\le \bb+\bA\bbeta\big),\quad \bbeta\sim\textsc{n}_{p}(\bmu,\bLambda),\ \bZ\sim\Normal_{n}(\bzero_{n},\bSigma),
\]
which equals $\Phi_{n}(\bgamma;\bGamma)$ by the same diagonal rescaling with $\bs=[(\bA\bLambda\bA^{\intercal}+\bSigma) \odot \bI_{n}]^{1/2}$.

Finally, we show that $\bGamma$ and 
$\bM = \begin{bmatrix} 
    \bar{\bOmega}       & \bDelta\\
    \bDelta^{\intercal} & \bGamma
\end{bmatrix}$ 
are full-rank correlation matrices (i.e., positive definite with unit diagonal), where $\bOmega = \bomega \bar{\bOmega} \bomega$ with $\bomega=(\bOmega\odot \bI_{p})^{1/2}$.

Since $\bLambda \succ 0$ and $\bSigma \succ 0$, we have
\[
    \bA\bLambda\bA^{\intercal}+\bSigma \succ 0.
\]
By definition $\bs=\big[(\bA\bLambda\bA^{\intercal}+\bSigma) \odot \bI_{n} \big]^{1/2}$ is diagonal with strictly positive entries, so $\bs^{-1}$ is well-defined.
Then
\[
    \bGamma = \bs^{-1} \big(\bA\bLambda\bA^{\intercal}+\bSigma\big) \bs^{-1}
\]
is symmetric and positive definite.
Moreover,
\[
    \diag(\bGamma) = \diag \big(\bs^{-1} (\bA\bLambda\bA^{\intercal}+\bSigma) \bs^{-1}\big) = \bs^{-1}\,\diag(\bA\bLambda\bA^{\intercal}+\bSigma) \,\bs^{-1} = \bI_{n},
\]
so $\bGamma$ has unit diagonal.
Hence $\bGamma$ is a correlation matrix.

Let $\bbeta\sim \textsc{n}_{p}(\bmu,\bLambda)$ and $\bZ\sim \Normal_{n}(\bzero_{n},\bSigma)$ be independent, and denote the scale-correlation decomposition $\bLambda=\blambda\,\bar{\bLambda}\,\blambda$ with $\blambda=(\bLambda\odot \bI_{p})^{1/2}$ and $\bar{\bLambda}=\bar{\bOmega}$.
Define the standardized Gaussian vectors
\[
\bX = \blambda^{-1}(\bbeta-\bmu)\in\bbR^{p},
\qquad
\bY = \bs^{-1} \big(\bA(\bbeta-\bmu)+\bZ\big)\in\bbR^{n}.
\]
Then $(\bX,\bY)$ is jointly Gaussian with mean zero, and its block covariance is
\[
\var(\bX) = \blambda^{-1}\bLambda\,\blambda^{-1} = \bar{\bLambda} = \bar{\bOmega},
\]
\[
\var(\bY) = \bs^{-1} \big(\bA\bLambda\bA^{\intercal}+\bSigma\big)\bs^{-1} = \bGamma,
\]
\[
    \cov(\bX,\bY) = \blambda^{-1}\,\cov \big(\bbeta-\bmu, \bA(\bbeta-\bmu)+\bZ\big)\,\bs^{-1} = \blambda^{-1}\, \bLambda\,\bA^{\intercal}\bs^{-1} = \bar{\bLambda}\, \blambda\,\bA^{\intercal}\bs^{-1} = \bDelta.
\]
Therefore, the joint covariance of $(\bX,\bY)$ is exactly
\[
\begin{bmatrix}
    \var(\bX) & \cov(\bX,\bY)\\
    \cov(\bX,\bY)^{\intercal} & \var(\bY)
\end{bmatrix}
 = 
\begin{bmatrix}
    \bar{\bOmega} & \bDelta\\
    \bDelta^{\intercal} & \bGamma
\end{bmatrix}.
\]
In addition, $\diag(\var(\bX))=\diag(\bar{\bOmega})= \bI_{p}$ and $\diag(\var(\bY))=\diag(\bGamma)= \bI_{n}$, so $\bM$ has unit diagonal.
Being a covariance matrix of a nondegenerate Gaussian vector, $\bM$ is symmetric and positive definite.
With unit diagonal, $\bM$ is in fact a correlation matrix.

This proves that $\bGamma$ is a correlation matrix and the block matrix $\bM$ is positive definite (so, a correlation matrix).
\end{proof}

\begin{Lem}\label{le: lin transf SUN}
Let $ \bbeta \sim \sun_{p,n}(\bxi,\bOmega,\bDelta,\bgamma,\bGamma)$.
If $\bA\in \bbR^{m\times p}$ has full row rank $m \le p$, then
\begin{equation*}
\bA\bbeta \sim \sun_{m,n}(\bA\bxi,\bA\bOmega\bA^{\intercal},\bDelta_{\bA},\bgamma,\bGamma),
\end{equation*}
with $\bDelta_{\bA}=\left[(\bA\bOmega\bA^{\intercal})\odot \bI_{m}\right]^{-1/2}\bA\bomega\bDelta$, $\bomega = (\bOmega\odot \bI_{p})^{1/2}$.
\end{Lem}
\begin{proof}
	See Section~7.1.2 of \citet{azzalini2014skew}.
\end{proof}

\begin{Lem}\label{le: SUN plus Gaussian}
Let $ \bbeta \sim \sun_{p,n}(\bxi,\bOmega,\bDelta,\bgamma,\bGamma)$ and $\bepsilon\sim \textsc{n}_{p}(\bb,\bW)$ independent.
Then
\begin{equation*}
\bbeta + \bepsilon \sim \sun_{p,n}(\tilde\bxi,\tilde\bOmega,\tilde\bDelta,\bgamma,\bGamma),
\end{equation*}
with $\tilde\bxi=\bxi+\bb$,  $\tilde\bOmega = \bOmega+\bW$,  $\tilde\bDelta = \tilde\bomega^{-1}\bomega\bDelta$.
Here $\bOmega=\bomega\,\bar\bOmega\,\bomega$ with $\bomega = (\bOmega\odot \bI_{p})^{1/2}$, and $\tilde\bOmega = \tilde\bomega\,\bar{\tilde\bOmega}\,\tilde\bomega$ with $\tilde\bomega = (\tilde\bOmega\odot \bI_{p})^{1/2}$.
\end{Lem}

\begin{proof}
For a random vector $\bx$ let $M_{\bx}(\bt)$ be the corresponding moment generating function (MGF) evaluated in $\bt$ (when it exists).
Using the MGFs of the SUN \citep[Section~7.1]{azzalini2014skew} and of the Gaussian distribution, we have
\[
    M_{\bbeta}(\bt) = \exp \Big(\bt^{\intercal}\bxi+\frac{1}{2}\,\bt^{\intercal}\bOmega\,\bt\Big)\,
\frac{\Phi_{n} \Big(\bgamma+\bDelta^{\intercal}\bar\bOmega^{-1}\bomega^{-1}\bOmega\,\bt\ ;\ \bGamma\Big)}{\Phi_{n}(\bgamma;\bGamma)}
\]
and 
\[
    M_{\bepsilon}(\bt)=\exp(\bt^{\intercal}\bb+\frac{1}{2}\,\bt^{\intercal}\bW\,\bt).
\]
 Independence gives
\[
M_{\bbeta+\bepsilon}(\bt)
= M_{\bbeta}(\bt)\,M_{\bepsilon}(\bt)
= \exp \Big(\bt^{\intercal}(\bxi+\bb)+\frac{1}{2}\,\bt^{\intercal}(\bOmega+\bW)\,\bt\Big)\,
\frac{\Phi_{n} \Big(\bgamma+\bDelta^{\intercal}\bar\bOmega^{-1}\bomega^{-1}\bOmega\,\bt\ ;\ \bGamma\Big)}{\Phi_{n}(\bgamma;\bGamma)}.
\]
Exploiting the identities $\bar\bOmega^{-1}\,\bomega^{-1}\,\bOmega=\bomega$ and 
$\bar{\tilde\bOmega}^{-1}\,\tilde\bomega^{-1}\,\tilde\bOmega=\tilde\bomega$,
we obtain
\[
\bDelta^{\intercal}\bar\bOmega^{-1}\bomega^{-1}\bOmega\,\bt =\bDelta^{\intercal}\bomega\,\bt =\big(\tilde\bDelta^{\intercal}\bar{\tilde\bOmega}^{-1}\tilde\bomega^{-1}\tilde\bOmega\big)\bt.
\]
Therefore
\[
\bgamma+\bDelta^{\intercal}\bar\bOmega^{-1}\bomega^{-1}\bOmega\,\bt =\bgamma+\tilde\bDelta^{\intercal}\bar{\tilde\bOmega}^{-1}\tilde\bomega^{-1}\tilde\bOmega\,\bt
\]
and so
\[
M_{\bbeta+\bepsilon}(\bt) =\exp \Big(\bt^{\intercal}\tilde\bxi+\frac{1}{2}\,\bt^{\intercal}\tilde\bOmega\,\bt\Big)\, \frac{\Phi_{n} \Big(\bgamma+\tilde\bDelta^{\intercal}\tilde{\bar\bOmega}^{-1}\tilde\bomega^{-1}\tilde\bOmega\,\bt\ ;\ \bGamma\Big)}{\Phi_{n}(\bgamma;\bGamma)},
\]
which is exactly the MGF of $\sun_{p,n}(\tilde\bxi,\tilde\bOmega,\tilde\bDelta,\bgamma,\bGamma)$ with the correct parametrization.
Since MGFs identify the distribution, the result follows.
\end{proof}

\begin{Lem}\label{le: sun from sun x cdf}
Let
$$p(\bbeta) \propto \mathrm{d} \sun_{p,m}(\bxi,\bOmega,\bDelta,\bgamma,\bGamma)\cdot
\Phi_{n}(\bb + \bA \bbeta; \bSigma),$$
with $\bb\in \bbR^{n}$, $\bA\in \bbR^{n\times p}$ and $\bSigma\in \bbR^{n\times n} \succ 0$ (a full-rank covariance matrix).
Then
\begin{equation*}
p(\bbeta) = \mathrm{d} \sun_{p,m+n}(\tilde\bxi,\tilde\bOmega,\tilde\bDelta,\tilde\bgamma,\tilde\bGamma), \text{ with}
\end{equation*}
\begin{align*}
&\tilde\bxi = \bxi, \quad \tilde\bOmega = \bOmega, \quad \tilde{\bDelta} = [\bDelta \quad \bar\bOmega \bomega \bA^{\intercal}\bs^{-1}] \in \bbR^{p\times(m+n)},\\
&\tilde\bgamma = \begin{bmatrix}
\bgamma\\
\bs^{-1}(\bb + \bA\bxi)
\end{bmatrix} \in \bbR^{m+n}, \quad \tilde\bGamma = \begin{bmatrix}
\bGamma & \bDelta^{\intercal} \bomega \bA^{\intercal}\bs^{-1}\\
\bs^{-1}\bA\bomega\bDelta & \bs^{-1}(\bA\bOmega\bA^{\intercal} + \bSigma)\bs^{-1}
\end{bmatrix} \in \bbR^{(m+n)\times (m+n)},
\end{align*}
where $\bs= [(\bA\bOmega\bA^{\intercal} + \bSigma)\odot\bI_{n}]^{1/2}$.
\end{Lem}

\begin{proof}
Recalling the $\sun_{p,m}$ density kernel we can rewrite
\[
p(\bbeta)\ \propto\ \phi_{p}(\bbeta-\bxi;\bOmega)\, \Phi_{m} \big(\bgamma+\bDelta^{\intercal}\bar\bOmega^{-1}\bomega^{-1}(\bbeta-\bxi);\ \bGamma-\bDelta^{\intercal}\bar\bOmega^{-1}\bDelta\big) \, \cdot \, \Phi_{n}(\bb+\bA\bbeta;\bSigma).
\]
By Lemma~\ref{le: sun from gaussian x cdf}, with $(\bmu,\bLambda)=(\bxi,\bOmega)$, we can rewrite the extra cdf as
\[
    \Phi_{n}(\bb+\bA\bbeta; \bSigma) =\Phi_{n} \big(\bs^{-1}(\bb+\bA\bxi) + \bs^{-1}\bA(\bbeta-\bxi);\ \bs^{-1}\bSigma\bs^{-1}\big),
\]
where $\bs=[(\bA\bOmega\bA^{\intercal}+\bSigma)\odot\bI_{n}]^{1/2}$.

Stack the two cdf factors into a single (m+n)-variate Gaussian cdf by defining
\[
    \tilde\bgamma=\begin{bmatrix}\bgamma\\ \bs^{-1}(\bb+\bA\bxi)\end{bmatrix},\qquad \tilde\bDelta=\begin{bmatrix}\bDelta & \bar\bOmega\,\bomega\,\bA^{\intercal}\,\bs^{-1}\end{bmatrix},
\]
and
\[
\tilde\bGamma=\begin{bmatrix}
    \bGamma & \bDelta^{\intercal}\bomega\,\bA^{\intercal}\bs^{-1}\\
    \bs^{-1}\bA\,\bomega\,\bDelta & \bs^{-1}(\bA\bOmega\bA^{\intercal}+\bSigma)\bs^{-1}
\end{bmatrix}.
\]

We claim that
\[
    \Phi_{m} \big(\bgamma+\bDelta^{\intercal}\bar\bOmega^{-1}\bomega^{-1}(\bbeta-\bxi);\ \bGamma-\bDelta^{\intercal}\bar\bOmega^{-1}\bDelta\big)  \Phi_{n} \big(\bs^{-1}(\bb+\bA\bxi)+\bs^{-1}\bA(\bbeta-\bxi); \, \bs^{-1}\bSigma\bs^{-1}\big)
\]
equals
\[
\Phi_{m+n} \big(\tilde\bgamma+\tilde\bDelta^{\intercal}\bar\bOmega^{-1}\bomega^{-1}(\bbeta-\bxi);\ \tilde\bGamma-\tilde\bDelta^{\intercal}\bar\bOmega^{-1}\tilde\bDelta\big).
\]

Indeed, with $\tilde\bDelta=[\bDelta,\ \bJ]$ where $\bJ=\bar\bOmega\,\bomega\,\bA^{\intercal}\bs^{-1}$, compute
\[
    \tilde\bDelta^{\intercal}\bar\bOmega^{-1}\tilde\bDelta
    =\begin{bmatrix}
    \bDelta^{\intercal}\bar\bOmega^{-1}\bDelta & \bDelta^{\intercal}\bar\bOmega^{-1} \bJ\\
    \bJ^{\intercal}\bar\bOmega^{-1}\bDelta & \bJ^{\intercal}\bar\bOmega^{-1} \bJ
    \end{bmatrix} =
    \begin{bmatrix}
    \bDelta^{\intercal}\bar\bOmega^{-1}\bDelta & \bDelta^{\intercal}\bomega\,\bA^{\intercal}\bs^{-1}\\
    \bs^{-1}\bA\,\bomega\,\bDelta & \bs^{-1}\bA\,\bOmega\,\bA^{\intercal}\bs^{-1}
    \end{bmatrix},
\]
since $\bar\bOmega^{-1} \bJ=\bomega\,\bA^{\intercal}\bs^{-1}$ and $\bJ^{\intercal}\bar\bOmega^{-1} \bJ=\bs^{-1}\bA\,\bOmega\,\bA^{\intercal}\bs^{-1}$.
Therefore
\[
    \tilde\bGamma-\tilde\bDelta^{\intercal}\bar\bOmega^{-1}\tilde\bDelta = \begin{bmatrix}
    \bGamma-\bDelta^{\intercal}\bar\bOmega^{-1}\bDelta & \bzero\\
    \bzero & \bs^{-1}\bSigma\bs^{-1}
    \end{bmatrix}
\]
and
\[
    \tilde\bgamma+\tilde\bDelta^{\intercal}\bar\bOmega^{-1}\bomega^{-1}(\bbeta-\bxi) =
    \begin{bmatrix}
        \bgamma+\bDelta^{\intercal}\bar\bOmega^{-1}\bomega^{-1}(\bbeta-\bxi)\\
        \bs^{-1}(\bb+\bA\bxi)+\bs^{-1}\bA(\bbeta-\bxi)
    \end{bmatrix}.
\]
Hence the product of the two lower dimensional cdfs equals the single $(m+n)$-variate cdf above.
Thus,
\[
    p(\bbeta) \propto \phi_{p}(\bbeta-\bxi;\bOmega) \, \Phi_{m+n} \big(\tilde\bgamma+\tilde\bDelta^{\intercal}\bar\bOmega^{-1}\bomega^{-1}(\bbeta-\bxi); \tilde\bGamma-\tilde\bDelta^{\intercal} \bar\bOmega^{-1} \tilde\bDelta\big),
\]
which is exactly the $\mathrm{d}\sun_{p,m+n}(\tilde\bxi,\tilde\bOmega,\tilde\bDelta,\tilde\bgamma,\tilde\bGamma)$ kernel with $\tilde\bxi=\bxi$ and $\tilde\bOmega=\bOmega$.
\end{proof}

\subsection{Proof of Proposition~\ref{prop:iid_filter_pdt_0}}
Recall that
\[
    \bbeta_{1}\sim \textsc{n}_{p} \big(\bxi_{1},\,\bOmega_{1}\big) \quad\text{with}\quad \bxi_{1}=\ba_{1\mid 0},\ \ \bOmega_{1}=\bP_{1\mid 0}.
\]
From the probit-type emission \eqref{eq:emission_MP}, we can write the likelihood kernel as a Gaussian cdf
\[
    p(\by_{1}\mid\bbeta_{1}) \ =\ \Phi_{m_{1}L} \big(\bar{\bX}_{1,\by_{1}}\,\bbeta_{1}; \,\bLambda_{1}\big),
\]
that is in the form $\Phi_{n}(\bb+\bA\bbeta; \,\bSigma)$ with $n=m_{1}L$, $\bb=\bzero$, $\bA=\bar{\bX}_{1,\by_{1}}$, and $\bSigma=\bLambda_{1}$.
Therefore, the posterior kernel is
\[
    p(\bbeta_{1}\mid\by_{1}) \ \propto\ \phi_{p} \big(\bbeta_{1}-\bxi_{1};\,\bOmega_{1}\big)\ \Phi_{m_{1}L} \big(\bar{\bX}_{1,\by_{1}}\,\bbeta_{1}; \,\bLambda_{1}\big).
\]
Applying Lemma~\ref{le: sun from gaussian x cdf} with $(\bmu,\bLambda)=(\bxi_{1},\bOmega_{1})$, $\bb=\bzero$, $\bA=\bar{\bX}_{1,\by_{1}}$, and $\bSigma=\bLambda_{1}$, we obtain
\[
    \bbeta_{1}\mid\by_{1}\ \sim\ \sun_{p,\,m_{1}L} \big(\bxi_{1},\,\bOmega_{1},\,\bDelta_{1},\,\bgamma_{1},\,\bGamma_{1}\big), \text{ with }
\]
\[
    \bDelta_{1}=\bar{\bOmega}_{1}\,\bomega_{1}\,\bar{\bX}_{1,\by_{1}}^{\intercal}\,\bs_{1}^{-1}, \qquad \bgamma_{1}=\bs_{1}^{-1}\,\bar{\bX}_{1,\by_{1}}\,\bxi_{1}, \qquad \bGamma_{1} = \bs_{1}^{-1}\,\big(\bar{\bX}_{1,\by_{1}}\,\bOmega_{1}\,\bar{\bX}_{1,\by_{1}}^{\intercal}+\bLambda_{1}\big)\,\bs_{1}^{-1},
\]
and $\bs_{1}$ as stated.
Finally, by Lemma~\ref{le: sun from gaussian x cdf} the normalizing constant of the posterior kernel equals the marginal likelihood
\[
    p(\by_{1}) = \Phi_{m_{1}L} \big(\bgamma_{1}; \, \bGamma_{1}\big).
\]

\subsection{Proof of Theorem~\ref{thm:iid_filter}}
First, we prove the prediction part.
From the transition $\bbeta_{t}=\bG_{t}\bbeta_{t-1}+\boldeta_{t}$ with $\boldeta_{t}\sim \textsc{n}_{p}(\mathbf{0},\bW_{t})$ independent of $\bbeta_{t-1}$, apply first the linear map of Lemma~\ref{le: lin transf SUN} with $\bA=\bG_{t}$ to obtain
\[
    \bG_{t}\bbeta_{t-1}\ \sim\ \sun_{p,\,M_{t-1}L} \big(\bG_{t}\bxi_{t-1},\,\bG_{t}\bOmega_{t-1}\bG_{t}^{\intercal},\,\bDelta',\,\bgamma_{t-1},\,\bGamma_{t-1}\big),
\]
where $\bDelta' = \left[(\bG_{t}\bOmega_{t-1}\bG_{t}^{\intercal})\odot \bI_{p}\right]^{-\frac{1}{2}}\,\bG_{t}\,\bomega_{t-1}\,\bDelta_{t-1}$.

Then, adding a Gaussian noise using Lemma~\ref{le: SUN plus Gaussian} with $\bW=\bW_{t}$ we get
\[
    \bbeta_{t}=\bG_{t}\bbeta_{t-1}+\boldeta_{t}\ \sim\ \sun_{p,\,M_{t-1}L} \big(\bxi_{t\mid t-1},\,\bOmega_{t\mid t-1},\,\bDelta_{t\mid t-1},\,\bgamma_{t-1},\,\bGamma_{t-1}\big), \text{ where}
\]
\[
    \bxi_{t\mid t-1}=\bG_{t}\bxi_{t-1},\qquad \bOmega_{t\mid t-1}=\bG_{t}\bOmega_{t-1}\bG_{t}^{\intercal}+\bW_{t},
\]
\[
    \bDelta_{t\mid t-1}=\bomega_{t\mid t-1}^{-1}\,\left[(\bG_{t}\bOmega_{t-1}\bG_{t}^{\intercal})\odot \bI_{p}\right]^{\frac{1}{2}}\,\bDelta' =\bomega_{t\mid t-1}^{-1}\,\bG_{t}\,\bomega_{t-1}\,\bDelta_{t-1},
\]
with $\bomega_{t\mid t-1}=(\bOmega_{t\mid t-1}\odot \bI_{p})^{1/2}$.
Set $\bgamma_{t\mid t-1}=\bgamma_{t-1}$ and $\bGamma_{t\mid t-1}=\bGamma_{t-1}$.

\medskip 

Now we derive the correction.
The likelihood at time $t$ can be written as a Gaussian cdf in the form
\[
    p(\by_{t}\mid \bbeta_{t}) \ =\ \Phi_{m_{t}L} \big(\bar{\bX}_{t,\by_{t}}\,\bbeta_{t}; \,\bLambda_{t}\big).
\]
Multiplying this cdf factor by the SUN prediction kernel and applying Lemma~\ref{le: sun from sun x cdf} with $(\bxi,\bOmega,\bDelta,\bgamma,\bGamma)=(\bxi_{t\mid t-1},\bOmega_{t\mid t-1},\bDelta_{t\mid t-1},\bgamma_{t\mid t-1},\bGamma_{t\mid t-1})$, $\bb=\mathbf{0}$, $\bA=\bar{\bX}_{t,\by_{t}}$, and $\bSigma=\bLambda_{t}$ we obtain
\[
    \bbeta_{t}\mid \by_{1:t}\ \sim\ \sun_{p,\,M_{t}L} \big(\bxi_{t},\bOmega_{t},\bDelta_{t},\bgamma_{t},\bGamma_{t}\big),
\]
with the stated updates
\[
    \bxi_{t}=\bxi_{t\mid t-1},\quad \bOmega_{t}=\bOmega_{t\mid t-1},\quad \bDelta_{t}=
    \begin{bmatrix}
    \bDelta_{t\mid t-1} & \bar{\bOmega}_{t}\,\bomega_{t}\,\bar{\bX}_{t,\by_{t}}^{\intercal}\, \bs_{t}^{-1}
    \end{bmatrix},
\]
\[
    \bgamma_{t}=\begin{bmatrix}\bgamma_{t\mid t-1}\\
    \bs_{t}^{-1}\,\bar{\bX}_{t,\by_{t}}\,\bxi_{t}\end{bmatrix},\qquad \bGamma_{t}=
    \begin{bmatrix}
    \bGamma_{t\mid t-1} & \bDelta_{t\mid t-1}^{\intercal}\,\bomega_{t}\,\bar{\bX}_{t,\by_{t}}^{\intercal}\,\bs_{t}^{-1}\\
    \bs_{t}^{-1}\,\bar{\bX}_{t,\by_{t}}\,\bomega_{t}\,\bDelta_{t\mid t-1} & \bs_{t}^{-1}\big(\bar{\bX}_{t,\by_{t}}\,\bOmega_{t}\,\bar{\bX}_{t,\by_{t}}^{\intercal}+\bLambda_{t}\big)\bs_{t}^{-1}
    \end{bmatrix},
\]
and, following the usual notation, $\bs_{t}=\big[(\bar{\bX}_{t,\by_{t}}\,\bOmega_{t}\,\bar{\bX}_{t,\by_{t}}^{\intercal}+\bLambda_{t})\odot \bI_{m_{t}L}\big]^{1/2}$ and $\bomega_{t}=(\bOmega_{t}\odot \bI_{p})^{1/2}$.

Finally, by Lemma~\ref{le: sun from sun x cdf} the normalizing constant of the posterior kernel equals the marginal likelihood.
Therefore, by the definition of conditional probability, we have
\[
p(\by_{t}\mid \by_{1:t-1})=\frac{\Phi_{M_{t}L} \big(\bgamma_{t};\bGamma_{t}\big)}{\Phi_{M_{t-1}L} \big(\bgamma_{t\mid t-1};\bGamma_{t\mid t-1}\big)}.
\]

\subsection{Proof of Proposition~\ref{prop:ADFcorrection}}
From derivations in Section~\ref{subsec:ADF}, at time $t$ one first projects the approximate ADF filtering of time $t-1$, $h_{t-1}^{\adf}(\bbeta_{t-1})$, on the space of Gaussian densities, obtaining $q_{t-1}^{\adf}(\bbeta_{t-1})=\phi_{p}\left (\bbeta_{t-1} -\ba_{t-1}; \bP_{t-1} \right)$, via a moment matching step.

This induces the Gaussian predictive distribution $q_{t\mid t-1}^{\adf}(\bbeta_{t})=\phi_{p}\left (\bbeta_{t} -\ba_{t\mid t-1}; \bP_{t\mid t-1} \right)$, with $\ba_{t\mid t-1}$ and $\bP_{t\mid t-1}$ defined as in \eqref{eq:predADF}.
Using $q_{t\mid t-1}^{\adf}(\bbeta_{t})$ to approximate $p(\bbeta_{t}\mid \by_{1:t-1})$, the filtering distribution of time $t$, which satisfies 
\begin{equation}
p(\bbeta_{t}\mid \by_{1:t})= \dfrac{p(\bbeta_{t}\mid \by_{1:t-1})p(\by_{t}\mid \bbeta_{t})}{p(\by_{t}\mid \by_{1:t-1})},
\end{equation}
is then approximated with $h_{t}^{\adf}(\bbeta_{t})\propto q_{t\mid t-1}^{\adf}(\bbeta_{t}) p(\by_{t}\mid\bbeta_{t})$, resulting in
\begin{equation}\label{eq:ADFfilterningKernel}
h_{t}^{\adf}(\bbeta_{t})\propto \phi_{p}\left (\bbeta_{t} -\ba_{t\mid t-1}; \bP_{t\mid t-1} \right) \Phi_{m_{t}L}\left(\bar{\bX}_{t,\by_{t}} \bbeta_{t}; \bLambda_{t} \right).
\end{equation}
By Lemma~\ref{le: sun from gaussian x cdf}, the right-hand side of \eqref{eq:ADFfilterningKernel} is the kernel of a $\sun_{p,m_{t}L}(\bxi_{t}^{\adf}, \bOmega_{t}^{\adf}, \bDelta_{t}^{\adf}, \bgamma_{t}^{\adf}, \bGamma_{t}^{\adf})$, with
\begin{equation}
    \begin{split}
        \bxi_{t}^{\adf} &=\ba_{t\mid t-1},
        \quad \bOmega_{t}^{\adf}=\bP_{t\mid t-1},
        \quad \bDelta_{t}^{\adf}=\bar{\bOmega}_{t}^{\adf}{\bomega}_{t}^{\adf} \bar{\bX}_{t,\by_{t}} ^{\intercal} (\bs_{t}^{\adf})^{-1},\\
        \quad \bgamma_{t}^{\adf} &= (\bs_{t}^{\adf})^{-1} \bar{\bX}_{t,\by_{t}} \bxi_{t}^{\adf}, \quad \bGamma_{t}^{\adf} = (\bs_{t}^{\adf})^{-1} (\bar{\bX}_{t,\by_{t}} \bOmega_{t}^{\adf}\bar{\bX}_{t,\by_{t}}^{\intercal}+\bLambda_{t})(\bs_{t}^{\adf})^{-1},
    \end{split}
\end{equation}
with $\bs_{t}^{\adf}= \left[ \left( \bar{\bX}_{t,\by_{t}} \bOmega_{t}^{\adf} \bar{\bX}_{t,\by_{t}}^{\intercal}+\bLambda_{t} \right) \odot \bI_{m_{t} L} \right]^{1/2}$, $\bOmega_{t}^{\adf} = \bomega_{t}^{\adf} \bar{\bOmega}_{t}^{\adf} \bomega_{t}^{\adf}$ and $\bomega_{t}^{\adf} =\left(\bOmega_{t}^{\adf} \odot \bI_{p}\right)^{1/2}$, as stated in Proposition~\ref{prop:ADFcorrection}.

Finally, to obtain an approximation of $p(\by_{t}\mid \by_{1:t-1})$, one simply needs to compute the normalizing constant of \eqref{eq:ADFfilterningKernel}: from the form of a SUN density \eqref{eq:dSUN} adapted to the parameters obtained for $h_{t}^{\adf}(\bbeta_{t})$, one gets $\widehat{p}_{\adf}(\by_{t}\mid \by_{1:t-1})=\Phi_{m_{t}L}(\bgamma_{t}^{\adf}; \bGamma_{t}^{\adf})$, which concludes the proof.

\subsection{Proof of Proposition~\ref{prop:EPcorrection}}
By definition, the hybrid density satisfies
\begin{equation*}
    \begin{split}
    h_{t,i}^{\ep}(\bbeta_{t}) &\propto q_{t,-i}^{\ep}(\bbeta_{t}) \, p(y_{t,i}\mid \bbeta_{t}) \\
    & = \phi_{p}(\bbeta_{t}-\bQ_{t,-i}^{-1}\br_{t,-i}; \bQ_{t,-i}^{-1})\, \Phi_{L}\left ( \bV_{[y_{t,i}]}\bX_{t,i} \bbeta_{t}; \bV_{[y_{t,i}]} \bSigma \bV_{[y_{t,i}]}^{\intercal} \right ),
    \end{split}
\end{equation*}
so, assuming $\bQ_{t,-i}$ is positive-definite, its kernel is proportional to the product of a $p$-variate Gaussian density, with mean $\bQ_{t,-i}^{-1}\br_{t,-i}$ and variance-covariance matrix $\bQ_{t,-i}^{-1}$, and the $L$-variate Gaussian cdf $\Phi_{L}\left ( \bV_{[y_{t,i}]}\bX_{t,i} \bbeta_{t}; \bV_{[y_{t,i}]} \bSigma \bV_{[y_{t,i}]}^{\intercal} \right )$.
The result then follows from Lemma~\ref{le: sun from gaussian x cdf} by taking $\bmu=\bQ_{t,-i}^{-1}\br_{t,-i}$, $\bLambda=\bQ_{t,-i}^{-1}$, $\bb=\bzero_L$, $\bA=\bV_{[y_{t,i}]}\bX_{t,i}$, and the cdf covariance matrix equal to
$\bV_{[y_{t,i}]}\bSigma\bV_{[y_{t,i}]}^{\intercal}$.

\subsection{Derivation of the EP site updates}
\label{sec:supp_EP_site_updates}

We derive the site updates \eqref{eq:EPmomentsZ_update} used in Algorithm~\ref{alg:EP}.
For a fixed time $t$ and site $i$, assume that $\bQ_{t,-i}$ is positive definite.
The cavity density is
\[
    q_{t,-i}^{\ep}(\bbeta_{t}) = \frac{1}{\Psi(\br_{t,-i},\bQ_{t,-i})} \exp\left\{ -\frac{1}{2}\bbeta_{t}^{\intercal}\bQ_{t,-i}\bbeta_{t} + \bbeta_{t}^{\intercal}\br_{t,-i} \right\},
\]
where $\bQ_{t,-i}=\sum_{j\neq i}\bQ_{t,j}$ and $\br_{t,-i}=\sum_{j\neq i}\br_{t,j}$.
Let the updated site be
\[
    q_{t,i}^{\ep,\new}(\bbeta_{t}) = \frac{1}{Z_{t,i}^{\new}} \exp\left\{ -\frac{1}{2}\bbeta_{t}^{\intercal}\bQ_{t,i}^{\new}\bbeta_{t} + \bbeta_{t}^{\intercal}\br_{t,i}^{\new} \right\}.
\]
The updated global approximation is proportional to
$q_{t,i}^{\ep,\new}(\bbeta_{t})q_{t,-i}^{\ep}(\bbeta_{t})$ and has natural parameters $\bQ_{t}^{\ep,\new}=\bQ_{t,-i}+\bQ_{t,i}^{\new}$, $\br_{t}^{\ep,\new}=\br_{t,-i}+\br_{t,i}^{\new}$.
Hence its normalized form is
\[
    q_{t}^{\ep,\new}(\bbeta_{t}) = \frac{1}{\Psi(\br_{t}^{\ep,\new},\bQ_{t}^{\ep,\new})} \exp\left\{ -\frac{1}{2}\bbeta_{t}^{\intercal}\bQ_{t}^{\ep,\new}\bbeta_{t} + \bbeta_{t}^{\intercal}\br_{t}^{\ep,\new} \right\}.
\]
EP updates the site so that the normalizing constant, mean, and covariance matrix of the updated global approximation match those of the hybrid density
\[
    h_{t,i}^{\ep}(\bbeta_{t}) \propto p(y_{t,i}\mid\bbeta_{t})q_{t,-i}^{\ep}(\bbeta_{t}).
\]
Let
\[
    Z_{h_{t,i}^{\ep}} = \int p(y_{t,i}\mid\bbeta_{t})q_{t,-i}^{\ep}(\bbeta_{t})d\bbeta_{t}, \quad \bmu_{h_{t,i}^{\ep}} = \mathbb{E}_{h_{t,i}^{\ep}}(\bbeta_{t}), \quad \bSigma_{h_{t,i}^{\ep}} =\var_{h_{t,i}^{\ep}}(\bbeta_{t}).
\]
The moment matching conditions give
\[
    (\bQ_{t}^{\ep,\new})^{-1} = \bSigma_{h_{t,i}^{\ep}}, \qquad (\bQ_{t}^{\ep,\new})^{-1}\br_{t}^{\ep,\new} = \bmu_{h_{t,i}^{\ep}}.
\]
Therefore $\bQ_{t,i}^{\new} = \bSigma_{h_{t,i}^{\ep}}^{-1} - \bQ_{t,-i}$, and $\br_{t,i}^{\new}= \bSigma_{h_{t,i}^{\ep}}^{-1} \bmu_{h_{t,i}^{\ep}}- \br_{t,-i}= \left( \bQ_{t,-i}+ \bQ_{t,i}^{\new}\right) \bmu_{h_{t,i}^{\ep}} - \br_{t,-i}$.
Finally, the normalizing constant condition is
\[
    \int q_{t,i}^{\ep,\new}(\bbeta_{t}) q_{t,-i}^{\ep}(\bbeta_{t}) d\bbeta_{t} = Z_{h_{t,i}^{\ep}}.
\]
The left hand side equals
\[
    \frac{ \Psi(\br_{t}^{\ep,\new},\bQ_{t}^{\ep,\new}) }{ Z_{t,i}^{\new}  \Psi(\br_{t,-i},\bQ_{t,-i}) },
\]
and hence
\[
    Z_{t,i}^{\new} = \frac{ \Psi(\br_{t}^{\ep,\new},\bQ_{t}^{\ep,\new})}{ \Psi(\br_{t,-i},\bQ_{t,-i}) Z_{h_{t,i}^{\ep}} }.
\]

\subsection{Derivation of the EP approximate marginal likelihood}

The form of the sequential EP approximation of the marginal likelihood \eqref{eq:EPmarginalLikelihood} follows from replacing the predictive and emission distributions with the corresponding approximated quantities in the exact formula
\[
p (\by_{t}\mid \by_{1:t-1}) = \int \prod_{i=1}^{m_{t}} p(y_{t,i}\mid \bbeta_{t}) p(\bbeta_{t}\mid \by_{1:t-1}) d\bbeta_{t},
\]
as in usual EP procedures \citep{vehtari2020expectation,fasano2025multivariate}, with the convention that $p (\by_{1}\mid \by_{1:0})=p (\by_{1})$.
Thus, one gets
\begin{equation}
\begin{split}
    \widehat{p}_{\ep} (\by_{t}\mid \by_{1:t-1})
    &=\int \prod_{i=1}^{m_{t}} q_{t,i}^{\ep} (\bbeta_{t}) q_{t,0}^{\ep} (\bbeta_{t}) d\bbeta_{t}\\
    &=\dfrac{1}{\Psi\left(\br_{t,0},\bQ_{t,0}\right)\prod_{i=1}^{m_{t}} Z_{t,i}}\int \exp\left\{-\tfrac{1}{2} \bbeta_{t}^{\intercal} \left(\sum_{i=0}^{m_{t}} \bQ_{t,i}\right)\bbeta_{t} + \bbeta_{t}^{\intercal}\left(\sum_{i=0}^{m_{t}} \br_{t,i}\right) \right\} d \bbeta_{t} \\
    &=\dfrac{1}{\Psi\left(\br_{t,0},\bQ_{t,0}\right)\prod_{i=1}^{m_{t}} Z_{t,i}}\int \exp\left\{-\tfrac{1}{2} \bbeta_{t}^{\intercal} \bQ_{t}^{\ep}\bbeta_{t} + \bbeta_{t}^{\intercal}\br_{t}^{\ep} \right\} d \bbeta_{t}\\
    &=\dfrac{\Psi\left(\br_{t}^{\ep},\bQ_{t}^{\ep}\right)}{\Psi\left(\br_{t,0},\bQ_{t,0}\right)\prod_{i=1}^{m_{t}} Z_{t,i}}.
\end{split}
\end{equation}

\section{Computational scaling of the filtering algorithms}
\label{sec:supp_computational_scaling}

We provide here a comparison of the computational scaling of the three filtering strategies considered in the manuscript.
Motivated by the online inference setting, the goal is to isolate the dependence on the time index $t$ and on the number of observations $m_{t}$ available at each time point.
Accordingly, in this discussion we regard $p$, $L$, the number $R$ of Monte Carlo samples, and the number of EP sweeps as fixed or bounded constants.

All SUN samplers used in the algorithms rely on the additive representation in \eqref{eq:additiveRepresentationSUN}.
Thus, sampling from a $\sun_{p,q}$ distribution requires sampling from a $p$-variate Gaussian distribution and a $q$-variate truncated normal distribution, with the latter representing the computational bottleneck.
In our implementation, this step is performed using the minimax-tilting accept-reject sampler of \citet{botev2017normal}, as implemented in the R package \texttt{TruncatedNormal}.
Moreover, whenever moments of a SUN distribution are required, as in ADF and sequential EP, we estimate them by Monte Carlo from i.i.d.\ SUN samples, rather than evaluating the exact moment formulae involving Gaussian cdfs and their derivatives.

Alternatively, one could compute SUN moments from the exact moment formulae \citep{arellano2022some}, which involve Gaussian cdfs and their derivatives, rather than by Monte Carlo SUN sampling.
This replaces the computation of the relevant moments via accept-reject sampling with repeated numerical evaluations of high-dimensional Gaussian probabilities having cost $\mathcal{O}(q^3)$ each \citep{botev2017normal, cao2021exploiting}.
However, because many such evaluations are required for the mean and covariance matrix, this route is slower in implementations in typical real-data settings.

Let $\alpha(q)$ denote the acceptance probability of the truncated normal accept-reject step.
In a conservative worst-case scenario, the inverse acceptance probability grows exponentially with $q$, namely $\alpha(q)^{-1} = \mathcal{O}(\exp(cq))$, $c>0$, although it can be improved in more favorable settings \citep{botev2017normal}.
Thus, the cost of obtaining $R$ accepted SUN draws, and hence also the cost of Monte Carlo moment estimation from $R$ SUN draws, is $\mathcal{O}(R\exp(cq))$.

For the exact i.i.d.\ filter (Algorithm~\ref{alg:IID}), the relevant truncated normal dimension is $q_{t}^{\mathrm{iid}} = M_{t}L$, with $M_{t}=\sum_{s=1}^{t}m_s$.
The leading computational cost of the i.i.d.\ filter at a given time point $t$ is therefore
\[
    \mathcal{C}_{\mathrm{iid}}(t) = \mathcal{O}\left( \exp(c M_{t}) \right),
\]
which simplifies to $\mathcal{C}_{\mathrm{iid}}(t) = \mathcal{O}\left( \exp(c m t) \right)$, in the balanced case $m_{t}=m$ for all $t$.
So the cost grows exponentially in both time $t$ and sample size $m$.

For the ADF filter (Algorithm~\ref{alg:ADF}), the relevant truncated normal dimension is $q_{t}^{\adf}=m_{t}L$, instead.
Thus, the leading computational cost of the ADF Monte Carlo moment calculations is
\[
    \mathcal{C}_{\adf}(t) = \mathcal{O}\left( \exp(c m_{t}) \right).
\]
which is constant in time $t$ if $m_{t}$ is uniformly bounded.
However, the cost still grows exponentially with the current number of observations $m_{t}$.

For the sequential EP filter (Algorithm~\ref{alg:EP}), each likelihood contribution is handled separately through a hybrid density.
By Proposition~\ref{prop:EPcorrection}, the truncated normal dimension of the hybrid density associated with each site update is $q_{t,i}^{\ep}=L$,
which does not depend on either $t$ or $m_{t}$.
Let $K_{t}$ denote the number of EP sweeps performed at time $t$, which is upper-bounded by the maximum number of iterations allowed in the implementation.
Thus, the computational cost is
\[
    \mathcal{C}_{\ep}(t) = \mathcal{O}\left( K_{t} m_{t} R\exp(cL) \right) = \mathcal{O}\left( m_{t} \right).
\]
As a consequence, sequential EP has constant cost per site update with respect to both $t$ and $m_{t}$, while its total cost at time $t$ is linear in the number of sites $m_{t}$.

The resulting comparison is summarized in Table~\ref{tab:computational_scaling}.
The table highlights that the exact i.i.d.\ sampler is affected by the cumulative growth of the SUN dimension, ADF removes the accumulation over time but still depends exponentially on the current sample size, while sequential EP keeps the SUN dimension fixed at the site level and therefore scales linearly in the number of observations processed.

\begin{table}[ht]
\centering
\caption{Computational scaling of the three filtering strategies.
Here $c$ is a positive constant.}
\label{tab:computational_scaling}
\begin{tabular}{ll}
\toprule
Method & Cost at time $t$ \\
\midrule
i.i.d.\ filter & $\mathcal{O}\left( \exp(c M_{t}) \right)$\\
ADF & $\mathcal{O}\left( \exp(c m_{t}) \right)$ \\
Sequential EP & $\mathcal{O}\left( m_{t} \right)$ \\
\bottomrule
\end{tabular}
\end{table}

\section{Sequential importance sampling for the dynamic multinomial probit}
\label{sec:sis_supp}
For completeness, we briefly summarize the standard sequential importance sampling (resampling) strategy considered in the manuscript for online inference in the dynamic MNP model \eqref{eq:DMNP_lik}--\eqref{eq:init_dist_MP}.
Such a strategy is also known as bootstrap particle filtering.
For more details on sequential importance sampling (resampling) algorithms, we refer, for instance, to \citet{gordon1993novel,doucet2001sequential, doucet2009tutorial,chopin2020introduction}.
The aim is to approximate recursively the filtering distribution $p(\bbeta_{t} \mid \by_{1:t})$ and the predictive distributions $p(\by_{t} \mid \by_{1:t-1})$.

Sequential importance sampling represents the filtering distribution at time $t$ through a weighted particle system
\[
    \{ \bbeta_{1:t}^{(r)}, w_{t}^{(r)} \}_{r=1}^{R},
\]
where $\bbeta_{1:t}^{(r)}$ denotes the $r$th sampled trajectory up to time $t$, and $w_{t}^{(r)}$ is the associated normalized importance weight.
In the particle filter used in the manuscript, particles are propagated according to the state transition
\[
p(\bbeta_{t} \mid \bbeta_{t-1}) = \phi_{p}(\bbeta_{t} - \bG_{t} \bbeta_{t-1}; \bW_{t}),
\]
while the weights are updated through the likelihood
\[
p(\by_{t} \mid \bbeta_{t}) = \Phi_{m_{t} L}\!\left(\bar{\bX}_{t,\by_{t}}\bbeta_{t}; \bLambda_{t}\right).
\]

More precisely, suppose that at time $t-1$ we have particles
\[
\{ \bbeta_{1:t-1}^{(r)}, w_{t-1}^{(r)} \}_{r=1}^{R}
\]
approximating $p(\bbeta_{1:t-1}\mid \by_{1:t-1})$.
Then, for each $r=1,\ldots,R$, the sequential importance sampling update proceeds as follows:
\begin{enumerate}
    \item sample
    \[
        \bbeta_{t}^{(r)} \sim p(\bbeta_{t} \mid \bbeta_{t-1}^{(r)}) = \phi_{p}(\bbeta_{t} - \bG_{t} \bbeta_{t-1}^{(r)}; \bW_{t}),
    \]
    and set $\bbeta_{1:t}^{(r)} = (\bbeta_{1:t-1}^{(r)}, \bbeta_{t}^{(r)})$;
    
    \item compute the unnormalized weights
    \[
    \tilde w_{t}^{(r)} = w_{t-1}^{(r)} \, p(\by_{t} \mid \bbeta_{t}^{(r)}) = w_{t-1}^{(r)} \prod_{i=1}^{m_{t}} \Phi_{L}\!\left(\bV_{[y_{t,i}]}\bX_{t,i}\bbeta_{t}^{(r)};\, \bV_{[y_{t,i}]}\bSigma\bV_{[y_{t,i}]}^{\intercal}\right);
    \]
    
    \item normalize
    \[
        w_{t}^{(r)} = \frac{\tilde w_{t}^{(r)}}{\sum_{j=1}^{R} \tilde w_{t}^{(j)}}.
    \]
\end{enumerate}

The resulting weighted empirical measure
\[
    \sum_{r=1}^{R} w_{t}^{(r)} \, \delta_{\bbeta_{t}^{(r)}}(d\bbeta_{t})
\]
provides a particle approximation of the filtering distribution $p(\bbeta_{t} \mid \by_{1:t})$.
In addition, the predictive distribution can be estimated by
\[
    \widehat p(\by_{t} \mid \by_{1:t-1}) = \int p(\by_{t}\mid \bbeta_{t})\sum_{r=1}^{R} w_{t-1}^{(r)} \delta_{\bbeta_{t}^{(r)}}(\bbeta_{t})\,d\bbeta_{t} = \sum_{r=1}^{R} \tilde w_{t}^{(r)},
\]
when normalized weights at time $t-1$ are used.
Hence, the full marginal likelihood can be approximated recursively as
\[
    \widehat p_{\textsc{PF}}(\by_{1:T}) = \prod_{t=1}^{T} \widehat p_{\textsc{PF}}(\by_{t} \mid \by_{1:t-1}).
\]

In practice, sequential importance sampling is typically combined with a resampling step in order to mitigate particle degeneracy.
In the implementation considered in the manuscript, resampling is performed after weight normalization, producing an equally weighted sample of particles to be propagated at the next iteration.
Although conceptually straightforward, this strategy becomes computationally demanding in the present framework, since each weight update (i.e., for each particle $r=1,\ldots, R$) requires the evaluation of a product of $m_{t}$ $L$-variate Gaussian cdfs in \eqref{eq:emission_MP}.

For the SMC implementation, we evaluate the $L$-variate Gaussian CDFs for all $R$ particles simultaneously using \texttt{lpmvnorm} from the \texttt{mvtnorm} package in \texttt{R} \citep{rcoreteam2024r}.
For each observation $i$, the covariance matrix entering the CDF is common to all particles, whereas the upper integration limits vary across particles.
We therefore compute its Cholesky factor once and reuse it across the $R$ evaluations.

\section{Details of the EM algorithm for parameter inference}
\label{sec:app_param_inf}
We report in this section the derivations of the EM algorithm for the estimation of $\bSigma$ introduced in Section~\ref{sec:5}.
In Section~\ref{subsec:S.4.1} we present details about the constrained maximization of $Q(\bSigma\mid \bSigma')$ subject to $\text{tr}(\bSigma)=L$.
This optimization requires $\bS = \sum_{t=1}^{T} \sum_{i=1}^{m_{t}} \bbE_{p(\bepsilon_{t,i} \mid \by_{1:T};\bSigma')} \left[\bepsilon_{t,i} \bepsilon_{t,i}^{\intercal} \right]$.
Section~\ref{subsec:S.4.3} is dedicated to the computation of such a quantity.
This is obtained by leveraging approximate Gaussian smoothing distributions for $\bbeta_{t}$, developed in Section~\ref{subsec:S.4.2}, obtained by backward recursions applied to the approximate Gaussian filtering distributions returned by either the ADF or EP filters.

\subsection{Constrained optimization details}
\label{subsec:S.4.1}
The M-step consists of maximizing $Q(\bSigma\mid\bSigma')$
over $\bSigma\in\mathcal{S}_{L}$, where
\[
\mathcal{S}_{L} = \left\{ \bSigma\in\bbR^{L\times L}: \bSigma=\bSigma^{\intercal},\ \bSigma\succ\boldsymbol{0},\ \operatorname{tr}(\bSigma)=L \right\}.
\]
The Lagrangian writes
\begin{equation}
\label{eq:lagrangian}
\mathcal{L}(\bSigma,\lambda) = \sum_{t=1}^{T} \sum_{i=1}^{m_{t}} \bbE_{p(\bepsilon_{t,i} \mid \by_{1:T};\bSigma')} \left [ \log p(\bepsilon_{t,i}; \bSigma) \right ] +\lambda(\text{tr}(\bSigma)-L).
\end{equation}

\paragraph{First order conditions.}

In order to obtain first-order conditions we need to compute the gradient of the Lagrangian and set it to zero.
We start by considering partial derivatives with respect to $\bSigma$ (the partial derivative with respect to $\lambda$ is trivial).
Focusing on the first part of the Lagrangian and exchanging derivative with expectation, we get:
\begin{equation*}
\begin{split}
   &\frac{\partial}{\partial \bSigma} \bbE_{p(\bepsilon_{1:T} \mid \by_{1:T};\bSigma^{\prime})} \left \{ \sum_{t=1}^{T} \sum_{i=1}^{m_{t}} \log p(\bepsilon_{t,i}; \bSigma) \right \} = \sum_{t=1}^{T} \sum_{i=1}^{m_{t}} \bbE_{p(\bepsilon_{t,i}\mid \by_{1:T};\bSigma^{\prime})} \left \{ \frac{\partial}{\partial \bSigma} \log p(\bepsilon_{t,i}; \bSigma) \right \}.
\end{split}
\end{equation*}
Note that
\begin{equation}
\label{eq:p_epsilon}
    \log p(\bepsilon_{t,i}; \bSigma) = \log\phi_{L}(\bepsilon_{t,i}; \bSigma) = -\frac{1}{2} \log \det(\bSigma) - \frac{1}{2} \bepsilon_{t,i}^{\intercal} \bSigma^{-1} \bepsilon_{t,i} +\mathrm{const}.
\end{equation}
Thus, we obtain (see equations (57) and (61) in \citet{petersen2012matrix})
\begin{align*}
    & \frac{\partial}{\partial \bSigma} \log p(\bepsilon_{t,i}; \bSigma) = -\frac{1}{2} \bSigma^{-1} + \frac{1}{2} \bSigma^{-1} \bepsilon_{t,i} \bepsilon_{t,i}^{\intercal} \bSigma^{-1},
\end{align*}
from which it follows
\begin{equation*}
\begin{split}
    & \bbE_{p(\bepsilon_{t,i}\mid \by_{1:T};\bSigma^{\prime})} \left \{ \frac{\partial}{\partial \bSigma} \log p(\bepsilon_{t,i}; \bSigma) \right \} = -\frac{1}{2} \bSigma^{-1} + \frac{1}{2} \bSigma^{-1} \bbE_{p(\bepsilon_{t,i}\mid \by_{1:T};\bSigma^{\prime})} \left [ \bepsilon_{t,i} \bepsilon_{t,i}^{\intercal} \right ] \bSigma^{-1}.
\end{split}
\end{equation*}

Trivially $\dfrac{\partial}{\partial \bSigma} \text{tr}(\bSigma) = \bI$, see (99) of \citet{petersen2012matrix}.
Thus, calling $M_{T}=\sum_{t=1}^{T} m_{t}$ and $\bS$ as above, first order conditions for partial derivatives with respect to $\bSigma$ write
\begin{equation*}
    \begin{split}
     \dfrac{\partial}{\partial \bSigma} \mathcal{L}(\bSigma,\lambda) = 0 &\iff \sum_{t=1}^{T} \sum_{i=1}^{m_{t}} \bbE_{p(\bepsilon_{t,i} \mid \by_{1:T};\bSigma')}\left[ -\frac{1}{2} \bSigma^{-1} + \frac{1}{2} \bSigma^{-1} \bepsilon_{t,i} \bepsilon_{t,i}^{\intercal} \bSigma^{-1} \right] +\lambda \bI =0\\
     &\iff -\frac{1}{2} M_{T} \bSigma^{-1} + \frac{1}{2} \bSigma^{-1} \bS \bSigma^{-1} +\lambda\bI=0\\
     &\iff -\frac{1}{2} M_{T} \bSigma + \frac{1}{2} \bS +\lambda \bSigma^{2}=0\\
     &\iff \lambda \bSigma^{2} -\frac{1}{2} M_{T} \bSigma + \frac{1}{2} \bS=0\\
    \end{split}
\end{equation*}
Since $\bSigma$ and $\bS$ are symmetric matrices and commute ($\bSigma\bS=\bS\bSigma$), we are guaranteed they are simultaneously diagonalizable by some orthogonal matrix $\bQ$, see Theorem 1.3.21 in \citet{horn2012matrix}.
That is, there exists an orthogonal matrix $\bQ$ such that $\bSigma = \bQ \, \diag(\sigma_{1}(\lambda),\ldots,\sigma_{L}(\lambda)) \, \bQ^{\intercal}$ and $\bS=\bQ \, \diag(s_{1},\ldots,s_{L}) \, \bQ^{\intercal}$, where $\sigma_{l}(\lambda)$ and $s_{l}$, $l = 1, \ldots, L$, are the eigenvalues of $\bSigma$ and $\bS$ respectively.
Thus,
\begin{equation}
\label{eq:gradSigma}
    \begin{split}
        \dfrac{\partial}{\partial \bSigma} \mathcal{L}(\bSigma,\lambda) = 0 &\iff \bQ[\lambda \diag(\sigma_{1}(\lambda),\ldots,\sigma_{L}(\lambda))^{2} -\frac{1}{2}M_{T}\diag(\sigma_{1}(\lambda),\ldots,\sigma_{L}(\lambda))\\
        &\quad + \frac{1}{2} \diag(s_{1},\ldots,s_{L}) ]\bQ^{\intercal}=0\\
        &\iff \lambda\sigma_{l}(\lambda)^{2} -\frac{1}{2}M_{T}\sigma_{l}(\lambda) + \frac{1}{2}s_{l} = 0,\quad \text{for all } l=1,\ldots,L.
    \end{split}
\end{equation}

Throughout the remainder of this subsection, we assume $\bS\succ\boldsymbol{0}$.
For the exact E-step, this follows because $\bSigma^{\prime}\succ\boldsymbol{0}$ and the conditional distribution of each $\bepsilon_{t,i}$ is absolutely continuous and hence is not supported on a proper linear subspace.
Therefore,
\[
\ba^{\intercal} \bbE\left[ \bepsilon_{t,i}\bepsilon_{t,i}^{\intercal} \mid\by_{1:T};\bSigma^{\prime} \right] \ba = \bbE\left[ (\ba^{\intercal}\bepsilon_{t,i})^{2} \mid\by_{1:T};\bSigma^{\prime} \right] >0
\]
for every nonzero $\ba\in\bbR^{L}$.
The Monte Carlo estimator of $\bS$ is also positive definite almost surely.
Indeed, since $R>L$ and the draws used to estimate each conditional second moment come from a full-dimensional continuous distribution, every estimated conditional second-moment matrix is positive definite almost surely, and so is their sum.

Solving \eqref{eq:gradSigma}, one gets
\begin{equation*}
  \sigma_{l}(\lambda) = \frac{\frac{M_{T}}{2}\pm\sqrt{\left(\frac{M_{T}}{2}\right)^{2} - 2\lambda s_{l}}}{2\lambda} = \frac{M_{T}}{4\lambda}\Big(1 \pm \sqrt{1 - \frac{8\lambda}{M_{T}}\,v_{l}}\Big),
\end{equation*}
with $v_{l}=s_{l}/M_{T}$.

\paragraph{The natural candidate solution.}
The ``$1-\sqrt{\cdot}$'' branch is the natural candidate for each $l=1,\ldots,L$, since it ensures that for $\lambda\to 0$ we obtain the unconstrained solution $\bSigma=\dfrac{1}{M_{T}}\bS=\bQ\diag(v_{1},\ldots,v_{L})\bQ^{\intercal}$.
Indeed, as $\lambda\to 0$ one gets:
\begin{align}
  \sigma_{l}(\lambda) &= \frac{M_{T}}{4\lambda}\Big(1 - \big[\,1-\frac{4\lambda}{M_{T}}v_{l}+o(\lambda)\,\big]\Big) \;=\; v_{l} + o(1),
\end{align}
which matches the unconstrained solution $\sigma_{l}=v_{l}$.
Thus, the natural candidate solution is
\begin{equation}\label{eq:sigma_lambda}
  \sigma_{l}(\lambda) = \frac{M_{T}}{4\lambda}\Big(1 - \sqrt{1 - \frac{8\lambda}{M_{T}}\,v_{l}}\Big), \quad l=1,\ldots,L,
\end{equation}
where $\lambda$ satisfies the feasibility condition
\begin{equation}
    1 - \frac{8\lambda}{M_{T}}\,v_{l} \;\ge\; 0\quad\text{for all } l \Longleftrightarrow \lambda \le \lambda_{\max} := \dfrac{M_{T}}{8 v_{\max}},
    \label{eq:disc}
\end{equation}
with $v_{\max}=\max_{l=1,\dots,L}\{v_{l}\}$, and is such that $\sum_{l=1} ^{L}\sigma_{l}(\lambda)=L$, as derived below.
Setting to zero the partial derivative of the Lagrangian with respect to $\lambda$ one indeed has
\begin{equation*}
     \dfrac{\partial}{\partial \lambda} \mathcal{L}(\bSigma,\lambda) = 0 \iff \text{tr}(\bSigma)=L \iff \sum_{l=1}^{L} \sigma_{l}(\lambda) = L
\end{equation*}

Using \eqref{eq:sigma_lambda} we call
\begin{equation}\label{eq:g_lambda}
  g(\lambda) \;=\; \sum_{l=1}^{L} \sigma_{l}(\lambda) - L \;=\; \sum_{l=1}^{L}\frac{M_{T}}{4\lambda}\Big(1 - \sqrt{1-\frac{8\lambda}{M_{T}}\,v_{l}}\Big) - L,
\end{equation}
and we want to find the roots of this equation to obtain a stationary candidate.
One can verify $g'(\lambda)>0$ on the domain defined by \eqref{eq:disc}, see Lemma~\ref{lemma:g_incr} below, so $g$ is strictly increasing.
At $\lambda=0$, we take the continuous limit $g(0) = \sum_{l=1}^{L} v_{l}-L$.
Moreover, direct inspection of \eqref{eq:sigma_lambda} gives $\lim_{\lambda\to-\infty}\sigma_{l}(\lambda)=0$, $l=1,\ldots,L$, and hence $\lim_{\lambda\to-\infty}g(\lambda)=-L<0$.
Since $g$ is strictly increasing and $g(0)=\sum_{l=1}^L v_{l}-L$, the following cases arise:
\begin{itemize}
    \item If $\sum_{l=1}^{L}v_{l}>L$, then $g(0)>0$, whereas $\lim_{\lambda\to-\infty}g(\lambda)=-L<0$.
    Therefore, continuity and strict monotonicity imply that there is a unique root $\lambda^{\star}<0$.
    \item If $\sum_{l=1}^L v_{l}=L$, the root is $\lambda^{\star}=0$.
    \item If $\sum_{l=1}^L v_{l}<L$, a positive root exists if and only if $g(\lambda_{\max})\geq0$.
    The root is interior when $g(\lambda_{\max})>0$ and lies at the boundary when $g(\lambda_{\max})=0$.
\end{itemize}
Solving $g(\lambda)=0$ gives $\lambda^{\star}$; the stationary candidate is then given by
\begin{equation}
  \bSigma^{\star} = \bQ\,\mathrm{diag}\!\big(\sigma_{1}(\lambda^{\star}),\ldots,\sigma_{L}
  (\lambda^{\star})\big)\,\bQ^{\intercal},\qquad
  \text{with }\sum_{l=1}^L \sigma_{l}(\lambda^{\star})=L.
\end{equation}

\begin{Lem}\label{lemma:g_incr}
The function $g(\lambda)$ defined in \eqref{eq:g_lambda}, continuously extended at $\lambda=0$, is strictly increasing on $(-\infty,\lambda_{\max}]$.
\end{Lem}

\begin{proof}
    The function $g(\lambda)$ is defined for those $\lambda$ where the discriminants are nonnegative, i.e.\ $1-\frac{8\lambda}{M_{T}}v_{l}\ge 0$ for all $l$, and with the continuous extension at $\lambda=0$ given by $\sigma_{l}(0)=v_{l}$.
    For $\lambda<\lambda_{\max}$, set $b_{l}=\frac{8}{M_{T}}v_{l}$ and $u_{l}(\lambda)=\sqrt{1-b_{l}\lambda}\in(0,\infty)$.
    Then
    \begin{equation}
      \sigma_{l}(\lambda)=\frac{M_{T}}{4\lambda}\,\big(1-u_{l}(\lambda)\big).
    \end{equation}
    Differentiating and using $u_{l}'(\lambda)=-\frac{b_{l}}{2u_{l}(\lambda)}$,
    \begin{align}
      \sigma_{l}'(\lambda) &= \frac{M_{T}}{4}\,\frac{\big(\frac{b_{l}}{2u_{l}(\lambda)}\big)\lambda - (1-u_{l}(\lambda))}{\lambda^{2}} \;=\; \frac{M_{T}}{4}\,\frac{\frac{1-u_{l}(\lambda)^{2}}{2u_{l}}- (1-u_{l}(\lambda))}{\lambda^{2}} \nonumber\\
       &= \frac{M_{T}}{4}\,\ \frac{1-u_{l}(\lambda)^{2} - 2u_{l}(\lambda) +2u_{l}(\lambda)^{2}}{2u_{l}(\lambda)\,\lambda^{2}} \nonumber\\
      &= \frac{M_{T}}{4}\,\frac{(1-u_{l}(\lambda))^{2}}{2u_{l}(\lambda)\,\lambda^{2}} \;>\; 0 \qquad \text{for all } \lambda\in(-\infty,\lambda_{\max})\setminus\{0\},
    \end{align}
    since $M_{T}>0$, $u_{l}(\lambda)>0$, and $\lambda^{2}>0$ for $\lambda\in(-\infty,\lambda_{\max})\setminus\{0\}$.
    Moreover, as $\lambda\to 0$ one has $1-u_{l}(\lambda)=\frac{b_{l}}{2}\lambda+o(\lambda)$, hence
    \begin{equation}
      \sigma_{l}'(\lambda)\;\xrightarrow[\lambda\to 0]{}\;\frac{M_{T}}{4}\cdot\frac{(b_{l}^{2}/4)\lambda^{2}}{2\cdot 1 \cdot \lambda^{2}} \;=\; \frac{M_{T} b_{l}^{2}}{32} \;=\; \frac{2 v_{l}^{2}}{M_{T}} \;>\; 0,
    \end{equation}
    since $v_{l}=s_{l}/M_{T}>0$ for every $l=1,\ldots,L$ as $\bS\succ0$.
    Therefore, each $\sigma_{l}(\lambda)$ is strictly increasing on $(-\infty,\lambda_{\max})$, using the continuity at $\lambda=0$.
    Hence, $g$ is strictly increasing on $(-\infty,\lambda_{\max})$.
    Since $g$ is continuous at $\lambda_{\max}$, the result extends to $(-\infty,\lambda_{\max}]$.
\end{proof}

\paragraph{Second order conditions for the natural candidate solution.}
We now verify that the candidate solution obtained from the ``$1-\sqrt{\cdot}$'' branch satisfies the second-order sufficient conditions.
Let $\bD$ be a nonzero symmetric perturbation satisfying $\operatorname{tr}(\bD)=0$.
The curve $\bSigma(u)=\bSigma^{\star}+u\bD$
then preserves the trace constraint for every $u$, since $\operatorname{tr}\{\bSigma(u)\} = \operatorname{tr}(\bSigma^{\star}) + u\operatorname{tr}(\bD) = L$.
Moreover, since $\bSigma^{\star}\succ0$, the matrix $\bSigma(u)$ remains positive definite for all sufficiently small $u$.

Recalling the form of the log-density $ \log p(\bepsilon_{t,i};\bSigma)$ in \eqref{eq:p_epsilon} and exploiting the identity $\bx^{\intercal}\bA\bx = \operatorname{tr}(\bA\bx\bx^{\intercal})$, the
Lagrangian \eqref{eq:lagrangian} is
\begin{equation}
    \mathcal{L}(\bSigma,\lambda) = -\frac{M_{T}}{2}\log\det(\bSigma) -\frac{1}{2}\operatorname{tr}(\bSigma^{-1}\bS) + \lambda\{\operatorname{tr}(\bSigma)-L\} +\mathrm{const}.
    \label{eq:Lagrangian_explicit}
\end{equation}
Define the restriction of the Lagrangian to the feasible curve
$\bSigma(u)=\bSigma^{\star}+u\bD$ as
\begin{equation}
    \ell_{\bD}(u) = \mathcal{L}(\bSigma(u),\lambda^{\star}).
\end{equation}
The second directional derivative of the Lagrangian along $\bD$ is
then defined as
\begin{equation*}
    \mathrm{D}_{\bD}^{2} \mathcal{L}(\bSigma^{\star},\lambda^{\star}) = \ell_{\bD}''(0) = \left. \frac{\mathrm{d}^{2}}{\mathrm{d}u^{2}} \mathcal{L}(\bSigma(u),\lambda^{\star}) \right|_{u=0}.
\end{equation*}
By standard properties of matrix differentials (see, e.g., equations (43), (36), and (40) in \citet{petersen2012matrix}), it holds
\begin{align*}
    &\frac{\mathrm{d}}{\mathrm{d}u} \log\det\{\bSigma(u)\} = \operatorname{tr}\{\bSigma(u)^{-1}\bD\}, \quad \frac{\mathrm{d}}{\mathrm{d}u} \operatorname{tr}\{\bSigma(u)^{-1} \bS\} = \operatorname{tr}\left\{\frac{\mathrm{d}}{\mathrm{d}u} \bSigma(u)^{-1} \bS\right\}\\
    &\frac{\mathrm{d}}{\mathrm{d}u} \bSigma(u)^{-1} = -\bSigma(u)^{-1}\bD\bSigma(u)^{-1}, \quad \frac{\mathrm{d}}{\mathrm{d}u}\operatorname{tr}\{\bSigma(u)\} = \operatorname{tr}\{\bD\} = 0.
\end{align*}
So one gets
\begin{align}
    \ell_{\bD}'(u) &= -\frac{M_{T}}{2} \operatorname{tr}\{\bSigma(u)^{-1}\bD\} + \frac{1}{2} \operatorname{tr} \left\{ \bSigma(u)^{-1}\bD \bSigma(u)^{-1}\bS \right\},
\end{align}
and therefore
\begin{align}
    \ell_{\bD}''(u) &= \frac{M_{T}}{2} \operatorname{tr} \left\{ \bSigma(u)^{-1}\bD \bSigma(u)^{-1}\bD \right\} - \operatorname{tr} \left\{ \bSigma(u)^{-1}\bD \bSigma(u)^{-1}\bD \bSigma(u)^{-1}\bS \right\}.
\end{align}
Evaluating at $u=0$ gives
\begin{align}
    \mathrm{D}_{\bD}^{2} \mathcal{L}(\bSigma^{\star},\lambda^{\star}) &= \frac{M_{T}}{2} \operatorname{tr} \left\{ (\bSigma^{\star})^{-1}\bD (\bSigma^{\star})^{-1}\bD \right\} - \operatorname{tr} \left\{ (\bSigma^{\star})^{-1}\bD (\bSigma^{\star})^{-1}\bD (\bSigma^{\star})^{-1}\bS \right\}.
    \label{eq:second_directional_general}
\end{align}

At the stationary point, the first-order condition implies $\bS=M_{T}\bSigma^{\star}-2\lambda^{\star}(\bSigma^{\star})^{2}$.
Substituting this into
\eqref{eq:second_directional_general} yields
\begin{align}
    \mathrm{D}_{\bD}^{2} \mathcal{L}(\bSigma^{\star},\lambda^{\star}) &= -\frac{M_{T}}{2} \operatorname{tr} \left\{ (\bSigma^{\star})^{-1}\bD (\bSigma^{\star})^{-1}\bD \right\} + 2\lambda^{\star} \operatorname{tr} \left\{ \bD(\bSigma^{\star})^{-1}\bD \right\}.
    \label{eq:second_directional_stationary}
\end{align}
Write now $\bSigma^{\star}=\bQ\bLambda^{\star}\bQ^{\intercal}$, with $\bQ$ an orthonormal eigenbasis as above and $\bLambda^{\star}=
\diag(\sigma_1^{\star},\ldots,\sigma_L^{\star})$, where, for brevity, $\sigma_{l}^{\star}=\sigma_{l}(\lambda^{\star})$.
Calling $\widetilde{\bD} = \bQ^{\intercal}\bD\bQ = (\widetilde d_{jk})_{j,k=1}^{L}$, one gets
\begin{align*}
\operatorname{tr} \left\{ (\bSigma^{\star})^{-1}\bD (\bSigma^{\star})^{-1}\bD \right\}
&= \operatorname{tr} \left\{ (\bLambda^{\star})^{-1}\widetilde{\bD}(\bLambda^{\star})^{-1}\widetilde{\bD} \right\} = \sum_{l=1}^{L} \frac{\widetilde d_{ll}^{\,2}}{(\sigma_{l}^{\star})^{2}} + 2\sum_{1\leq j<k\leq L} \frac{\widetilde d_{jk}^{\,2}}{\sigma_{j}^{\star}\sigma_{k}^{\star}},\\
\operatorname{tr}\left\{ \bD(\bSigma^{\star})^{-1}\bD \right\} &= \operatorname{tr} \left\{ \widetilde{\bD} (\bLambda^{\star})^{-1} \widetilde{\bD} \right\} =
\sum_{l=1}^{L} \frac{\widetilde d_{ll}^{\,2}}{\sigma_{l}^{\star}} + \sum_{1\leq j<k\leq L} \left( \frac{1}{\sigma_{j}^{\star}} + \frac{1}{\sigma_{k}^{\star}} \right) \widetilde d_{jk}^{\,2}.
\end{align*}
Substituting in \eqref{eq:second_directional_stationary}, one obtains
\begin{align}
    \mathrm{D}_{\bD}^{2} \mathcal{L}(\bSigma^{\star},\lambda^{\star}) &= \sum_{l=1}^{L} \frac{ 4\lambda^{\star}\sigma_{l}^{\star}-M_{T} }{ 2(\sigma_{l}^{\star})^{2} } \widetilde d_{ll}^{\,2} + \sum_{1\leq j<k\leq L} \frac{ 2\lambda^{\star} (\sigma_{j}^{\star}+\sigma_{k}^{\star})-M_{T} }{ \sigma_{j}^{\star}\sigma_{k}^{\star} } \widetilde d_{jk}^{\,2}.
    \label{eq:second_directional_eigenbasis}
\end{align}
If $\lambda^{\star}=0$, \eqref{eq:second_directional_eigenbasis} directly gives
\begin{align}
\mathrm{D}_{\bD}^{2} \mathcal{L}(\bSigma^{\star},0) &= -\sum_{l=1}^{L} \frac{M_{T}}{2(\sigma_{l}^{\star})^{2}} \widetilde d_{ll}^{\,2} - \sum_{1\leq j<k\leq L} \frac{M_{T}}{\sigma_{j}^{\star}\sigma_{k}^{\star}} \widetilde d_{jk}^{\,2} <0
\end{align}
for every nonzero feasible perturbation $\bD$.
Indeed, since $\bQ$ is orthogonal, $\bD\neq\boldsymbol{0}$ if and only if $\widetilde{\bD}\neq\boldsymbol{0}$.
Consequently, for every nonzero symmetric perturbation $\bD$ satisfying $\operatorname{tr}(\bD)=0$, at least one element $\widetilde d_{jk}$ is nonzero, and hence $\mathrm{D}_{\bD}^{2} \mathcal{L}(\bSigma^{\star},0) <0$.

Suppose now that $\lambda^{\star}\neq0$ and that the all-minus solution is interior to the admissible domain.
From \eqref{eq:sigma_lambda}, one has
\begin{equation}
    4\lambda^{\star}\sigma_{l}^{\star}-M_{T} = -M_{T}\sqrt{ 1-\frac{8\lambda^{\star}}{M_{T}}v_{l} } <0, \qquad l=1,\ldots,L.
    \label{eq:negative_diagonal_coefficient}
\end{equation}
Similarly, for every $j\neq k$,
\begin{align}
    2\lambda^{\star} (\sigma_{j}^{\star}+\sigma_{k}^{\star})-M_{T} &= -\frac{M_{T}}{2} \left( \sqrt{ 1-\frac{8\lambda^{\star}}{M_{T}}v_{j} }+\sqrt{ 1-\frac{8\lambda^{\star}}{M_{T}}v_{k} } \right) <0.
    \label{eq:negative_offdiagonal_coefficient}
\end{align}

Thus, all the coefficients in \eqref{eq:second_directional_eigenbasis} are strictly negative and $\mathrm{D}_{\bD}^{2} \mathcal{L}(\bSigma^{\star},\lambda^{\star}) <0$ for every nonzero symmetric perturbation $\bD$ satisfying $\operatorname{tr}(\bD)=0$, since, for the same argument above, if $\bD\neq\boldsymbol{0}$ at least one element
$\widetilde d_{jk}$ is nonzero.
Therefore, the all-minus candidate satisfies the second-order
sufficient conditions and is a strict local maximizer of
$Q(\bSigma\mid\bSigma')$ over $\mathcal{S}_L$.

\paragraph{Choice of the stationary candidate and alternative branches.}

The preceding results provide a theoretical justification for the update used in our implementation.
Whenever the all-minus trace equation admits an interior root, its strict monotonicity guarantees that this root is unique within the all-minus branch, while the second-order sufficient conditions derived above guarantee that the corresponding $\bSigma^{\star}$ is a strict local maximizer of $Q(\bSigma\mid\bSigma')$ under the trace constraint.
The all-minus trace equation admitted an interior root at every iteration in all our numerical experiments, and the corresponding local maximizer was therefore used as the covariance update.

More generally, stationary candidates involving the ``$1+\sqrt{\cdot}$'' branch may also be considered, either as a fallback when the all-minus trace equation has no admissible root or as part of a more exhaustive search over the stationary points of the constrained problem.
The second-order necessary conditions imply that an interior local maximizer can contain at most one eigenvalue selected from the plus branch.
Indeed, suppose that two eigenvalues, indexed by $j$ and $k$, are selected from the plus branch.
A positive plus root can occur only when $\lambda^{\star}>0$, and is given by
\[
    \sigma_{l}^{+}(\lambda^{\star}) = \frac{M_{T}}{4\lambda^{\star}} \left( 1+ \sqrt{ 1-\frac{8\lambda^{\star}}{M_{T}}v_{l} } \right).
\]
For an interior solution, the corresponding discriminants are strictly positive, and therefore
\begin{align}
    2\lambda^{\star} \left\{ \sigma_{j}^{+}(\lambda^{\star}) + \sigma_{k}^{+}(\lambda^{\star}) \right\} -M_{T} &= \frac{M_{T}}{2} \left(\sqrt{1-\frac{8\lambda^{\star}}{M_{T}}v_{j} } + \sqrt{ 1-\frac{8\lambda^{\star}}{M_{T}}v_{k}}\right)>0.
    \label{eq:two_plus_positive_curvature}
\end{align}

Consider a symmetric perturbation $\bD$ whose representation $\widetilde \bD$ in the common eigenbasis of $\bSigma^{\star}$ and $\bS$ has $\widetilde d_{jk}=\widetilde d_{kj}\neq0$ and all remaining entries equal to zero.
This perturbation has zero trace since $\operatorname{tr}(\bD)=\operatorname{tr}(\bQ\widetilde\bD\bQ^{\intercal})=\operatorname{tr}(\bQ^{\intercal}\bQ\widetilde\bD)=\operatorname{tr}(\widetilde\bD)=0$ and therefore defines a feasible direction.
It follows from \eqref{eq:second_directional_eigenbasis} that
\begin{equation}
    \mathrm{D}_{\bD}^{2} \mathcal{L}(\bSigma^{\star},\lambda^{\star}) = \frac{ 2\lambda^{\star} (\sigma_{j}^{\star}+\sigma_{k}^{\star})-M_{T} }{ \sigma_{j}^{\star}\sigma_{k}^{\star}}\widetilde d_{jk}^{\,2} >0.
\end{equation}
This violates the second-order necessary condition for a local maximum, which requires the second directional derivative to be nonpositive along every feasible direction.
Hence, an interior local maximizer can contain at most one eigenvalue selected from the plus branch.
Consequently, in addition to the all-minus candidate, it is sufficient to consider the $L$ configurations in which exactly one eigenvalue uses the plus root and all remaining eigenvalues use the minus root.
For a fixed index $k$, the corresponding trace equation is
\begin{equation}
    \sigma_{k}^{+}(\lambda) + \sum_{l\neq k}\sigma_{l}^-(\lambda) = L.
    \label{eq:one_plus_trace_equation}
\end{equation}
Since $\sigma_{k}^{+}(\lambda)>0$ only for $\lambda\in(0,\lambda_{\max}]$, with $\lambda_{\max}=M_{T}/(8v_{\max})$ as in \eqref{eq:disc}, admissible roots of \eqref{eq:one_plus_trace_equation} must be sought in this interval.
Unlike the all-minus trace function, the left-hand side of \eqref{eq:one_plus_trace_equation} is not necessarily monotone.
Therefore, a generalized approximate EM update can be obtained by evaluating all the admissible candidate solutions described above and selecting the one with the largest value of $Q(\bSigma\mid\bSigma^{\prime})$.

\subsection[Approximate Gaussian smoothing]{Approximate Gaussian smoothing of $\bbeta_{t}$}
\label{subsec:S.4.2}
The maximization procedure developed in Section~\ref{subsec:S.4.1} requires the computation of 
$\bS=\sum_{t=1}^{T} \sum_{i=1}^{m_{t}}\bbE_{p(\bepsilon_{t,i}\mid \by_{1:T};\bSigma^{\prime})} \left [ \bepsilon_{t,i} \bepsilon_{t,i}^{\intercal} \right ]$.
A crucial quantity is then given by the smoothing distributions $p(\bepsilon_{t,i}\mid \by_{1:T})$, $t=1,\ldots,T$, $i=1,\ldots,m_{t}$.
An approximation of such a distribution is developed in Section~\ref{subsec:S.4.3} below, leveraging a distribution closely related to the smoothing distribution of $\bbeta_{t}$, for which a Gaussian approximation is derived in this section.
This Gaussian smoothing approximation is obtained starting from the Gaussian filtering approximations obtained with ADF or EP filters and going back in time, as explained below.

Specifically, denote with $\ba_{t}^{\dagger},\bP_{t}^{\dagger}$, the mean and variance-covariance matrix of the Gaussian approximation of the filtering distribution at time $t$, i.e., $p(\bbeta_{t}\mid \by_{1:t}) \approx q_{t}^{\dagger}(\bbeta_{t}) = \phi_{p}(\bbeta_{t} - \ba_{t}^{\dagger};\bP_{t}^{\dagger})$, where $\dagger$ is either $\adf$ or $\ep$.
See Sections \ref{subsec:ADF} and \ref{sec:EP} for details.
The smoothing calculations then proceed by treating the Gaussian filtering approximation $q_{t}^{\dagger}(\bbeta_{t})$ as the true filtering distribution and applying the Kalman smoothing procedure.
Recalling the state dynamics $\bbeta_{t+1} = \bG_{t+1} \bbeta_{t} + \boldeta_{t+1}$ with $\boldeta_{t+1}\sim \textsc{n}_{p}(\mathbf{0},\bW_{t+1})$, see \eqref{eq:trans_kern_MP}, we obtain the Gaussian approximation
\begin{equation*}
    p(\bbeta_{t:t+1} \mid \by_{1:t}) \approx \phi_{2p} \left( \begin{bmatrix} \bbeta_{t}\\
    \bbeta_{t+1}
    \end{bmatrix} -
    \begin{bmatrix}
    \ba_{t}^{\dagger}\\
    \ba_{t+1 \mid t}^{\dagger}
    \end{bmatrix};
    \begin{bmatrix}
    \bP_{t}^{\dagger} & \bP_{t}^{\dagger} \bG_{t+1}^{\intercal}\\
    \bG_{t+1} \bP_{t}^{\dagger} & \bP_{t+1 \mid t}^{\dagger}
\end{bmatrix} \right),
\end{equation*}
where $\ba_{t+1 \mid t}^{\dagger} = \bG_{t+1} \ba_{t}^{\dagger}$ and $\bP_{t+1 \mid t}^{\dagger}= \bG_{t+1} \bP_{t}^{\dagger} \bG_{t+1}^{\intercal} + \bW_{t+1}$.
Thus, by conditioning properties of multivariate Gaussian distributions, such an approximation entails
\begin{equation} \label{eq:inverse_dynamic}
    p(\bbeta_{t} \mid \bbeta_{t+1}, \by_{1:t}) \approx \phi_{p} \left( \bbeta_{t} -\ba_{t}^{\dagger} - \bJ_{t} (\bbeta_{t+1} - \ba_{t+1 \mid t}^{\dagger}); \bP_{t}^{\dagger} -\bJ_{t} \bG_{t+1} \bP_{t}^{\dagger} \right),
\end{equation}
where $\bJ_{t} = \bP_{t}^{\dagger} \bG_{t+1}^{\intercal} (\bP^{\dagger}_{t+1 \mid t})^{-1}$.
Hence, one can obtain approximate smoothing moments recursively as
\begin{align*}
    \bbE(\bbeta_{t} \mid \by_{1:T}) &= \bbE[\bbE( \bbeta_{t} \mid \bbeta_{t+1}, \by_{1:T}) \mid \by_{1:T}] = \bbE[\bbE(\bbeta_{t} \mid \bbeta_{t+1}, \by_{1:t}) \mid \by_{1:T}] \\
    &\approx \bbE[\ba_{t}^{\dagger} + \bJ_{t} (\bbeta_{t+1} - \ba_{t+1 \mid t}^{\dagger}) \mid \by_{1:T}] \approx \ba_{t}^{\dagger} + \bJ_{t} (\ba_{t+1 \mid T}^{\dagger} - \ba_{t+1 \mid t}^{\dagger}),
\end{align*}
where $\ba_{t+1 \mid T}^{\dagger} = \bbE(\bbeta_{t+1} \mid \by_{1:T})$, and
\begin{align*}
    \var(\bbeta_{t} \mid \by_{1:T}) &= \var[ \bbE(\bbeta_{t} \mid \by_{1:T}, \bbeta_{t+1}) \mid \by_{1:T} ] + \bbE[\var (\bbeta_{t} \mid \by_{1:T}, \bbeta_{t+1}) \mid \by_{1:T} ]\\
    &\approx \var[\ba_{t}^{\dagger} + \bJ_{t} (\bbeta_{t+1} - \ba_{t+1 \mid t}^{\dagger}) \mid \by_{1:T} ] + \bbE[ \bP_{t}^{\dagger} - \bJ_{t} \bG_{t+1} \bP_{t}^{\dagger} \mid \by_{1:T} ]\\
    &\approx \bJ_{t} \bP_{t+1 \mid T}^{\dagger} \bJ_{t}^{\intercal} + \bP_{t}^{\dagger} - \bJ_{t} \bG_{t+1} \bP_{t}^{\dagger},
\end{align*}
where $\bP_{t+1 \mid T}^{\dagger} = \var(\bbeta_{t+1} \mid \by_{1:T})$.
Thus, a natural Gaussian approximation of the smoothing is obtained by considering a Gaussian distribution with the moments as above, namely
\begin{align*}
    p(\bbeta_{t} \mid \by_{1:T}) \approx \phi_{p}(\bbeta_{t}-\ba_{t \mid T}^{\dagger}; \bP_{t \mid T}^{\dagger}),
\end{align*}
where $\ba_{t \mid T}^{\dagger}$ and $\bP_{t \mid T}^{\dagger}$ solve the recursive equations:
\begin{equation}
    \ba_{t \mid T}^{\dagger} = \ba_{t}^{\dagger} + \bJ_{t} (\ba_{t+1 \mid T}^{\dagger} - \ba_{t+1 \mid t}^{\dagger}) \text{ and } \bP_{t \mid T}^{\dagger} = \bJ_{t} \bP_{t+1 \mid T}^{\dagger} \bJ_{t}^{\intercal} + \bP_{t}^{\dagger} - \bJ_{t} \bG_{t+1} \bP_{t}^{\dagger},
\end{equation}
with terminal conditions $\ba_{T\mid T}^{\dagger}=\ba_{T}^{\dagger}$ and $\bP_{T\mid T}^{\dagger}=\bP_{T}^{\dagger}$.

\subsection[Computation of S]{Computation of $\bS$}
\label{subsec:S.4.3}
We finally derive an approximation for the smoothing distribution $p(\bepsilon_{t,i}\mid \by_{1:T})$, which is used to compute the corresponding second moments of $\bepsilon_{t,i}$, needed to compute the matrix $\bS$ in Section~\ref{subsec:S.4.1}.
It holds
\begin{equation}
\label{eq:smoothing_epsilon}
    \begin{split}
    p(\bepsilon_{t,i}\mid \by_{1:T}) \propto p(y_{t,i}\mid \bepsilon_{t,i}, \by_{1:T \setminus t,i})p(\bepsilon_{t,i}\mid \by_{1:T \setminus t,i}) = p(y_{t,i}\mid \bepsilon_{t,i}, \by_{1:T \setminus t,i})p(\bepsilon_{t,i}),
    \end{split}
\end{equation}
where $\by_{1:T \setminus t,i}$ is obtained removing $y_{t,i}$ from $\by_{1:T}$ and the last equality in \eqref{eq:smoothing_epsilon} follows from:
\begin{equation*}
    \begin{split}
        p(\bepsilon_{t,i}\mid \by_{1:T \setminus t,i}) &= \frac{p(\by_{1:T \setminus t,i}\mid \bepsilon_{t,i})p(\bepsilon_{t,i})}{p(\by_{1:T \setminus t,i})}\\
        &= p(\bepsilon_{t,i}) \frac{\int p(\by_{1:T \setminus t,i}\mid \bepsilon_{1:T\setminus t,i}, \bbeta_{1:T}, \bepsilon_{t,i})p(\bepsilon_{1:T\setminus t,i}, \bbeta_{1:T}) d \bepsilon_{1:T\setminus t,i} \, d \bbeta_{1:T}}{p(\by_{1:T \setminus t,i})}\\
        &= p(\bepsilon_{t,i}) \frac{\int p(\by_{1:T \setminus t,i}\mid \bepsilon_{1:T\setminus t,i}, \bbeta_{1:T})p(\bepsilon_{1:T\setminus t,i})p(\bbeta_{1:T}) d \bepsilon_{1:T\setminus t,i} \,
        d \bbeta_{1:T}}{p(\by_{1:T \setminus t,i})} =p(\bepsilon_{t,i}).
    \end{split}
\end{equation*}
Considering now $p(y_{t,i}\mid \bepsilon_{t,i}, \by_{1:T \setminus t,i})$, we can observe that:
\begin{equation}
\label{eq:pred_lik_y_ti}
    \begin{split}
        p(y_{t,i}\mid \bepsilon_{t,i}, \by_{1:T \setminus t,i}) &= \mathbb{P}(- \bV_{[y_{t,i}]} \bX_{t,i} \bbeta_{t} \leq \bV_{[y_{t,i}]} \bepsilon_{t,i} \ \mid \bepsilon_{t,i}, \by_{1:T \setminus t,i}) \\
        &= \int \mathbbm{1}(- \bV_{[y_{t,i}]} \bX_{t,i} \bbeta_{t} \leq \bV_{[y_{t,i}]} \bepsilon_{t,i} \ ) p(\bbeta_{t}\mid \bepsilon_{t,i}, \by_{1:T \setminus t,i}) d \bbeta_{t},
    \end{split} 
\end{equation}
where $\mathbbm{1}(\cdot)$ is the indicator function and $p(\bbeta_{t}\mid \bepsilon_{t,i},\by_{1:T \setminus t,i}) = p(\bbeta_{t}\mid \by_{1:T \setminus t,i})$ since
\begin{equation*}
    p(\bbeta_{t}\mid \bepsilon_{t,i},\by_{1:T \setminus t,i})= \frac{p(\by_{1:T \setminus t,i}\mid \bbeta_{t},\bepsilon_{t,i})p(\bbeta_{t}\mid \bepsilon_{t,i})}{p(\by_{1:T \setminus t,i}\mid \bepsilon_{t,i})} = \frac{p(\by_{1:T \setminus t,i}\mid \bbeta_{t})p(\bbeta_{t})}{p(\by_{1:T \setminus t,i})} = p(\bbeta_{t}\mid \by_{1:T \setminus t,i}).
\end{equation*}
Now, one can write
\begin{equation}
\label{eq:smoothing_beta_t_minus_ti}
    p(\bbeta_{t}\mid \by_{1:T \setminus t,i})\propto \dfrac{p(\bbeta_{t}\mid \by_{1:T})}{p(y_{t,i} \mid \bbeta_{t})}.
\end{equation}
In general, \eqref{eq:smoothing_beta_t_minus_ti} might be computationally demanding.
However, the EP approach approximates both the numerator and denominator with Gaussian kernels.
Indeed, in the EP framework, $p(\bbeta_{t}\mid \by_{1:T})\approx \phi_p(\bbeta_{t}- \ba_{t\mid T}^{\ep}; \bP_{t\mid T}^{\ep})$, with $\ba_{t\mid T}^{\ep}$ and $\bP_{t\mid T}^{\ep}$ as in Section~\ref{subsec:S.4.2}, while $p(y_{t,i} \mid \bbeta_{t})\approx q_{t,i}^{\ep} (\bbeta_{t}) \propto \exp\left\{ -\tfrac{1}{2}\bbeta_{t}^{\intercal} \bQ_{t,i}\bbeta_{t} + \bbeta_{t}^{\intercal} \br_{t,i} \right\}$, as in Section~\ref{sec:EP}.
One can substitute the above-mentioned EP approximations in \eqref{eq:smoothing_beta_t_minus_ti}.
Defining $\bQ_{t\mid T}^{\ep} = \left(\bP_{t\mid T}^{\ep}\right)^{-1}$ and $\br_{t\mid T}^{\ep} = \bQ_{t\mid T}^{\ep}\ba_{t\mid T}^{\ep}$, this entails the approximation
\begin{equation}
\label{eq:approx_smoothing_beta_t_minus_ti}
    p(\bbeta_{t}\mid \by_{1:T \setminus t,i})\approx \phi_p\left(\bbeta_{t} - \ba_{t\mid T \setminus t,i}^{\ep}; \bP_{t\mid T \setminus t,i}^{\ep} \right),
\end{equation}
with $\bP_{t\mid T \setminus t,i}^{\ep} = (\bQ_{t\mid T}^{\ep}-\bQ_{t,i})^{-1}$ and $\ba_{t\mid T \setminus t,i}^{\ep} = \bP_{t\mid T \setminus t,i}^{\ep}(\br_{t\mid T}^{\ep}-\br_{t,i})$.
It thus follows from \eqref{eq:pred_lik_y_ti}, that
\begin{equation}
\label{eq:approx_pred_lik_y_ti}
    \begin{split}
        p(y_{t,i}\mid \bepsilon_{t,i}, \by_{1:T \setminus t,i}) &\approx \int \mathbbm{1}(- \bV_{[y_{t,i}]} \bX_{t,i} \bbeta_{t} \leq \bV_{[y_{t,i}]} \bepsilon_{t,i} \ ) \phi_p\left(\bbeta_{t} - \ba_{t\mid T \setminus t,i}^{\ep}; \bP_{t\mid T \setminus t,i}^{\ep} \right) d \bbeta_{t}\\
        &= \Phi_L( \bV_{[y_{t,i}]} \bepsilon_{t,i} +\bV_{[y_{t,i}]} \bX_{t,i} \ba_{t\mid T \setminus t,i}^{\ep}; \bV_{[y_{t,i}]}\bX_{t,i} \bP_{t\mid T \setminus t,i}^{\ep} \bX_{t,i}^{\intercal} \bV_{[y_{t,i}]}^{\intercal}).
    \end{split} 
\end{equation}
Replacing this in \eqref{eq:smoothing_epsilon}, one gets the following result, whose proof follows from Lemma \ref{le: sun from gaussian x cdf}.
\begin{Prop}
\label{prop:smoothing_eps}
    Assume that $\bX_{t,i}$ has full row rank $L$.
    If the approximation \eqref{eq:approx_pred_lik_y_ti} of $p(y_{t,i}\mid \bepsilon_{t,i},\by_{1:T\setminus t,i})$ is used in \eqref{eq:smoothing_epsilon}, the resulting approximation of $p(\bepsilon_{t,i}\mid \by_{1:T})$ satisfies
    \begin{equation*}
    p(\bepsilon_{t,i}\mid\by_{1:T}) \approx \mathrm{d}\sun_{L,L} \left( \bepsilon_{t,i}; \bxi_{t,i\mid T}^{\epsilon}, \bOmega_{t,i\mid T}^{\epsilon}, \bDelta_{t,i\mid T}^{\epsilon}, \bgamma_{t,i\mid T}^{\epsilon}, \bGamma_{t,i\mid T}^{\epsilon} \right).
    \end{equation*}
    where:
    \begin{align}
        & \bxi_{t,i\mid T}^{\epsilon}=\bzero_L, \quad \bOmega_{t,i\mid T}^{\epsilon} = {\bSigma}, \quad \bDelta_{t,i\mid T}^{\epsilon} = \bar{\bOmega}_{t,i\mid T}^{\epsilon} \bomega_{t,i\mid T}^{\epsilon} \bV_{[y_{t,i}]}^{\intercal} (\bs_{t,i\mid T}^{\epsilon})^{-1},\\
        &\bgamma_{t,i\mid T}^{\epsilon} = (\bs_{t,i\mid T}^{\epsilon})^{-1} \bV_{[y_{t,i}]} \bX_{t,i} \ba_{t\mid T \setminus t,i}^{\ep},\\
        & \bGamma_{t,i\mid T}^{\epsilon} = (\bs_{t,i\mid T}^{\epsilon})^{-1} \left[ \bV_{[y_{t,i}]}\left( \bSigma + \bX_{t,i} \bP_{t\mid T \setminus t,i}^{\ep} \bX_{t,i}^{\intercal} \right) \bV_{[y_{t,i}]}^{\intercal} \right] (\bs_{t,i\mid T}^{\epsilon})^{-1},
    \end{align}
    having defined
    \begin{align}
        &\bomega_{t,i\mid T}^{\epsilon} = (\bOmega_{t,i\mid T}^{\epsilon} \odot \bI_{L})^{1/2}, \quad \bar{\bOmega}_{t,i\mid T}^{\epsilon} = (\bomega_{t,i\mid T}^{\epsilon})^{-1} \bOmega_{t,i\mid T}^{\epsilon} (\bomega_{t,i\mid T}^{\epsilon})^{-1}\\
        &\bs_{t,i\mid T}^{\epsilon} = \left[ \bV_{[y_{t,i}]}\left( \bSigma + \bX_{t,i} \bP_{t\mid T \setminus t,i}^{\ep} \bX_{t,i}^{\intercal} \right) \bV_{[y_{t,i}]}^{\intercal} \odot\bI_{L} \right]^{1/2}.
    \end{align}
\end{Prop}

One can then use this result to compute $\bbE_{p(\bepsilon_{t,i}\mid \by_{1:T};\bSigma^{\prime})} \left [ \bepsilon_{t,i} \bepsilon_{t,i}^{\intercal} \right ]$, e.g.\ via Monte Carlo sampling, in the M-step, where $\bSigma$ is fixed to $\bSigma'$.
Notice that EP is particularly convenient to approximate $p(\bbeta_{t}\mid \by_{1:T \setminus t,i})$, due to availability of Gaussian approximations for both the numerator and denominator in \eqref{eq:smoothing_beta_t_minus_ti}, whose ratio is again a Gaussian density.
In case the ADF filter is considered, one could instead consider the approximation $p(\bbeta_{t}\mid\by_{1:T\setminus t,i})\approx\phi_p\left(\bbeta_{t}-\ba_{t\mid T}^{\adf};\bP_{t\mid T}^{\adf}\right)$, with $\ba_{t\mid T}^{\adf}$ and $\bP_{t\mid T}^{\adf}$ as in Section~\ref{subsec:S.4.2}, under the assumption that the inclusion of the single observation $y_{t,i}$ does not have a strong impact on the smoothing distribution of $\bbeta_{t}$.
If such approximation is used, the resulting approximation for $p(\bepsilon_{t,i}\mid\by_{1:T})$ would be analogous to the one presented in Proposition \ref{prop:smoothing_eps}, with $\ba_{t\mid T\setminus t,i}^{\ep}$ and $\bP_{t\mid T\setminus t,i}^{\ep}$ replaced by $\ba_{t\mid T}^{\adf}$ and $\bP_{t\mid T}^{\adf}$, respectively.

\section{Supplementary Information on the Experiments}
\label{sec:app_exper}

\subsection{Synthetic-Data Generation and Implementation Details}
\label{sec:app_simulation_design}

Both simulation studies in Sections~\ref{sec:exp_filtering} and \ref{sec:EM_constrained} use $L=3$ nonbaseline alternatives and a $p=12$ dimensional state $\bbeta_{t}$.
For each observation, we consider two individual-specific covariates whose coefficients are allowed to vary across the three alternatives, three alternative-specific covariates with coefficients shared across alternatives, and one intercept for each alternative.
Let $\bu_{t,i}\in\bbR^{2}$ denote the individual-specific covariates and $\bw_{t,i,\ell}\in\bbR^{3}$ the covariates associated with alternative $\ell=1,2,3$.
The resulting design matrix is
\[
\bX_{t,i} = \begin{bmatrix}
\bu_{t,i}^{\intercal} & \bzero_2^{\intercal} & \bzero_2^{\intercal} & \bw_{t,i,1}^{\intercal} & 1 & 0 & 0\\
\bzero_2^{\intercal} & \bu_{t,i}^{\intercal} & \bzero_2^{\intercal} & \bw_{t,i,2}^{\intercal} & 0 & 1 & 0\\
\bzero_2^{\intercal} & \bzero_2^{\intercal} & \bu_{t,i}^{\intercal} & \bw_{t,i,3}^{\intercal} & 0 & 0 & 1
\end{bmatrix},
\]
so that $\bX_{t,i}\in\bbR^{3\times 12}$.
Both individual-specific covariate vectors and alternative-specific covariate vectors are generated independently from $\Unif(-0.5,0.5)$.
At each time point, the sample size is generated from a $\Poi(\lambda_m)$ distribution censored to $[2,4\lambda_m]$, which avoids degenerate sample sizes of $0$ or $1$ and excludes implausibly large values.

For each scenario, the resulting covariate arrays and realized sample sizes are held fixed across replications.
Conditional on this common design and data-generating truth, the complete latent state trajectory and all responses are independently regenerated in every replication using the state parameters and $\bSigma$ reported in Section~\ref{sec:6}.
This design incorporates sampling variability in the latent states and responses while maintaining paired comparisons among filtering methods within each replication.

\subsection{Filtering Accuracy and Computational Performance}
\label{sec:app_filtering}

For the exact i.i.d.\ sampler, $10,000$ draws are generated from each filtering distribution.
In ADF and EP, $10,000$ Monte Carlo draws are used to estimate the moments of the relevant SUN distributions, while SMC uses $10,000$ particles.
For EP, each time-point correction stops when $| \widehat\ell_{t}^{(k)} - \widehat\ell_{t}^{(k-1)}| / | \widehat\ell_{t}^{(k-1)}| <10^{-3}$, where $\widehat\ell_{t}^{(k)}$ is the EP approximation to the log marginal likelihood contribution after sweep $k$.

The $48$ replications are distributed across $48$ cluster workers, while no parallelization is nested within an individual filtering run.
Consequently, the elapsed times in Table~\ref{tab:rt_a} compare the methods under the same per-replication computational allocation.
For the two regimes in which exact filtering is unavailable, ADF is used as a common feasible reference.

\begin{figure}[t]
\centering
\includegraphics[width=0.82\linewidth]{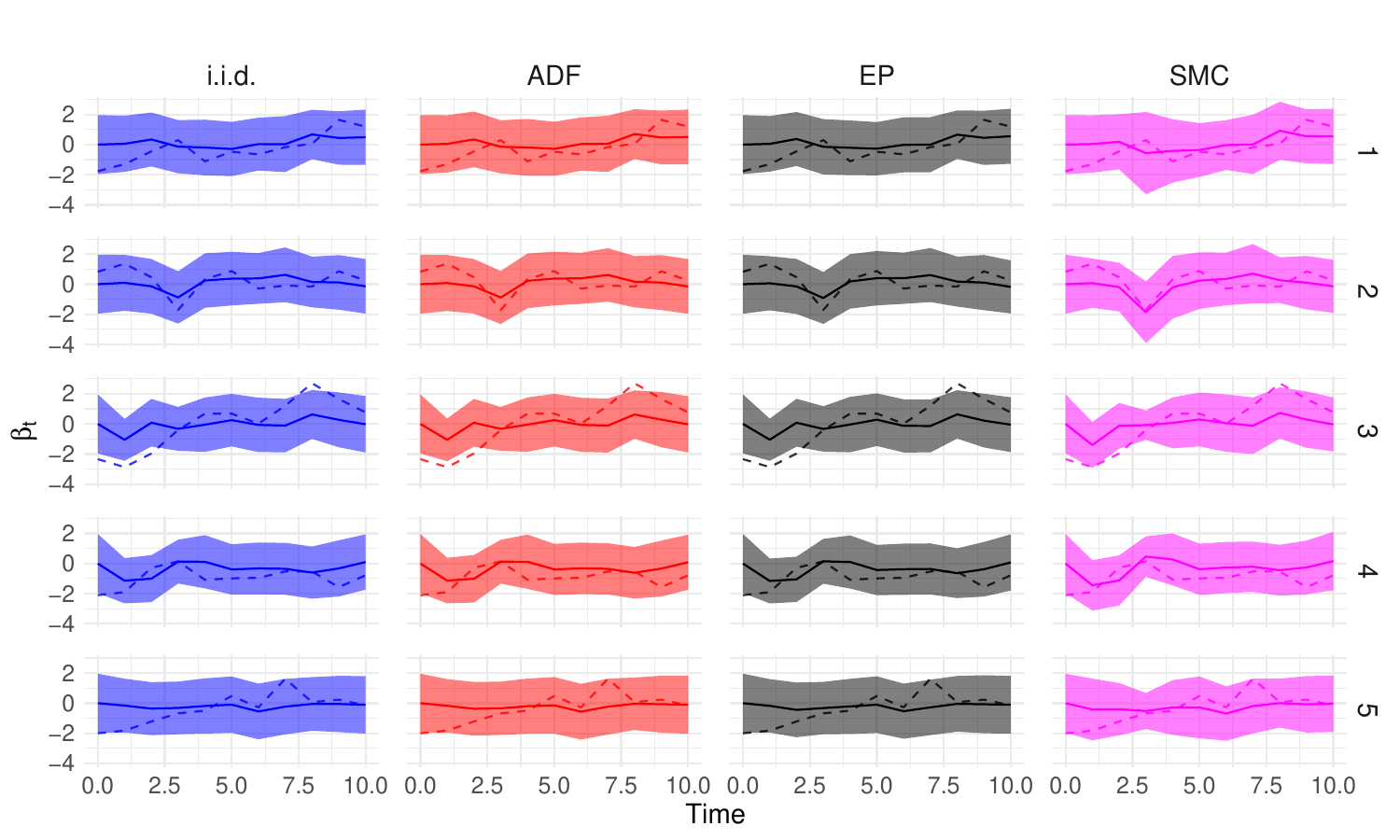}
\caption{Filtering inference for a representative dataset with $(T,\lambda_m)=(10,10)$.
Columns correspond to the exact i.i.d.\ filter, ADF, EP, and SMC, and rows correspond to the first five state coordinates.
Solid curves are filtering means, shaded regions are filtering means $\pm 1.96$ posterior standard deviations, and dashed curves are the simulated latent states.}
\label{fig:traj_T10_m10}
\end{figure}

\begin{figure}[t]
\centering
\includegraphics[width=0.82\linewidth]{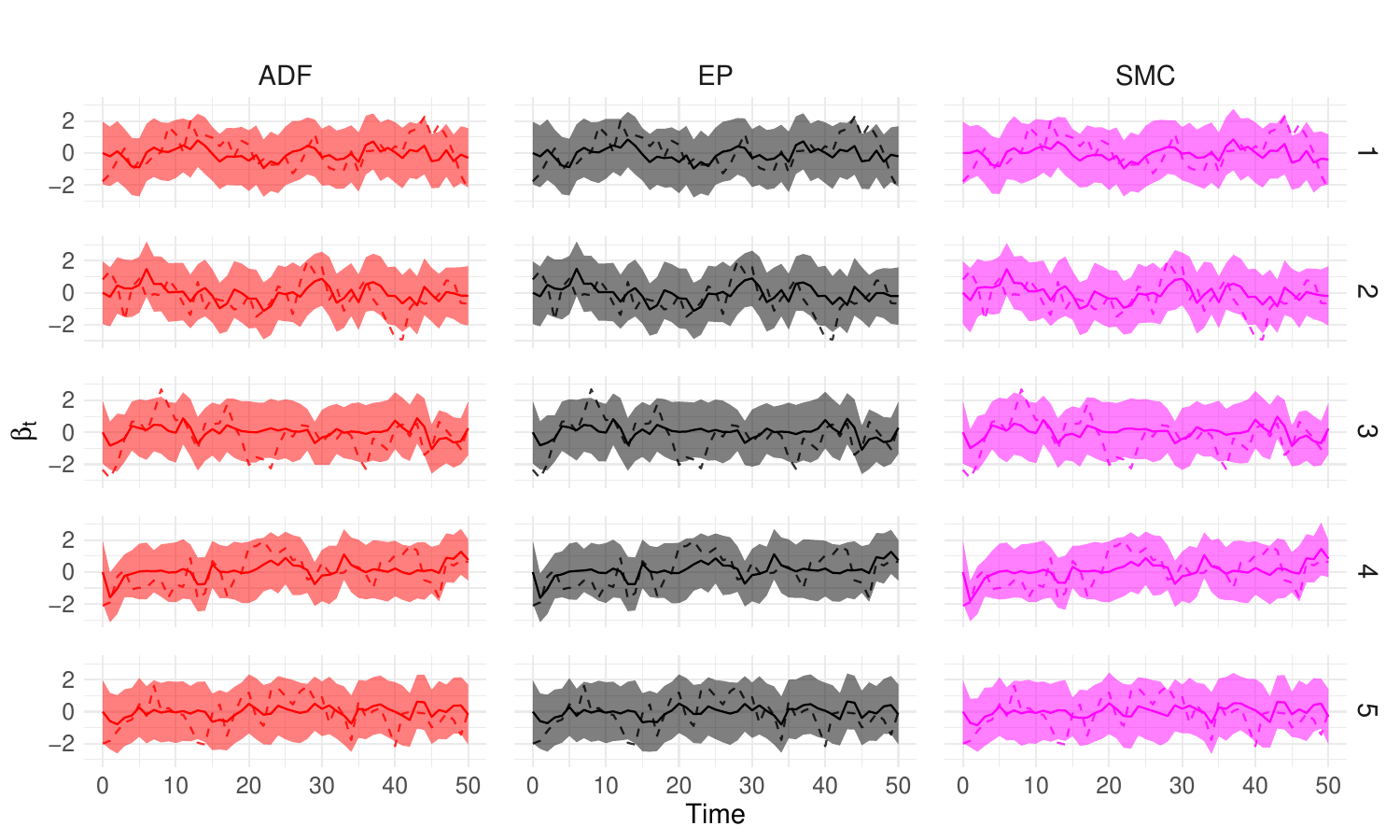}
\caption{Filtering inference for a representative dataset with $(T,\lambda_m)=(50,10)$.
Columns correspond to ADF, EP, and SMC, and rows correspond to the first five state coordinates.
Solid curves are filtering means, shaded regions are filtering means $\pm 1.96$ posterior standard deviations, and dashed curves are the simulated latent states.}
\label{fig:traj_T50_m10}
\end{figure}

\begin{figure}[t]
\centering
\includegraphics[width=0.82\linewidth]{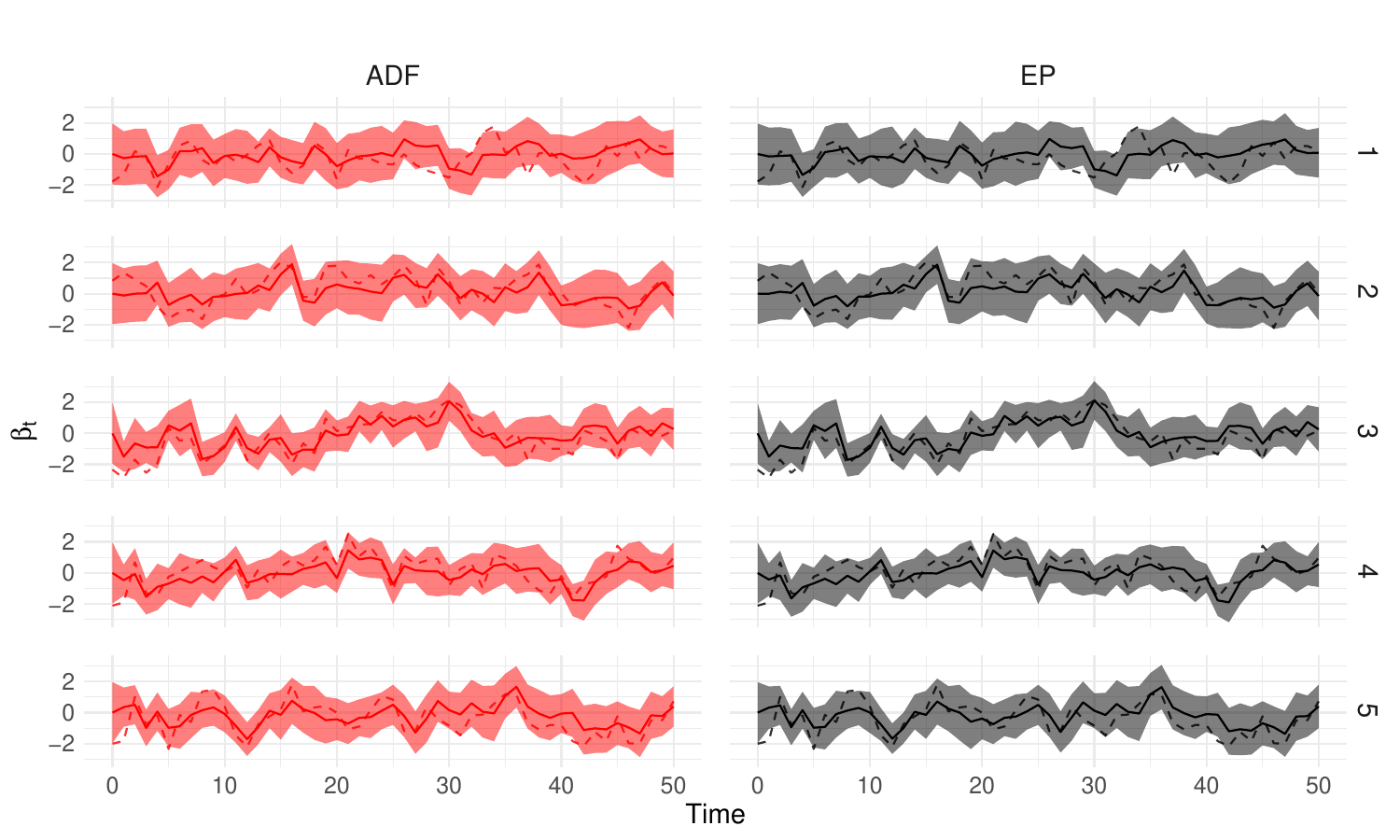}
\caption{Filtering inference for a representative dataset with $(T,\lambda_m)=(50,50)$.
Columns correspond to ADF and EP, and rows correspond to the first five state coordinates.
Solid curves are filtering means, shaded regions are filtering means $\pm 1.96$ posterior standard deviations, and dashed curves are the simulated latent states.}
\label{fig:traj_T50_m50}
\end{figure}

Figures~\ref{fig:traj_T10_m10}--\ref{fig:traj_T50_m50} complement the aggregate results in Table~\ref{tab:rt_a} by displaying the first five state coordinates for one representative dataset from each regime.
The methods yield practically indistinguishable filtering inference across the three regimes.

\FloatBarrier

\subsection{EM Convergence in the Covariance-Recovery Experiment}
\label{sec:app_em_convergence}

\begin{figure}[t]
\centering
\includegraphics[width=\textwidth]{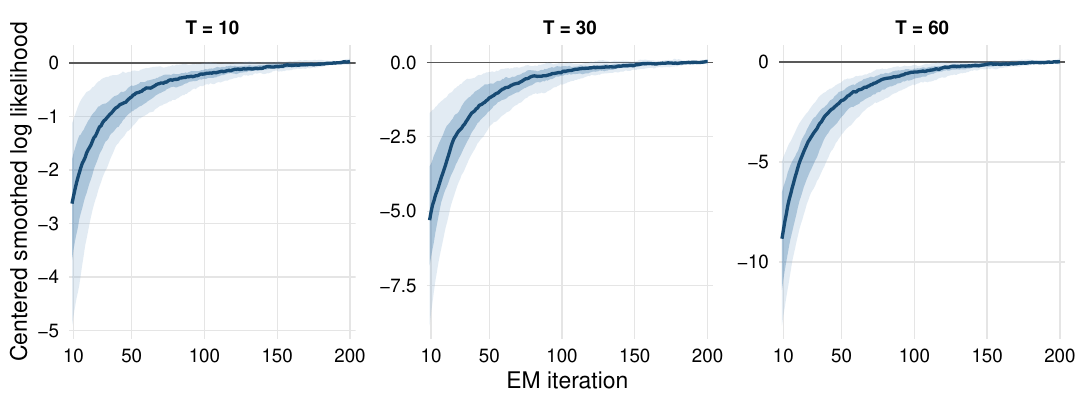}
\caption{Convergence diagnostic for the EM covariance-recovery experiment.
The solid curves are median centered log marginal likelihood histories across the $48$ replications.
Dark and light bands are the interquartile and $5$--$95\%$ ranges, respectively.
Histories are smoothed using a $10$-iteration trailing average and centered at their replication-specific means over the final $20$ iterations.}
\label{fig:em-recovery-convergence}
\end{figure}

At each EM iteration, the approximate E-step is implemented using the sequential EP filter, with $10,000$ Monte Carlo draws for estimating the moments of the EP hybrid distributions and a convergence tolerance of $10^{-3}$.
The conditional second moments of the latent utility errors required to construct $\bS$ are estimated separately using $10,000$ Monte Carlo draws.
The resulting estimate of $\bS$ is then used in the constrained M-step to update $\bSigma$.
Because the $48$ replications are based on independently simulated latent trajectories and responses, their log marginal likelihood levels are not directly comparable.
For each replication, we therefore smooth the log marginal likelihood history using a $10$-iteration trailing average and center it at its mean over the final $20$ iterations.
Figure~\ref{fig:em-recovery-convergence} summarizes the centered trajectories across replications: the plateau near zero suggests convergence of the procedure and thus support the use of the estimates obtained after $200$ EM iterations.

\printbibliography

\end{refsection}

\end{document}